\documentclass[preprint,3p,times]{elsarticle}
\RequirePackage{multirow,booktabs,subfigure,color,array,hhline,makecell}
\usepackage{amssymb}
\usepackage{amsmath}
\usepackage{graphicx}
\usepackage{subfigure}
\usepackage{amsthm}
\usepackage{mathrsfs}
\usepackage{indentfirst}
\usepackage[colorlinks,citecolor=blue,urlcolor=blue]{hyperref}
\allowdisplaybreaks
\usepackage[table]{xcolor}
\usepackage{tikz-network}
\usepackage{pgf}
\usepackage{tikz}
\usepackage{subfigure}
\usetikzlibrary{arrows, decorations.pathmorphing, backgrounds, positioning, fit, petri, automata}
\usetikzlibrary{shadows,arrows,positioning}
\RequirePackage{algorithm,algpseudocode}
\allowdisplaybreaks
\usepackage{changes}

\newtheorem{thm}{Theorem}
\newtheorem{defin}{Definition}

\newtheorem{assum}{Assumption}
\newtheorem{rem}{Remark}
\newtheorem{cor}{Corollary}
\newtheorem{con}{Condition}

\newtheorem{Ex}{Example}
\allowdisplaybreaks[4]
\AtBeginDocument{%
	\providecommand\BibTeX{{%
			\normalfont B\kern-0.5em{\scshape i\kern-0.25em b}\kern-0.8em\TeX}}}

\journal{Expert Systems with Applications}
\begin{document}
\begin{frontmatter}
\title{Bipartite Mixed Membership Distribution-Free Model. A novel model for community detection in overlapping bipartite weighted networks}

\author[label1]{Huan Qing\corref{cor1}}
\ead{qinghuan@u.nus.edu}

\author[label2]{Jingli Wang}
\ead{jlwang@nankai.edu.cn}
\cortext[cor1]{Corresponding author.}
\address[label1]{School of Mathematics, China University of Mining and Technology, Xuzhou, 221116, Jiangsu, China}
\address[label2]{School of Statistics and Data Science, KLMDASR, LEBPS, and LPMC, Nankai University, Tianjin, 300071, Tianjin, China}
\begin{abstract}
Modeling and estimating mixed memberships for overlapping unipartite un-weighted networks has been well studied in recent years. However, to our knowledge, there is no model for a more general case, the overlapping bipartite weighted networks. To close this gap, we introduce a novel model, the Bipartite Mixed Membership Distribution-Free (BiMMDF) model. Our model allows an adjacency matrix to follow any distribution as long as its expectation has a block structure related to node membership. In particular, BiMMDF can model overlapping bipartite signed networks and it is an extension of many previous models, including the popular mixed membership stochastic blcokmodels. An efficient algorithm with a theoretical guarantee of consistent estimation is applied to fit BiMMDF. We then obtain the separation conditions of BiMMDF for different distributions. Furthermore, we also consider missing edges for sparse networks. The advantage of BiMMDF is demonstrated in extensive synthetic networks and eight real-world networks.
\end{abstract}


\end{frontmatter}
\section{Introduction}\label{sec1}
Complex networks are ubiquitous in our daily life \citep{2002Food,palla2007quantifying,barabasi2004network,guimera2005functional,rubinov2010complex}. A network is composed of a set of nodes and edges which represent the relationship between nodes. Many real-world networks of interest may be described by bipartite graphs, and they are bipartite networks or two-mode networks \citep{borgatti1997network,latapy2008basic}. In a bipartite network, nodes are decomposed into two disjoint sets such that edges can only connect nodes from different sets. Let us cite some bipartite networks for examples. In the actors-movies network \citep{watts1998collective, latapy2008basic}, each actor is linked to the movies he/she played in; in the author-paper network \citep{newman2001structure,newman2001scientific}, each author is linked to the paper he/she signed; in the country-language network \cite{kunegis2013konect}, each country is linked to the language it hosts and edge weight denotes the proportion of the population of a given country speaking a given language; in the character-work network \cite{alberich2002marvel}, each character is linked to the movie he/she appears in; in the user-movie network \cite{guo2014etaf}, movies are rated by users; in the user-item network \cite{wang2010latent}, individual items are rated by users. More examples of bipartite networks can be found in \cite{latapy2008basic}. In the above bipartite networks, a node may belong to multiple clusters rather than a single cluster. For example, in the actors-movies network, Jackie Chan can be classified as an action \& comedy actor and a movie like ``Rush Hour" belongs to action and comedy. Such overlapping membership is common in real-world networks \citep{palla2005uncovering,MMSB,mao2020estimating}. This paper focuses on inferring each node's community membership to have a better understanding of the community structure of overlapping bipartite weighted networks.

Community detection is one of the most powerful tools for learning the latent structure of complex networks. The main goal of community detection is to find group of nodes \citep{fortunato2010community, fortunato2016community}. To solve the community detection problem, researchers usually follow three steps for model-based methods. In the first step, a null statistical model is used to generate networks with community structure \citep{goldenberg2010survey,jin2021survey}. In the second step, an algorithm is designed to fit the model and it is expected to infer communities satisfactorily for networks generated from the model. In the third step, the algorithm is applied to detect communities for real-world networks. For different types of networks, different statistical models should be proposed and so are the algorithms. Generally speaking, for static networks considered in this article, there are four cases: un-directed un-weighted networks, bipartite un-weighted networks, un-directed weighted networks, and bipartite weighted networks, where the first three cases are special cases of bipartite weighted networks. Note that directed networks \cite{malliaros2013clustering} can be regarded as a special case of bipartite networks since directed networks are unipartite or one-mode \citep{zhang2022identifiability,DISIM}. To have a better knowledge of statistical models for these cases, we will briefly introduce several representative statistical models for each case.

For community detection of un-directed un-weighted networks, it has been widely studied for decades \citep{fortunato2010community,papadopoulos2012community,fortunato2016community,javed2018community,jin2021survey,fortunato202220}. The Stochastic Blockmodel (SBM) \citep{SBM} is one of the most popular generative models to describe community structure for such networks. SBM assumes that the probability of a link between two nodes is determined by node communities. Though SBM is mathematically simple and easy to analyze, it performs poorly on real-world networks because it assumes that nodes in the same community have the same expected degree. The Degree-Corrected Stochastic Blockmodels (DCSBM) \citep{DCSBM} was proposed to address this limitation by considering the node heterogeneity parameters. To detect communities of networks generated from SBM and DCSBM, substantial works have been proposed in recent years, including maximizing the likelihood methods \citep{DCSBM,bickel2009nonparametric,zhao2012consistency}, low-rank approximation method \citep{le2016optimization}, convexified modularity maximization method \citep{chen2018convexified}, and spectral clustering algorithms \citep{rohe2011spectral,SCORE,lei2015consistency,joseph2016impact}. For a comprehensive review of recent developments about SBM, see \cite{abbe2017community}. One main limitation of SBM and DCSBM is that each node only belongs to a sole community. The classical Mixed Membership Stochastic Blockmodels (MMSB) \citep{MMSB} was proposed as an extension of SBM by allowing nodes to belong to multiple communities. The Degree-Corrected Mixed Membership (DCMM) model \citep{mixedSCORE} extends DCSBM from non-overlapping networks to overlapping networks. The Overlapping Continuous Community Assignment Model (OCCAM) \citep{OCCAM}, the Stochastic Blockmodel with Overlaps (SBMO) \citep{kaufmann2018spectral}, and the Overlapping Stochastic Block Models (OSBM) \citep{latouche2011overlapping} can also model overlapping networks. To estimate mixed memberships of networks generated from MMSB and DCMM, some methods are proposed, including MCMC \citep{MMSB},  variational approximation method \citep{Gopalan2013eff}, nonnegative matrix factorization inference methods \citep{ball2011efficient,psorakis2011overlapping,wang2011community},  tensor-based method \citep{anandkumar2013tensor}, and spectral methods \citep{mixedSCORE,MaoSVM,OCCAM,mao2020estimating}. For a general review on overlapping community detection, see \cite{xie2013overlapping}.

For community detection of bipartite un-weighted networks, \cite{DISIM} proposed Stochastic co-Blockmodel (ScBM) and Degree-Corrected Stochastic co-Blockmodel (DCScBM). ScBM and DCScBM can be seen as direct extensions of SBM and DCSBM from un-directed un-weighted networks to bipartite un-weighted networks, respectively. Spectral algorithms with theoretical guarantees on consistent estimation have been designed to estimate groups of nodes under ScBM and DCScBM, see algorithms proposed in \cite{DISIM,zhou2019analysis,DSCORE}.  Similar to SBM, ScBM also can not model overlapping networks. The Directed Mixed Membership Stochastic Blockmodels (DiMMSB) \citep{qing2021DiMMSB} was proposed to address this limitation by allowing nodes to belong to multiple communities. DiMMSB can be seen as a direct extension of MMSB from un-directed un-weighted networks to bipartite un-weighted networks. To estimate memberships under DiMMSB, \cite{qing2021DiMMSB} designed a spectral algorithm with a theoretical guarantee of estimation consistency.

For community detection of un-directed weighted networks, it has been an appealing topic in recent years. Edge weights are important and meaningful in a network since they can improve community detection \citep{newman2004analysis,barrat2004the}. \cite{newman2004analysis} studied a weighted network in which edge weights are nonnegative integers. To study a weighted network in which edge weights are more than nonnegative integers, many models extend SBM from un-directed un-weighted networks to un-directed weighted networks. Some Weighted Stochastic Blockmodels (WSBM) are developed in recent years \citep{aicher2015learning,ahn2018hypergraph,palowitch2017significance,xu2020optimal,ng2021weighted}. However, these WSBMs always assume that each node only belongs to one single community. To model overlapping un-directed weighted networks, the Weighted version of the MMSB (WMMSB) model \citep{dulac2020mixed} was proposed as an extension of MMSB by allowing edge weights to come from Poisson distribution.

For community detection of bipartite weighted networks, the Bipartite Distribution-Free models (BiDFM) and its extension BiDCDFM were proposed by \cite{BiDFMs} to model non-overlapping bipartite weighted networks. The two-way blockmodels \citep{airoldi2013multi} can model overlapping bipartite weighted networks. However, one main limitation of the two-way blockmodels is that edge weights are limited to follow Normal or Bernoulli distribution which causes the two-way blockmodels can not to model some real-world bipartite weighted networks. For example, in the country-language network \citep{kunegis2013konect}, edge weight ranges in $[0,1]$; in the user-movie network \citep{guo2014etaf} and user-item network \citep{wang2010latent}, edge weight ranges in $\{0, 1, 2, 3, 4, 5\}$; in a bipartite signed network, edge weights ranges in $\{0, 1, -1\}$ \citep{tang2016survey}. Such edge weights can not be generated from Normal or Bernoulli distributions. All aforementioned models fail to handle overlapping bipartite weighted networks in which edge weights can be any finite value and nodes can belong to multiple communities. Meanwhile, though the variational expectation-maximization (vEM) algorithm \citep{airoldi2013multi} was proposed to estimate community memberships for overlapping bipartite networks generated from the two-way blockmodels, it does not have any guarantees of consistency. In this article, our goal is to close these gaps and build a general model with a theoretical guarantee for overlapping bipartite weighted networks.

Our main contributions are summarized as follows:
\begin{itemize}
   \item [a)] We propose a novel model for overlapping bipartite weighted networks, the Bipartite Mixed Membership Distribution-Free (BiMMDF for short) model. BiMMDF allows elements of the adjacency matrix to follow any distribution as long as the expectation adjacency matrix has a block structure. MMSB and the two-way blockmodels are sub-models of BiMMDF. Overlapping bipartite signed networks can also be modeled by BiMMDF. To the best of our knowledge, our BiMMDF is the first model for overlapping bipartite weighted networks in which edge weights can be generated from any distribution.
   \item [b)] We use an efficient spectral algorithm to estimate node memberships for overlapping bipartite weighted networks generated from BiMMDF. Theoretically, we show that the algorithm is asymptotically consistent under BiMMDF. We also derive the separation conditions of BiMMDF for different distributions. To our knowledge, we are the first to reveal the difference in separation conditions for different distributions.
   \item [c)] To model real-world large-scale bipartite weighted networks in which many nodes have no connections, we propose a strategy by combining BiMMDF with a model for bipartite un-weighted networks to generate adjacency matrices with missing edges.
   \item [d)] We conduct substantial simulated networks to verify our theoretical results. Our experiments on eight real-world networks demonstrate the effectiveness of our model in detecting and understanding community structure.
\end{itemize}
The rest of the paper is organized as follows. Section \ref{sec2} introduces the model. Section \ref{sec3} introduces the algorithm. Section \ref{sec4} shows the consistency of the algorithm and provides some examples for different distributions. Section \ref{sec5} introduces the strategy to generate missing edges. Section \ref{sec6} conducts extensive experiments. Section \ref{sec7} concludes.
\section{The Bipartite Mixed Membership Distribution-Free model}\label{sec2}
\begin{table}[ht]
\centering
\scriptsize
\rowcolors{1}{gray!22}{gray!22}
\resizebox{\columnwidth}{!}{
\begin{tabular}{cc|cc}
\hline
Symbol& Description&Symbol& Description\\
\hline
$\mathbb{N}$&Set of nonnegative integers&$\mathbb{N}_{+}$&Set of positive integers\\
$\mathbb{R}$&Set of real numbers&$\mathbb{R}_{+}$&Set of nonnegative real numbers\\
$\mathcal{N}$&Bipartite weighted network&$\mathcal{V}_{r}$&Set of row nodes\\
$\mathcal{V}_{c}$&Set of column nodes&$P\in\mathbb{R}^{K\times K}$& Block matrix\\
$K$&Number of row (column) communities&$\|x\|_{q}$&$\ell_{q}$-norm for vector $x$\\
$n_{r}$&Number of row nodes&$M'$& Transpose of matrix $M$\\
$n_{c}$&Number of column nodes&$\mathbb{E}[M]$&Expectation of $M$\\
$A\in\mathbb{R}^{n_{r}\times n_{c}}$& $\mathcal{N}$'s adjacency matrix&$\|M\|_{2\rightarrow\infty}$&Maximum $\ell_{2}$-norm of $M$\\
$\Pi_{r}\in[0,1]^{n_{r}\times K}$&Membership matrix of row nodes&$M(i,:)$&$i$-th row of $M$\\
$\Pi_{c}\in[0,1]^{n_{c}\times K}$&Membership matrix of column nodes&$M(:,j)$&$j$-th column of $M$\\
$[m]$&$\{1,2,\ldots,m\}$ for positive integer $m$&$M(S_{r},:)$&Rows in the index set $S_{r}$ of $M$\\
$\rho$&Scaling parameter&$\mathrm{rank}(M)$&Rank of $M$\\
$\sigma_{k}(M)$&$k$-th largest singular value of $M$&$\lambda_{k}(M)$&$M$'s $k$-th largest eigenvalue in magnitude\\
$\mathcal{F}$&Distribution&$\mathcal{W}$&Set of edge weights\\
$\Omega\in\mathbb{R}^{n_{r}\times n_{c}}$&$A$'s expectation matrix $\rho\Pi_{r}P\Pi'_{c}$&$\kappa(M)$&Condition number of $M$\\
$U\in\mathbb{R}^{n_{r}\times K}$&Top $K$ left singular vectors of $\Omega$&$\mathcal{P}_{r}\in\{0,1\}^{K\times K}$&Permutation matrix\\
$V\in\mathbb{R}^{n_{c}\times K}$&Top $K$ right singular vectors of $\Omega$&$\mathcal{P}_{c}\in\{0,1\}^{K\times K}$&Permutation matrix\\
$\hat{U}\in\mathbb{R}^{n_{r}\times K}$&Top $K$ left singular vectors of $A$&$|a|$&Absolute value for real value $a$\\
$\hat{V}\in\mathbb{R}^{n_{c}\times K}$&Top $K$ right singular vectors of $A$&$e_{i}$&$e_{i}(j)=1(i=j)$\\
$\Lambda\in\mathbb{R}_{+}^{K\times K}$&Diagonal matrix of top $K$ singular values of $\Omega$&$\alpha_{\mathrm{in}}$&$\rho P(1,1)=\alpha_{\mathrm{in}}\frac{\mathrm{log}(n_{r})}{n_{r}}$ when $K\geq2$\\
$\hat{\Lambda}\in\mathbb{R}_{+}^{K\times K}$&Diagonal matrix of top $K$ singular values of $A$&$\alpha_{\mathrm{out}}$&$\rho P(1,2)=\alpha_{\mathrm{out}}\frac{\mathrm{log}(n_{r})}{n_{r}}$ when $K\geq2$\\
    $\hat{\Pi}_{r}\in[0,1]^{n_{r}\times K}$&Estimated membership of row nodes&$\hat{\Pi}_{c}\in[0,1]^{n_{r}\times K}$&Estimated membership of column nodes\\
$\tau$&$\mathrm{max}_{i\in[n_{r}],j\in[n_{c}]}|A(i,j)-\Omega(i,j)|$&$\gamma$&$\mathrm{max}_{i\in[n_{r}],j\in[n_{c}]}\mathbb{E}((A(i,j)-\Omega(i,j))^{2})/\rho$\\
$\mathrm{diag}(M)$&Diagonal matrix with $(i,i)$-th entry $M(i,i)$&$\mathrm{max}(0,M)$&Matrix with $(i,j)$-th entry $\mathrm{max}(0,M(i,j))$\\
$M^{-1}$&Inverse of matrix $M$&$I$&Identity matrix of compatible dimension\\
$p$&Sparsity parameter&$\mathcal{M}$&Model for bipartite unweighted networks\\
$\eta_{r}$&Proportion of highly mixed row nodes&$\eta_{c}$&Proportion of highly mixed column nodes\\
$\zeta_{r}$&Proportion of highly pure row nodes&$\zeta_{c}$&Proportion of highly pure column nodes\\
\hline
\end{tabular}
}
\caption{Main symbols of the paper.}
\label{table-symbol}
\end{table}
The main symbols involved in this paper are summarized in Table \ref{table-symbol}. Given a bipartite weighted network $\mathcal{N}=(\mathcal{V}_{r}, \mathcal{V}_{c}, \mathcal{W})$ with $n_{r}$ row nodes and $n_{c}$ column nodes, where $\mathcal{V}_{r}=\{1,2,\ldots, n_{r}\}$ is the set of row nodes, $\mathcal{V}_{c}=\{1,2,\ldots, n_{c}\}$ is the set of column nodes, and $\mathcal{W}$ represents the set of edge weights. Let $A\in\mathbb{R}^{n_{r}\times n_{c}}$ be $\mathcal{N}'s$ bi-adjacency matrix. In this paper, we allow $\mathcal{V}_{r}\neq \mathcal{V}_{c}$, i.e., row nodes can be different from column nodes and a bipartite setting case. We also allow $A(i,j)$ to be any finite real values for $i\in[n_{r}], j\in[n_{c}]$ instead of only nonnegative values. For convenience, we call $\mathcal{N}$ directed weighted network when $\mathcal{V}_{r}=\mathcal{V}_{c}$ (i.e., row nodes are the same as column nodes).

For an overlapping bipartite weighted network $\mathcal{N}$, our Bipartite Mixed Membership Distribution-Free model proposed in Definition \ref{BiMMDF} can model such $\mathcal{N}$.
\begin{defin}\label{BiMMDF}
Let $\Pi_{r}\in\mathbb{R}^{n_{r}\times K}, \Pi_{c}\in\mathbb{R}^{n_{c}\times K}$ such that $\mathrm{rank}(\Pi_{r})=K, \mathrm{rank}(\Pi_{c})=K, \Pi_{r}(i,k)\geq 0, \Pi_{c}(j,k)\geq0,\|\Pi_{r}(i,:)\|_{1}=1$, and $\|\Pi_{c}(j,:)\|_{1}=1$ for $i\in[n_{r}],j\in[n_{c}], k\in[K]$, where $\Pi_{r}(i,:)\in\mathbb{R}^{K\times 1}$ and $\Pi_{c}(j,:)\in\mathbb{R}^{K\times 1}$ are the community membership vector for row node $i$ and column node $j$, respectively. Let $P\in\mathbb{R}^{K\times K}$ satisfy $\mathrm{max}_{k,l\in[K]}|P(k,l)|=1$ and $\mathrm{rank}(P)=K$. Let $\rho>0$ and call it the scaling parameter. Let $A\in\mathbb{R}^{n_{r}\times n_{c}}$ be the bi-adjacency matrix of $\mathcal{N}$. For all pairs of $(i,j)$, our Bipartite Mixed Membership Distribution-Free (BiMMDF) model assumes that for any distribution $\mathcal{F}$, $A(i,j)$ are independent random variables generated from the distribution $\mathcal{F}$ satisfying
\begin{align}\label{AFOmega}
\mathbb{E}[A(i,j)]=\Omega(i,j), \mathrm{where~}\Omega:=\rho \Pi_{r}P\Pi'_{c}.
\end{align}
\end{defin}
For convenience, denote our model by $BiMMDF(n_{r}, n_{c}, K, P,\rho, \Pi_{r}, \Pi_{c},\mathcal{F})$. Figure \ref{dibiExample} summarizes sketches of different types of overlapping networks modeled by our BiMMDF.
\begin{figure}
\centering
\resizebox{\columnwidth}{!}{
\subfigure[]
{
\begin{tikzpicture}[scale=2.4]
\Vertices{./ex_complex_01_vertices.csv}
\Edges{./ex_undirectedunweighted_01_edges.csv}
\end{tikzpicture}
}
\subfigure[]
{
\begin{tikzpicture}[scale=2.4]
\Vertices{./ex_complex_01_vertices.csv}
\Edges{./ex_undirectedweighted_01_edges.csv}
\end{tikzpicture}
}
\subfigure[]
{
\begin{tikzpicture}[scale=2.4]
\Vertices{./ex_complex_01_vertices.csv}
\Edges{./ex_undirectedweighted_02_edges.csv}
\end{tikzpicture}
}
\subfigure[]
{
\begin{tikzpicture}[scale=2.4]
\Vertices{./ex_complex_01_vertices.csv}
\Edges{./ex_undirectedsigned_edges.csv}
\end{tikzpicture}
}
}
\resizebox{\columnwidth}{!}{
\subfigure[]
{
\begin{tikzpicture}[scale=2.4]
\Vertices{./ex_complex_01_vertices.csv}
\Edges{./ex_directedunweighted_01_edges.csv}
\end{tikzpicture}
}
\subfigure[]
{
\begin{tikzpicture}[scale=2.4]
\Vertices{./ex_complex_01_vertices.csv}
\Edges{./ex_directedweighted_01_edges.csv}
\end{tikzpicture}
}
\subfigure[]
{
\begin{tikzpicture}[scale=2.4]
\Vertices{./ex_complex_01_vertices.csv}
\Edges{./ex_directedweighted_02_edges.csv}
\end{tikzpicture}
}
\subfigure[]
{
\begin{tikzpicture}[scale=2.4]
\Vertices{./ex_complex_01_vertices.csv}
\Edges{./ex_directedsigned_edges.csv}
\end{tikzpicture}
}
}
\resizebox{\columnwidth}{!}{
\subfigure[]
{
\begin{tikzpicture}[scale=2.4]
\Vertices{./ex_complex_02_vertices.csv}
\Edges{./ex_bipartiteunweighted_edges.csv}
\end{tikzpicture}
}
\subfigure[]
{
\begin{tikzpicture}[scale=2.4]
\Vertices{./ex_complex_02_vertices.csv}
\Edges{./ex_bipartiteweighted1_edges.csv}
\end{tikzpicture}
}
\subfigure[]
{
\begin{tikzpicture}[scale=2.4]
\Vertices{./ex_complex_02_vertices.csv}
\Edges{./ex_bipartiteweighted2_edges.csv}
\end{tikzpicture}
}
\subfigure[]
{
\begin{tikzpicture}[scale=2.4]
\Vertices{./ex_complex_02_vertices.csv}
\Edges{./ex_bipartiteweighted3_edges.csv}
\end{tikzpicture}
}
}
\caption{Illustrations for networks modeled by BiMMDF. Panel (a): un-directed un-weighted network. Panel (b): un-directed weighted network with positive weights. Panel (c): un-directed weighted network with positive and negative weights. Panel (d): un-directed signed network. Panel (e): directed un-weighted network. Panel (f): directed weighted network with positive weights. Panel (g): directed weighted network with positive and negative weights. Panel (h): directed signed network. Panel (i): bipartite un-weighted network. Panel (j): bipartite weighted network with positive weights. Panel (k): bipartite weighted network with positive and negative weights. Panel (l): bipartite signed network.}\label{dibiExample}
\end{figure}
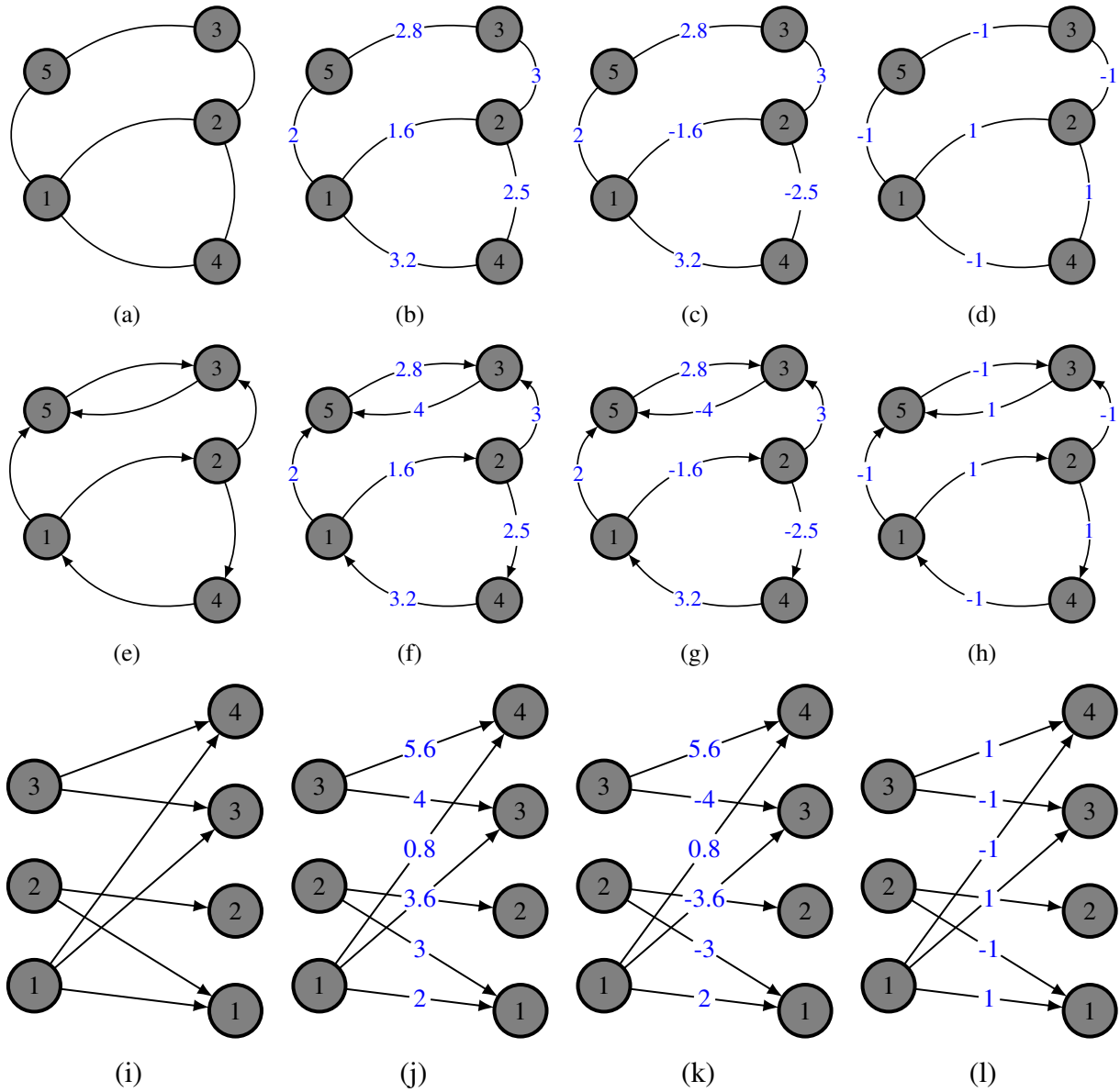
\begin{rem}
This remark provides some explanations and understandings on $K,\Pi_{r}, \Pi_{c}, P, \rho, \Omega$, and distribution $\mathcal{F}$ under our model $BiMMDF(n_{r}, n_{c}, K, P,\rho, \Pi_{r}, \Pi_{c},\mathcal{F})$.
\begin{itemize}
  \item For $K$, it is the number of row (and column) communities and it is much smaller than $\mathrm{min}(n_{r}, n_{c})$. When modeling a bipartite network in which both row and column nodes can belong to multiple communities, to make the model identifiable, the number of row communities must be the same as the number of column communities.
  \item For $\Pi_{r}$, it is the membership matrix for all row nodes. $\Pi_{r}(i,k)$ denotes the weight (which can also be seen as probability) of row node $i$ on row community $k$ for $i\in[n_{r}]$ and $k\in[K]$. The probability of row node $i$ belonging to all the $K$ row communities is 1, so BiMMDF requires $\sum_{k=1}^{K}\Pi_{r}(i,k)=1$ for $i\in[n_{r}]$. Meanwhile, we require the rank of $\Pi_{r}$ to be $K$ because we need to make BiMMDF identifiable. Similar explanations hold for $\Pi_{c}$.
  \item For $P$, it controls the block structure of $\Omega$ to make BiMMDF more applicable. Otherwise, if $P$ is an identity matrix, we have $\mathbb{E}[A(i,j)]=\Omega(i,j)=\rho\Pi_{r}(i,:)\Pi'_{c}(j,:)$, which is much simpler than $\rho\Pi_{r}(i,:)P\Pi'_{c}(j,:)$. Meanwhile, $\rho P$ in BiMMDF is not a probability matrix unless $\mathcal{F}$ is Bernoulli distribution, and whether $P$ can have negative elements depends on distribution $\mathcal{F}$. For detail, see Examples \ref{Bernoulli}-\ref{Signed}. We set $P$'s maximum absolute element as 1 mainly for theoretical convenience because we have considered the scaling parameter $\rho$ to make $\mathrm{max}_{k,l\in[K]}|\rho P(k,l)|$ be $\rho$. $P$ is an asymmetric matrix, so BiMMDF can model bipartite networks (sure, $P$ can be symmetric). Furthermore, we need $P$ to be full rank to make BiMMDF well-defined and identifiable. For models like MMSB, OCCAM, DCMM, DiMMSB, and MMDF for overlapping networks, their identifiability also requires $P$ to be full rank.
  \item For $\rho$, its range depends on distribution $\mathcal{F}$. For example, when $\mathcal{F}$ is Bernoulli distribution such that $\rho P$ is a probability matrix, $\rho$ ranges in $(0,1]$ because $\mathrm{max}_{k,l\in[K]}|P(k,l)|=1$; when $\mathcal{F}$ is Normal distribution, $\rho$ ranges in $(0,+\infty)$. For detail, see Examples \ref{Bernoulli}-\ref{Signed}.
  \item For $\Omega$, it is the expectation of $A$ under BiMMDF and we call it population adjacency matrix. Benefitted from the fact that $\mathrm{rank}(\Pi_{r})=K, \mathrm{rank}(\Pi_{c})=K, \mathrm{rank}(P)=K$ and $K\ll\mathrm{min}(n_{r}, n_{c})$, the rank of $\Omega$ is $K$ by Equation (\ref{AFOmega}), i.e., $\Omega$ has a low-dimensional structure with only $K$ nonzero singular values. We benefit a lot from $\Omega$'s low-dimensional structure when we design an algorithm to fit BiMMDF in Section \ref{sec3}.
  \item For $\mathcal{F}$, it can be any distribution as long as Equation (\ref{AFOmega}) holds. Several distributions are considered in Examples \ref{Bernoulli}-\ref{Signed}. It is possible that Equation (\ref{AFOmega}) does not hold for some distributions. For example, $\mathcal{F}$ can not be $t$-distribution whose mean is 0; $\mathcal{F}$ can not be Cauchy distribution whose mean does not exist.
\end{itemize}
\end{rem}
Table \ref{table-network-model} summarizes the comparisons of our BiMMDF with some previous models. In particular, BiMMDF can reduce to some previous models with some conditions.
\begin{itemize}
\item When $\mathcal{F}$ is Bernoulli distribution such that $A\in\{0,1\}^{n_{r}\times n_{c}}$, BiMMDF reduces to DiMMSB \citep{qing2021DiMMSB}.
\item When $\mathcal{F}$ is Normal or Bernoulli distribution, BiMMDF reduces to the two-way blockmodels \citep{airoldi2013multi}.
\item When $\Pi_{r}=\Pi_{c}, P=P'$, and $\mathcal{F}$ is Bernoulli distribution, BiMMDF reduces to MMSB \citep{MMSB}.
\end{itemize}

\begin{table}[ht]
\centering
\scriptsize
\rowcolors{1}{gray!22}{gray!22}
\resizebox{\columnwidth}{!}{
\begin{tabular}{cccccc}
\hline
Model&Adjacency matrix A&Distribution $\mathcal{F}$&Overlapping&Networks can be modeled\\
\hline
SBM \cite{SBM}&$A=A'$ and $A\in\{0,1\}^{n\times n}$&Bernoulli&No&Panel (a) of Figure \ref{dibiExample}\\
MMSB \cite{MMSB}&$A=A'$ and $A\in\{0,1\}^{n\times n}$&Bernoulli&Yes&Panel (a) of Figure \ref{dibiExample}\\
DCSBM \cite{DCSBM}&$A=A'$ and $A\in\mathbb{N}^{n\times n}$&Bernoulli and Poisson&No&Panel (b) with nonnegative integer weights of Figure \ref{dibiExample}\\
OSBM \cite{latouche2011overlapping}&$A=A'$ and $A\in\{0,1\}^{n\times n}$&Bernoulli&No&Panels (a), (e) of Figure \ref{dibiExample}\\
Two-way blockmodels \cite{airoldi2013multi}&$A\in\mathbb{R}^{n_{r}\times n_{c}}$&Normal and Bernoulli&Yes&Panels (a), (c), (e), (g), (i), (k) of Figure \ref{dibiExample}\\
WSBM \cite{aicher2015learning}&$A=A', A\in\mathbb{R}^{n\times n}$&Exponential family&No&Panels  (a)-(d) of Figure \ref{dibiExample}\\
ScBM and DCScBM \cite{DISIM}&$A\in\{0,1\}^{n_{r}\times n_{c}}$&Bernoulli&No&Panels (a), (e), (i) of Figure \ref{dibiExample}\\
OCCAM \cite{OCCAM}&$A=A'$ and $A\in\{0,1\}^{n\times n}$&Bernoulli&Yes&Panel (a) of Figure \ref{dibiExample}\\
DCMM \cite{mixedSCORE}&$A=A'$ and $A\in\{0,1\}^{n\times n}$&Bernoulli&Yes&Panel (a) of Figure \ref{dibiExample}\\
WSBM \cite{palowitch2017significance}&$A=A'$ and $A\in\mathbb{R}_{+}^{n\times n}$&Distributions defined on $\mathbb{R}_{+}$&No&Panels (a), (b) of Figure \ref{dibiExample}\\
WSBM \cite{ahn2018hypergraph}&$A=A'$ and $A\in\mathbb{R}^{n\times n}$&Arbitrary&No&Panels (a)-(d) of Figure \ref{dibiExample}\\
SBMO \cite{kaufmann2018spectral}&$A=A'$ and $A\in\{0,1\}^{n\times n}$&Bernoulli&Yes&Panel (a) of Figure \ref{dibiExample}\\
WSBM \cite{xu2020optimal}&$A=A'$ and $A\in\mathbb{R}^{n\times n}$&Arbitrary&No&Panels (a)-(d) of Figure \ref{dibiExample}\\
WMMSB \cite{dulac2020mixed}&$A=A'$ and $A\in\mathbb{N}^{n\times n}$&Poisson&Yes&Panel (b) with nonnegative integer weights of Figure \ref{dibiExample}\\
DiMMSB \cite{qing2021DiMMSB}&$A\in\{0,1\}^{n_{r}\times n_{c}}$&Bernoulli&Yes&Panels (a), (e), (i) of Figure \ref{dibiExample}\\
WSBM \cite{ng2021weighted}&$A=A'$ and $A\in\mathbb{R}_{+}^{n\times n}$&Gamma&No&Panel (b) of Figure \ref{dibiExample}\\
BiDFM and BiDCDFM \cite{BiDFMs}&$A\in\mathbb{R}^{n_{r}\times n_{c}}$&Arbitrary&No&Panels (a)-(l) of Figure \ref{dibiExample}\\
BiMMDF (this paper)&$A\in\mathbb{R}^{n_{r}\times n_{c}}$&Arbitrary&Yes&Panels (a)-(l) of Figure \ref{dibiExample}\\
\hline
\end{tabular}
}
\caption{Summary of comparisons of BiMMDF with some previous models.}
\label{table-network-model}
\end{table}
Similar to \cite{mao2020estimating, mixedSCORE}, call row node $i$ `pure' if $\Pi_{r}(i,:)$ degenerates (i.e., one entry is 1, all others $K-1$ entries are 0) and `mixed' otherwise. The same definitions hold for column nodes. In this article, we assume that for every $k\in[K]$, there exists at least one pure row node $i$ such that $\Pi_{r}(i,k)=1$ and at least one pure column node $j$ such that $\Pi_{c}(j,k)=1$, and these two assumptions are known as pure node assumption \citep{mao2020estimating, MaoSVM, OCCAM, qing2021DiMMSB,mixedSCORE}. The requirements  $\mathrm{rank}(\Pi_{r})=K$ and $\mathrm{rank}(\Pi_{c})=K$ in Definition \ref{BiMMDF} hold immediately as long as each row (column) community has at least one pure node.  Since we assume that $P$ is full rank and the pure node assumption holds, Proposition 1 of \cite{qing2021DiMMSB} guarantees that BiMMDF is identifiable. Meanwhile, the full rank condition on $P$ and pure node assumption on membership matrices are necessary for the identifiability of models for overlapping networks, to name a few, MMSB \citep{mao2020estimating}, DCMM \citep{MaoSVM,mixedSCORE}, and OCCAM \citep{OCCAM}.
\section{A spectral algorithm for fitting the model}\label{sec3}
Since the rank of $P$ is $K$, we have $\mathrm{rank}(\Omega)=K$. Let $\Omega=U\Lambda V'$ be the top-$K$ singular value decomposition (SVD) of $\Omega$ such that $U\in\mathbb{R}^{n_{r}\times K}, \Lambda\in\mathbb{R}_{+}^{K\times K}, V\in\mathbb{R}^{n_{c}\times K}$, $U'U=I, V'V=I$. Lemma 1 of \cite{qing2021DiMMSB} which is distribution-free guarantees the existences of simplex structures inherent in $U$ and $V$, i.e., there exist two $K\times K$ matrices $B_{r}$ and $B_{c}$ such that $U=\Pi_{r}B_{r}$ and $V=\Pi_{c}B_{c}$. Similar to \cite{mao2020estimating}, for simplex structures, applying the successive projection algorithm (SPA) \citep{gillis2015semidefinite} to $U$ (and $V$) with $K$ row (and column) communities obtains $B_{r}$ (and $B_{c}$). Thus, with given $\Omega$, we can exactly return $\Pi_{r}$ and $\Pi_{c}$ by setting $\Pi_{r}=UB^{-1}_{r}$ and  $\Pi_{c}=VB^{-1}_{c}$.

In practice, $\Omega$ is unknown but the adjacency matrix $A$ is given and we aim at estimating $\Pi_{r}$ and $\Pi_{c}$ based on $A$. Let $\hat{A}=\hat{U}\hat{\Lambda}\hat{V}'$ be the top-$K$ SVD of $A$ corresponding to the top-$K$ singular values of $A$ where $\hat{A}, \hat{U}, \hat{\Lambda}$, and $\hat{V}$ can be seen as approximations of $\Omega, U, \Lambda$, and $V$, respectively. Then one should be able to obtain a good estimation of $\Pi_{r}$ (and $\Pi_{c}$) by applying SPA on the rows of $\hat{U}$ (and $\hat{V}$) assuming there are $K$ row (and column) clusters. The spectral clustering algorithm considered to fit BiMMDF is summarized in Algorithm \ref{alg:DiSP}, which is the DiSP algorithm of \cite{qing2021DiMMSB} actually. In Algorithm \ref{alg:DiSP}, $\hat{\mathcal{I}}_{r}$ is the index set of pure nodes returned by SPA with input $\hat{U}$ when there are $K$ row communities. Similar explanation holds for $\hat{\mathcal{I}}_{c}$. Meanwhile, the algorithm for fitting BiMMDF is the same as that of DiMMSB because DiSP enjoys the distribution-free property since Lemma 1 of \cite{qing2021DiMMSB} always holds without dependence on distribution $\mathcal{F}$. Note that there are no tuning parameters required by the DiSP algorithm.
\begin{algorithm}
\caption{\textbf{DiSP}}
\label{alg:DiSP}
\begin{algorithmic}[1]
\Require Adjacency matrix $A\in \mathbb{R}^{n_{r}\times n_{c}}$ of a bipartite weighted network $\mathcal{N}$, number of row (and column) clusters $K$.
\Ensure $\hat{\Pi}_{r}$ and $\hat{\Pi}_{c}$.
\State Get the top-$K$ SVD of $A$ as $\hat{U}\hat{\Lambda}\hat{V}'$.
\State $\mathcal{\hat{I}}_{r}=\mathrm{SPA}(\hat{U})$ and  $\mathcal{\hat{I}}_{c}=\mathrm{SPA}(\hat{V})$.
\State $\hat{B}_{r}=\hat{U}(\mathcal{\hat{I}}_{r},:)$ and $\hat{B}_{c}=\hat{V}(\mathcal{\hat{I}}_{c},:)$.
\State $\hat{Y}_{r}=\hat{U}\hat{B}^{-1}_{r}$ and $\hat{Y}_{c}=\hat{V}\hat{B}^{-1}_{c}$.
\State $\hat{Y}_{r}=\mathrm{max}(0, \hat{Y}_{r})$ and  $\hat{Y}_{c}=\mathrm{max}(0, \hat{Y}_{c})$.
\State $\hat{\Pi}_{r}=\mathrm{diag}(\hat{Y}_{r}\textbf{1}_{K})^{-1}\hat{Y}_{r}$ and
$\hat{\Pi}_{c}=\mathrm{diag}(\hat{Y}_{c}\textbf{1}_{K})^{-1}\hat{Y}_{c}$.
\end{algorithmic}
\end{algorithm}

The time cost of DiSP mainly comes from the SVD step and the SPA step. The SVD step is also known as PCA \citep{mixedSCORE} and it is manageable even for a matrix with a large size. The complexity of SVD is $O(\mathrm{max}(n^{2}_{r},n^{2}_{c})K)$. The time cost of SPA is $O(\mathrm{max}(n_{r},n_{c})K^{2})$ \citep{mixedSCORE}. Since the number of clusters $K$ is much smaller than $n_{r}$ and $n_{c}$ in this article, as a result, the total time cost of DiSP is $O(\mathrm{max}(n^{2}_{r},n^{2}_{c})K)$. Results in Section \ref{RealDataSection} show that, for a real-world bipartite network with 16726 rows nodes and 22015 column nodes, DiSP takes around 20 seconds to process a standard personal computer (Thinkpad X1 Carbon Gen 8) using MATLAB R2021b.
\section{Main results for DiSP}\label{sec4}
In this section, we show that the sample-based estimates $\hat{\Pi}_{r}$ and $\hat{\Pi}_{c}$ concentrate around the true mixed membership matrix $\Pi_{r}$ and $\Pi_{c}$, respectively. Throughout this paper, $K$ is a known positive integer.

Set $\tau=\mathrm{max}_{i\in[n_{r}],j\in[n_{c}]}|A(i,j)-\Omega(i,j)|$ and  $\gamma=\frac{\mathrm{max}_{i\in[n_{r}],j\in[n_{c}]}\mathbb{E}[(A(i,j)-\Omega(i,j))^{2}]}{\rho}$, where $\tau$ and $\gamma$ are two parameters depending on distribution $\mathcal{F}$. For theoretical convenience, we need the following assumption.
\begin{assum}\label{a1}
Assume that $\rho\gamma\mathrm{max}(n_{r},n_{c})\geq \tau^{2}\mathrm{log}(n_{r}+n_{c}).$
\end{assum}
Assumption \ref{a1} controls the lower bound of $\rho\gamma$ for our theoretical analysis. Because $\gamma$ varies for different $\mathcal{F}$ and depends on $\rho$, the exact form of Assumption \ref{a1} can be obtained immediately for a specific distribution $\mathcal{F}$. For detail, see Examples \ref{Bernoulli}-\ref{Signed}. Meanwhile, theoretical guarantees for spectral methods studied in \citep{lei2015consistency,SCORE,DISIM,mixedSCORE,mao2020estimating,MaoSVM,zhou2019analysis,DSCORE} also need requirements like Assumption \ref{a1}. Similar to  conditions in Corollary 3.1 of \cite{mao2020estimating}, to simplify DiSP's theoretical upper bound, we use the following condition.
\begin{con}\label{c1}
$\kappa(P)=O(1),\frac{n_{r}}{n_{c}}=O(1), \lambda_{K}(\Pi'_{r}\Pi_{r})=O(\frac{n_{r}}{K})$, and $\lambda_{K}(\Pi'_{c}\Pi_{c})=O(\frac{n_{c}}{K})$.
\end{con}
In Condition \ref{c1}, $\kappa(P)=O(1)$ means that $P$ is well-conditioned; $\frac{n_{r}}{n_{c}}=O(1)$ means that $n_{r}$ is in the same order as $n_{c}$; $\lambda_{K}(\Pi'_{r}\Pi_{r})=O(\frac{n_{r}}{K})$ means that the ``size'' of each row community is in the same order. We are ready to present the main theorem.
\begin{thm}\label{Main}
(Error of DiSP) Under $BiMMDF(n_{r}, n_{c}, K, P,\rho,  \Pi_{r}, \Pi_{c},\mathcal{F})$, let $\hat{\Pi}_{r}$ an $\hat{\Pi}_{c}$ be obtained from Algorithm \ref{alg:DiSP}, suppose Assumption \ref{a1} and Condition \ref{c1} hold, and furthermore, $\sigma_{K}(\Omega)\gg \sqrt{\rho\gamma(n_{r}+n_{c})\mathrm{log}(n_{r}+n_{c})}$, there exists two permutation matrices $\mathcal{P}_{r}, \mathcal{P}_{c}\in\mathcal{R}^{K\times K}$ such that with probability at least $1-o((n_{r}+n_{c})^{-5})$, for $i\in[n_{r}],j\in[n_{c}]$, we have
\begin{align*}	&\|e'_{i}(\hat{\Pi}_{r}-\Pi_{r}\mathcal{P}_{r})\|_{1}=O(\frac{K^{2}\sqrt{\gamma \mathrm{log}(n_{r}+n_{c})}}{\sigma_{K}(P)\sqrt{\rho n_{c}}}),\|e'_{j}(\hat{\Pi}_{c}-\Pi_{c}\mathcal{P}_{c})\|_{1}=O(\frac{K^{2}\sqrt{\gamma \mathrm{log}(n_{r}+n_{c})}}{\sigma_{K}(P)\sqrt{\rho n_{r}}}).
\end{align*}
Especially, when $n_{r}=O(n), n_{c}=O(n)$, we have
\begin{align*}	&\|e'_{i}(\hat{\Pi}_{r}-\Pi_{r}\mathcal{P}_{r})\|_{1}=O(\frac{K^{2}\sqrt{\gamma \mathrm{log}(n)}}{\sigma_{K}(P)\sqrt{\rho n}}),\|e'_{j}(\hat{\Pi}_{c}-\Pi_{c}\mathcal{P}_{c})\|_{1}=O(\frac{K^{2}\sqrt{\gamma \mathrm{log}(n)}}{\sigma_{K}(P)\sqrt{\rho n}}).
\end{align*}
\end{thm}

From Theorem \ref{Main}, we see that our DiSP enjoys consistent estimation under our BiMMDF, i.e., theoretical upper bounds of DiSP's error rates go to zero as $n_{r}$ and $n_{c}$ go to infinity when $P, K, \rho$ and distribution $\mathcal{F}$ are fixed.

Especially, under the same settings of Theorem \ref{Main}, when  $n_{r}=n_{c}=n$, $K=O(1)$ (i.e., $K$ is a small positive integer), and $n$ is not too small, Theorem \ref{Main} says that DiSP's error rates are small with high probability when $\sigma_{K}(P)\gg\sqrt{\frac{\gamma\mathrm{log}(n)}{\rho n}}$. For convenience, we need the following definition for our further analysis.
\begin{defin}\label{DefinBiMMDFBalancedNet}
Let $BiMMDF(n,2,\Pi_{r},\Pi_{c},\alpha_{\mathrm{in}},\alpha_{\mathrm{out}},\mathcal{F})$ be a special case of $BiMMDF(n_{r}, n_{c}, K, P,\rho,  \Pi_{r}, \Pi_{c},\mathcal{F})$ when $n_{r}=n_{c}=n, K=2$, Condition \ref{c1} holds, and $\rho P$ has diagonal entries $p_{\mathrm{in}}=\alpha_{\mathrm{in}}\frac{\mathrm{log}(n)}{n}$ and non-diagonal entries $p_{\mathrm{out}}=\alpha_{\mathrm{out}}\frac{\mathrm{log}(n)}{n}$, where $|\alpha_{\mathrm{in}}|$ should not equal to $|\alpha_{\mathrm{out}}|$ because BiMMDF's identifiability requires $P$ to be full rank.
\end{defin}
The following corollary provides conditions on $\alpha_{\mathrm{in}}$ and $\alpha_{\mathrm{out}}$ to make DiSP's error rates small with high probability.
\begin{cor}\label{PhaseTransistion}
(Separation condition) Under $BiMMDF(n,2,\Pi_{r},\Pi_{c},\alpha_{\mathrm{in}},\alpha_{\mathrm{out}},\mathcal{F})$, when $n$ is not too small, with probability at least $1-o(n^{-5})$, DiSP's error rates are small and close to zero as long as
\begin{align}\label{CondAlphaInOut}
\gamma\mathrm{max}(|\alpha_{\mathrm{in}}|,
|\alpha_{\mathrm{out}}|)\geq \tau^{2}+o(1)\mathrm{~and~}||\alpha_{\mathrm{in}}|-|\alpha_{\mathrm{out}}||\gg\tau.
\end{align}
\end{cor}
\begin{rem}
When the network is undirected, all nodes are pure, each community has an equal size, and $\mathcal{F}$ is Bernoulli distribution, $BiMMDF(n,2,\Pi_{r},\Pi_{c},\alpha_{\mathrm{in}},\alpha_{\mathrm{out}},\mathcal{F})$ reduces to the SBM case such that nodes connect with probability $p_{\mathrm{in}}$ within clusters and $p_{\mathrm{out}}$ across clusters. This special case of SBM has been extensively studied, see \citep{abbe2015exact,hajek2016achieving,abbe2017community}. The main finding in \cite{abbe2015exact} says that exact recovery is possible if $|\sqrt{\alpha_{\mathrm{in}}}-\sqrt{\alpha_{\mathrm{out}}}|>\sqrt{2}$ and impossible if $|\sqrt{\alpha_{\mathrm{in}}}-\sqrt{\alpha_{\mathrm{out}}}|<\sqrt{2}$, where exact recovery means recovering the partition correctly with high probability when $n\rightarrow\infty$. Corollary \ref{PhaseTransistion} says that DiSP's error rates are small with high probability as long as Equation (\ref{CondAlphaInOut}) holds when $A$ is generated from different distribution $\mathcal{F}$ under $BiMMDF(n,2,\Pi_{r},\Pi_{c},\alpha_{\mathrm{in}},\alpha_{\mathrm{out}},\mathcal{F})$. For comparison, exact recovery requires that all nodes are pure, the network is undirected, and $\mathcal{F}$ is Bernoulli distribution while small error rates with high probability considered in this paper allow nodes to be mixed, the network to be bipartite, and $\mathcal{F}$ to be any distribution.
\end{rem}
For all pairs $(i,j)$ with $i\in[n_{r}], j\in[n_{c}]$, Examples \ref{Bernoulli}-\ref{Signed} provide $\gamma$'s upper bound and show that the explicit form of Equation (\ref{CondAlphaInOut}) is different for different distribution $\mathcal{F}$ under BiMMDF.
\begin{Ex}\label{Bernoulli}
When $\mathcal{F}$ is \textbf{Bernoulli distribution} such that $A(i,j)\sim \mathrm{Bernoulli}(\Omega(i,j))$, i.e., $A(i,j)\in\{0,1\}$.  For this case, BiMMDF degenerates to DiMMSB for bipartite un-weighted networks. For Bernoulli distribution, $P$ should have nonnegative elements, $\mathbb{E}[A(i,j)]=\Omega(i,j)$ satisfies Equation (\ref{AFOmega}), $\mathbb{P}(A(i,j)=1)=\Omega(i,j)$, and $\frac{\mathbb{E}[(A(i,j)-\Omega(i,j))^{2}]}{\rho}=\frac{\Omega(i,j)(1-\Omega(i,j))}{\rho}\leq\frac{\Omega(i,j)}{\rho}\leq 1$, so we have $\tau\leq 1$ and $\gamma\leq 1$, i.e., $\tau$ and $\gamma$ are finite. Then, Assumption \ref{a1} means $\rho\geq\frac{\mathrm{log}(n_{r}+n_{c})}{\mathrm{max}(n_{r},n_{c})}$, a lower bound requirement on $\rho$ for theoretical analysis. Setting $\gamma$ as $1$ in Theorem \ref{Main} obtains theoretical upper bounds of error rates of DiSP and we see that increasing $\rho$ decreases error rates. For $BiMMDF(n,2,\Pi_{r},\Pi_{c},\alpha_{\mathrm{in}},\alpha_{\mathrm{out}},\mathcal{F})$, $\rho P$ is a probability matrix when $\mathcal{F}$ is Bernoulli distribution, so $\rho$ ranges in $(0,1]$ since we require $\mathrm{max}_{k,l}|P(k,l)|=1$, and $\alpha_{\mathrm{in}}$ and $\alpha_{\mathrm{out}}$ range in $[0,\frac{n}{\mathrm{log}(n)}]$. Setting $\gamma=1$ and $\tau=1$ in Equation (\ref{CondAlphaInOut}) gives
\begin{align}\label{CondAlphaInOutBernoulli}
\mathrm{max}(\alpha_{\mathrm{in}},
\alpha_{\mathrm{out}})\geq 1+o(1)\mathrm{~and~}|\alpha_{\mathrm{in}}-\alpha_{\mathrm{out}}|\gg1.
\end{align}
\end{Ex}
\begin{Ex}\label{Poisson}
When $\mathcal{F}$ is \textbf{Poisson distribution} such that $A(i,j)\sim \mathrm{Poisson}(\Omega(i,j))$, i.e., $A(i,j)\in\mathbb{N}$. For Poisson distribution, $P$ should have positive elements, $\mathbb{E}[A(i,j)]=\Omega(i,j)$ satisfies Equation (\ref{AFOmega}), $\mathbb{P}(A(i,j)=m)=\frac{\Omega(i,j)^{m}}{m!}e^{-\Omega(i,j)}$ for any nonnegative integer $m$ and $\mathbb{E}[(A(i,j)-\Omega(i,j))^{2}]=\Omega(i,j)\leq \rho$, so we have $\gamma$ is finite and $\gamma\leq 1$. Therefore, for Poisson distribution, Assumption \ref{a1} means $\rho\geq\frac{\tau^{2}\mathrm{log}(n_{r}+n_{c})}{\mathrm{max}(n_{r},n_{c})}$. Setting  $\gamma$ as $1$ in Theorem \ref{Main} when $\mathcal{F}$ is Poisson distribution, we find that increasing $\rho$ decreases error rates. For $\tau$, it is an unknown finite positive integer. Since the mean of Poisson distribution can be any positive value, $\rho$ ranges in $(0,+\infty)$. For $BiMMDF(n,2,\Pi_{r},\Pi_{c},\alpha_{\mathrm{in}},\alpha_{\mathrm{out}},\mathcal{F})$, $\alpha_{\mathrm{in}}$ and $\alpha_{\mathrm{out}}$ range in $(0,+\infty)$ when $\mathcal{F}$ is Poisson distribution. Setting $\gamma=1$ in Equation (\ref{CondAlphaInOut}) gives
\begin{align}\label{CondAlphaInOutPoisson}
\mathrm{max}(\alpha_{\mathrm{in}},
\alpha_{\mathrm{out}})\geq \tau^{2}+o(1)\mathrm{~and~}|\alpha_{\mathrm{in}}-\alpha_{\mathrm{out}}|\gg\tau.
\end{align}
\end{Ex}
\begin{Ex}\label{Binomial}
When $\mathcal{F}$ is \textbf{Binomial distribution} such that $A(i,j)\sim\mathrm{Binomial}(m,\frac{\Omega(i,j)}{m})$ for any positive integer $m$, i.e.,  $A(i,j)\in\{0,1,2,\ldots,m\}$. For Binomial distribution, all elements of $P$ should be nonnegative, $\mathbb{E}[A(i,j)]=\Omega(i,j)$ satisfies Equation (\ref{AFOmega}), and $\mathbb{E}[(A(i,j)-\Omega(i,j))^{2}]=m\frac{\Omega(i,j)}{m}(1-\frac{\Omega(i,j)}{m})=\Omega(i,j)(1-\frac{\Omega(i,j)}{m})\leq \rho$. So, $\tau=m$ and $\gamma\leq1$. Then, Assumption \ref{a1} means $\rho\geq\frac{m^{2}\mathrm{log}(n_{r}+n_{c})}{\mathrm{max}(n_{r},n_{c})}$. Setting  $\gamma$ as $1$ in Theorem \ref{Main} gets theoretical upper bounds of error rates of DiSP when $\mathcal{F}$ is Binomial distribution and we see that increasing $\rho$ decreases error rates. Meanwhile, since $\frac{\Omega(i,j)}{m}$ is a probability, $\rho$ should be less than $m$ for this case. For $BiMMDF(n,2,\Pi_{r},\Pi_{c},\alpha_{\mathrm{in}},\alpha_{\mathrm{out}},\mathcal{F})$,  $\alpha_{\mathrm{in}}$ and $\alpha_{\mathrm{out}}$ range in $(0,\frac{mn}{\mathrm{log}(n)}]$ when $\mathcal{F}$ is Binomial distribution. Setting $\gamma=1, \tau=m$ in Equation (\ref{CondAlphaInOut}) when $\mathcal{F}$ is Binomial distribution, we have
\begin{align}\label{CondAlphaInOutBinomial}
\mathrm{max}(\alpha_{\mathrm{in}},
\alpha_{\mathrm{out}})\geq m^{2}+o(1)\mathrm{~and~}|\alpha_{\mathrm{in}}-\alpha_{\mathrm{out}}|\gg m.
\end{align}
Note that when $m$ is 1, the Binomial distribution reduces to the Bernoulli distribution, and we see that Equation (\ref{CondAlphaInOutBinomial}) matches Equation (\ref{CondAlphaInOutBernoulli}).
\end{Ex}
\begin{Ex}\label{Normal}
When $\mathcal{F}$ is \textbf{Normal distribution} such that $A(i,j)\sim \mathrm{Normal}(\Omega(i,j),\sigma^{2}_{A})$, i.e., $A(i,j)\in\mathbb{R}$, where $\sigma^{2}_{A}$ is the variance term of Normal distribution. For this case, BiMMDF reduces to the two-way blockmodels introduced in \cite{airoldi2013multi}. For Normal distribution, all elements of $P$ are real values, $\mathbb{E}[A(i,j)]=\Omega(i,j)$ satisfies Equation (\ref{AFOmega}), and $\mathbb{E}[(A(i,j)-\Omega(i,j))^{2}]=\sigma^{2}_{A}$. So, $\gamma=\frac{\sigma^{2}_{A}}{\rho}$. For $\tau$, it is an unknown finite value. Then, Assumption \ref{a1} means $\frac{\sigma^{2}_{A}\mathrm{max}(n_{r},n_{c})}{\mathrm{log}(n_{r}+n_{c})}\geq\tau^{2}$. Setting  $\gamma$ as $\frac{\sigma^{2}_{A}}{\rho}$ in Theorem \ref{Main}, we see that increasing $\rho$ (or decreasing $\sigma^{2}_{A}$) decreases error rates. Here, $\rho$ ranges in $(0,+\infty)$ because the mean of Normal distribution can be any value. Therefore, for $BiMMDF(n,2,\Pi_{r},\Pi_{c},\alpha_{\mathrm{in}},\alpha_{\mathrm{out}},\mathcal{F})$, $\alpha_{\mathrm{in}}$ and $\alpha_{\mathrm{out}}$ range in $(-\infty,+\infty)$, and we also have $\rho=\mathrm{max}(|p_{\mathrm{in}}|, |p_{\mathrm{out}}|)=\frac{\mathrm{log}(n)}{n}\mathrm{max}(|\alpha_{\mathrm{in}}|,
|\alpha_{\mathrm{out}}|)$. Setting $\gamma=\frac{\sigma^{2}_{A}}{\rho}=\frac{\sigma^{2}_{A}n}{\mathrm{max}(|\alpha_{\mathrm{in}}|,|\alpha_{\mathrm{out}}|)\mathrm{log}(n)}$ in Equation (\ref{CondAlphaInOut}) gives
\begin{align}\label{CondAlphaInOutNormal}
\frac{\sigma^{2}_{A}n}{\mathrm{log}(n)}\geq \tau^{2}+o(1)\mathrm{~and~}||\alpha_{\mathrm{in}}|-|\alpha_{\mathrm{out}}||\gg\tau.
\end{align}
Equation (\ref{CondAlphaInOutNormal}) differs a lot from Equations (\ref{CondAlphaInOutBernoulli})-(\ref{CondAlphaInOutBinomial}) because $\alpha_{\mathrm{in}}$ (and $\alpha_{\mathrm{out}}$) can be negative and there is no requirement on $\mathrm{max}(|\alpha_{\mathrm{in}}|,|\alpha_{\mathrm{out}}|)$ for Normal distribution.
\end{Ex}
\begin{Ex}\label{Exponential}
When $\mathcal{F}$ is \textbf{Exponential distribution} such that $A(i,j)\sim \mathrm{Exponential}(\frac{1}{\Omega(i,j)})$, i.e., $A(i,j)\in\mathbb{R}_{+}$. For Exponential distribution,  all elements of $P$ should be positive,  $\mathbb{E}[A(i,j)]=\Omega(i,j)$ satisfies Equation (\ref{AFOmega}), and $\mathbb{E}[(A(i,j)-\Omega(i,j))^{2}]=\Omega^{2}(i,j)\leq \rho^{2}$. So, $\gamma\leq\rho$. For $\tau$, it is an unknown finite value. Then, Assumption \ref{a1} means $\rho^{2}\geq\frac{\tau^{2}\mathrm{log}(n_{r}+n_{c})}{\mathrm{max}(n_{r},n_{c})}$. When setting $\gamma$ as $\rho$ in Theorem \ref{Main}, $\rho$ vanishes in the theoretical bounds, and this suggests that increasing $\rho$ does not influence DiSP's error rates. For this case, $\rho$ ranges in $(0,+\infty)$ because $\rho P$ is not a probability matrix. For $BiMMDF(n,2,\Pi_{r},\Pi_{c},\alpha_{\mathrm{in}},\alpha_{\mathrm{out}},\mathcal{F})$, $\alpha_{\mathrm{in}}$ and $\alpha_{\mathrm{out}}$ range in $(0,+\infty)$ because all elements of $P$ are positive for Exponential distribution. Setting  $\gamma=\rho=\mathrm{max}(|p_{\mathrm{in}}|, |p_{\mathrm{out}}|)=\frac{\mathrm{log}(n)}{n}\mathrm{max}(\alpha_{\mathrm{in}},\alpha_{\mathrm{out}})$ in Equation (\ref{CondAlphaInOut}) gives
\begin{align}\label{CondAlphaInOutExponential}
\mathrm{max}(\alpha^{2}_{\mathrm{in}},\alpha^{2}_{\mathrm{out}})\frac{\mathrm{log}(n)}{n}\geq \tau^{2}+o(1)\mathrm{~and~}|\alpha_{\mathrm{in}}-\alpha_{\mathrm{out}}|\gg\tau.
\end{align}
Equation (\ref{CondAlphaInOutExponential}) means that even when the second inequality holds, to make DiSP's error rates be small with high probability, $\mathrm{max}(\alpha_{\mathrm{in}},\alpha_{\mathrm{out}})$ must be large enough to make the first inequality hold because of the $\frac{\mathrm{log}(n)}{n}$ term.
\end{Ex}
\begin{Ex}\label{Uniform}
When $\mathcal{F}$ is \textbf{Uniform distribution} such that $A(i,j)\sim\mathrm{Uniform}(0,2\Omega(i,j))$, i.e., $A(i,j)\in(0,2\rho)$. For Uniform distribution, all elements of $P$ should be nonnegative, $\mathbb{E}[A(i,j)]=\frac{0+2\Omega(i,j)}{2}=\Omega(i,j)$ satisfies Equation (\ref{AFOmega}), $\tau$ is no larger than $2\rho$, and $\mathbb{E}[(A(i,j)-\Omega(i,j))^{2}]=\frac{4\Omega^{2}(i,j)}{12}\leq \frac{\rho^{2}}{3}$, i.e., $\gamma\leq\frac{\rho}{3}$. Therefore, Assumption \ref{a1} means $\rho^{2}\geq\frac{3\tau^{2}\mathrm{log}(n_{r}+n_{c})}{\mathrm{max}(n_{r},n_{c})}$. Since $\rho$ vanishes in bounds of error rates when setting $\gamma$ as $\frac{\rho}{3}$ in Theorem \ref{Main}, increasing $\rho$ does not influence DiSP's performance. Meanwhile, $\rho$ ranges in $(0,+\infty)$ because $\mathrm{Uniform}(0,2\Omega(i,j))$ has no upper bound requirement on $\Omega(i,j)$ when it is positive. Therefore, $\alpha_{\mathrm{in}}$ and $\alpha_{\mathrm{out}}$ range in $[0,+\infty)$ because all elements of $P$ are nonnegative for this case. Setting  $\gamma=\rho/3=\mathrm{max}(|p_{\mathrm{in}}|, |p_{\mathrm{out}}|)/3=\frac{\mathrm{log}(n)}{3n}\mathrm{max}(\alpha_{\mathrm{in}},\alpha_{\mathrm{out}})$ in Equation (\ref{CondAlphaInOut}) gives
\begin{align}\label{CondAlphaInOutUniform}
\mathrm{max}(\alpha^{2}_{\mathrm{in}},\alpha^{2}_{\mathrm{out}})\frac{\mathrm{log}(n)}{3n}\geq \tau^{2}+o(1)\mathrm{~and~}|\alpha_{\mathrm{in}}-\alpha_{\mathrm{out}}|\gg\tau.
\end{align}
\end{Ex}
\begin{Ex}\label{Logistic}
When $\mathcal{F}$ is \textbf{Logistic distribution} such that $A(i,j)\sim\mathrm{Logistic}(\Omega(i,j),\beta)$, i.e., $A(i,j)\in\mathbb{R}$ , where $\beta>0$. For Logistic distribution, all elements of $P$ are real values, $\mathbb{E}[A(i,j)]=\Omega(i,j)$ satisfies Equation (\ref{AFOmega}), and $\mathbb{E}[(A(i,j)-\Omega(i,j))^{2}]=\frac{\pi^{2}\beta^{2}}{3}$, i.e., $\gamma=\frac{\pi^{2}\beta^{2}}{3\rho}$. Therefore, Assumption \ref{a1} means $\frac{\pi^{2}\beta^{2}\mathrm{max}(n_{r},n_{c})}{3\mathrm{log}(n_{r}+n_{c})}\geq\tau^{2}$. Setting $\gamma$ as $\frac{\pi^{2}\beta^{2}}{3\rho}$ in Theorem \ref{Main} obtains theoretical upper bounds of DiSP's error rates, and we find that increasing $\rho$ (or decreasing $\beta$) decreases error rates. Meanwhile, $\rho$ ranges in $(0,+\infty)$, and $\alpha_{\mathrm{in}}$ (and $\alpha_{\mathrm{out}}$) ranges in $(-\infty,+\infty)$ because the mean of Logistic distribution can be any value. Setting  $\gamma=\frac{\pi^{2}\beta^{2}}{3\rho}=\frac{\pi^{2}\beta^{2}}{3\mathrm{max}(|p_{\mathrm{in}}|,|p_{\mathrm{out}}|)}=\frac{\pi^{2}\beta^{2}n}{3\mathrm{max}(|\alpha_{\mathrm{in}}|,|\alpha_{\mathrm{out}}|)\mathrm{log}(n)}$ in Equation (\ref{CondAlphaInOut}) gives
\begin{align}\label{CondAlphaInOutLogistic}
\frac{\pi^{2}\beta^{2}n}{3\mathrm{log}(n)}\geq \tau^{2}+o(1)\mathrm{~and~}||\alpha_{\mathrm{in}}|-|\alpha_{\mathrm{out}}||\gg\tau.
\end{align}
\end{Ex}
\begin{Ex}\label{Signed}
BiMMDF can also generate \textbf{bipartite signed networks} by setting $\mathbb{P}(A(i,j)=1)=\frac{1+\Omega(i,j)}{2}$ and $\mathbb{P}(A(i,j)=-1)=\frac{1-\Omega(i,j)}{2}$, i.e., $A(i,j)\in\{-1,1\}$. For this case, all elements of $P$ are real values, $\mathbb{E}[A(i,j)]=\Omega(i,j)$ satisfies Equation (\ref{AFOmega}), and $\mathbb{E}[(A(i,j)-\Omega(i,j))^{2}]=1-\Omega^{2}(i,j)\leq1$, i.e., $\gamma\leq \frac{1}{\rho}$. For $\tau$, its upper bound is 2. Then Assumption \ref{a1} means $\frac{\mathrm{max}(n_{r},n_{c})}{\mathrm{log}(n_{r}+n_{c})}\geq 4$. Setting $\gamma$ as $\frac{1}{\rho}$ in Theorem \ref{Main}, we see that increasing $\rho$ decreases error rates. Meanwhile, $\rho$ ranges in $(0,1)$ because $\frac{1+\Omega(i,j)}{2}$ and $\frac{1-\Omega(i,j)}{2}$ are probabilities. $\alpha_{\mathrm{in}}$ and $\alpha_{\mathrm{out}}$ range in $(-\frac{n}{\mathrm{log}(n)},\frac{n}{\mathrm{log}(n)})$ because all elements of $P$ are real values. Setting  $\gamma=\frac{1}{\rho}=\frac{1}{\mathrm{max}(|p_{\mathrm{in}}|,|p_{\mathrm{out}}|)}=\frac{n}{\mathrm{max}(|\alpha_{\mathrm{in}}|,|\alpha_{\mathrm{out}}|)\mathrm{log}(n)}$ and $\tau=2$ in Equation (\ref{CondAlphaInOut}) gives
\begin{align}\label{CondAlphaInOutSigned}
\frac{n}{\mathrm{log}(n)}\geq4+o(1)\mathrm{~and~}||\alpha_{\mathrm{in}}|-|\alpha_{\mathrm{out}}||\gg2.
\end{align}
\end{Ex}
Other choices of $\mathcal{F}$ are also possible as long as Equation (\ref{AFOmega}) holds under distribution $\mathcal{F}$ for our BiMMDF. For example, $\mathcal{F}$ can be Geometric, Laplace, and Gamma distributions in \url{http://www.stat.rice.edu/~dobelman/courses/texts/distributions.c&b.pdf} (accessed on 9 June 2023), where this link also provides details on probability mass function or probability density function for distributions in Examples \ref{Bernoulli}-\ref{Logistic}.
\section{Missing edge}\label{sec5}
From Examples \ref{Normal}-\ref{Signed}, we see that $A(i,j)$ is always nonzero for any node pair $(i,j)$ when $A$ is generated from our BiMMDF, which suggests that there is always an edge between nodes $i$ and $j$. However, real-world large-scale networks are usually sparse based on the fact that the total number of edges is usually small compared to the number of nodes \cite{lei2015consistency}. Similar to \cite{xu2020optimal}, an edge with weight 0 is deemed as a missing edge in this paper. To generate missing edges for bipartite weighted networks from our BiMMDF, we introduce the following strategy.

Let $\mathcal{M}$ be a model for bipartite unweighted networks and let $\mathcal{A}\in\{0,1\}^{n_{r}\times n_{c}}$ be a bi-adjacency matrix generated from $\mathcal{M}$. $\mathcal{M}$ can be models like ScBM and  DCScBM \cite{DISIM} as long as $\mathcal{A}$ is the bi-adjacency matrix of a bipartite unweighted network. To model real-world large-scale bipartite weighted networks with missing edges, for $i\in[n_{r}], j\in[n_{c}]$, we update $A(i,j)$ by $A(i,j)\mathcal{A}(i,j)$.

In particular, when $\mathcal{M}$ is the bipartite Erd\"os-R\'enyi random graph $G(n_{r}, n_{c},p)$ \cite{erdos1960evolution} such that $\mathbb{P}(\mathcal{A}(i,j)=1)=p$ and $\mathbb{P}(\mathcal{A}(i,j)=0)=1-p$ for $i\in[n_{r}], j\in[n_{c}]$, the number of missing edges in $A$ increases as $p$ decreases. $p$ controls the sparsity of $A$, and we call $p$ sparsity parameter. In Section \ref{sec6}, we will study $p$'s influence on DiSP's performance.
\begin{rem}
This remark provides the difference between the scaling parameter $\rho$ and the sparsity parameter $p$. Since $\mathbb{P}(A(i,j)=1)=\Omega(i,j)=\rho \Pi_{r}(i,:)P\Pi'_{c}(j,:)$ and $\mathbb{P}(A(i,j)=0)=1-\rho \Pi_{r}(i,:)P\Pi'_{c}(j,:)$ when $\mathcal{F}$ is Bernoulli distribution, $\rho$ controls the number of zeros in $A$ and it also controls network sparsity for this case. However, for distributions in Examples \ref{Poisson}-\ref{Signed}, $\rho$ does not control network sparsity anymore. Instead, $p$ always controls network sparsity.
\end{rem}
\section{Experimental Results}\label{sec6}
In this section, we present example applications of our method, first to simulated networks and then to real-world networks. For simulated networks, to verify our theoretical analysis in Examples \ref{Bernoulli}-\ref{Signed}, we investigate the performance of DiSP to the scaling parameter $\rho$ and the separation parameters $\alpha_{\mathrm{in}}$ and $\alpha_{\mathrm{out}}$. We also consider missing edges in our simulations by changing the sparsity parameter $p$. We will also show the power of DiSP in revealing and understanding the latent community structure of real-world networks by introducing several indices, visualizing $\hat{\Pi}_{r}$ and $\hat{\Pi}_{c}$, and depicting the row and column communities detected by DiSP.
\subsection{Baseline methods}
For simulated networks, we compare DiSP with three overlapping community detection approaches:
\begin{itemize}
  \item SVM-cone-DCMMSB (SVM-cD for short) \citep{MaoSVM} and Mixed-SCORE \citep{mixedSCORE} are two algorithms for the DCMM model. The original SVM-cD and Mixed-SCORE are designed to estimate mixed memberships for overlapping undirected networks, to make them function for bipartite networks, we modify them in the following way: First, we use the $K\times K$ singular value matrix $\hat{\Lambda}$ to replace their $K\times K$ eigenvalue matrix. Second, we use $\hat{U}$ (and $\hat{V}$) to replace their eigenvector matrix to estimate membership matrix $\Pi_{r}$ (and $\Pi_{c}$) for row (and column) nodes. SVM-cD does not require any tuning parameters. For Mixed-SCORE, we set the threshold $T$ in \cite{mixedSCORE} as $\mathrm{log}(n_{r}+n_{c})$ in our experimental studies.
  \item DiMSC \citep{DiMSC} estimates community memberships for overlapping bipartite un-weighted networks generated from the bipartite version of the DCMM model. DiMSC does not require any tuning parameters.
\end{itemize}
\subsection{Evaluation metric}
For simulated networks with known $\Pi_{r}$ and $\Pi_{c}$, we use Hamming error \citep{mixedSCORE} and Relative Error \citep{mao2020estimating} to evaluate the performance of the algorithms for overlapping community detection with known $\Pi_{r}$ and $\Pi_{c}$. These two criteria for a bipartite network are defined as
\begin{align*}
&\mathrm{Hamming~Error}=\mathrm{max}(\frac{\mathrm{min}_{\mathcal{P}\in S}\|\hat{\Pi}_{r}\mathcal{P}-\Pi_{r}\|_{1}}{n_{r}},\frac{\mathrm{min}_{\mathcal{P}\in S}\|\hat{\Pi}_{c}\mathcal{P}-\Pi_{c}\|_{1}}{n_{c}}),\\
&\mathrm{Relative~Error}=\mathrm{max}(\frac{\mathrm{min}_{\mathcal{P}\in S}\|\hat{\Pi}_{r}\mathcal{P}-\Pi_{r}\|_{F}}{\|\Pi_{r}\|_{F}},\frac{\mathrm{min}_{\mathcal{P}\in S}\|\hat{\Pi}_{c}\mathcal{P}-\Pi_{c}\|_{F}}{\|\Pi_{c}\|_{F}}),
\end{align*}
where $S$ is the set of $K\times K$ permutation matrices. In the definition of Hamming Error, $\frac{\mathrm{min}_{\mathcal{P}\in S}\|\hat{\Pi}_{r}\mathcal{P}-\Pi_{r}\|_{1}}{n_{r}}$ means the $l_{1}$ difference between $\hat{\Pi}_{r}$ and $\Pi_{r}$, where we consider the permutation of community label in $\hat{\Pi}_{r}$ since the difference between $\hat{\Pi}_{r}$ and $\Pi_{r}$ should not depend on how we tag each of the $K$ row communities. $\frac{\mathrm{min}_{\mathcal{P}\in S}\|\hat{\Pi}_{r}\mathcal{P}-\Pi_{r}\|_{1}}{n_{r}}$ ranges in $[0,1]$ and it is the smaller the better. $\frac{\mathrm{min}_{\mathcal{P}\in S}\|\hat{\Pi}_{c}\mathcal{P}-\Pi_{c}\|_{1}}{n_{c}}$ measures the $l_{1}$ difference between $\hat{\Pi}_{c}$ and $\Pi_{c}$. We let Hamming Error be the maximum of $\frac{\mathrm{min}_{\mathcal{P}\in S}\|\hat{\Pi}_{r}\mathcal{P}-\Pi_{r}\|_{1}}{n_{r}}$ and $\frac{\mathrm{min}_{\mathcal{P}\in S}\|\hat{\Pi}_{c}\mathcal{P}-\Pi_{c}\|_{1}}{n_{c}}$ to measure the performance of a method over both row and column nodes. Thus, Hamming Error ranges in $[0,1]$, and smaller is better. Similar arguments hold for the definition of Relative Error except that Relative error measures the $l_{2}$ difference between the estimated membership matrix and the ground-truth membership matrix. Relative Error is nonnegative, and smaller is better. We do not use metrics like NMI \citep{danon2005comparing,bagrow2008evaluating,luo2017community}, ARI \citep{hubert1985comparing,luo2017community}, and overlapping NMI \citep{lancichinetti2009detecting,OCCAM} that require binary overlapping membership vectors \citep{MaoSVM} because entries of $\Pi_{r}$ and $\Pi_{c}$ considered in this paper may not be binary.
\subsection{Synthetic bipartite weighted networks}\label{sec5Synthetic}
In this section, we investigate the sensitivity of DiSP and competing approaches to the scaling parameter $\rho$ and sparsity parameter $p$ when the simulated overlapping bipartite weighted networks are generated from different distribution $\mathcal{F}$ under our BiMMDF model. For simulated networks, we aim at validating our theoretical analysis in Examples \ref{Bernoulli}-\ref{Signed} that DiSP has different behaviors when the scaling parameter $\rho$ is changed under different distributions, and investigating the behavior of DiSP when the sparsity parameter $p$ is changed. We also conduct some simulations by changing $\alpha_{\mathrm{in}}$ and $\alpha_{\mathrm{out}}$ to show that DiSP achieves the threshold in Equation (\ref{CondAlphaInOut}) for different distribution $\mathcal{F}$ under $BiMMDF(n,2,\Pi_{r},\Pi_{c},\alpha_{\mathrm{in}},\alpha_{\mathrm{out}},\mathcal{F})$. To facilitate comparisons, we summarize simulations conducted in this paper in Table \ref{ALLSIMs}.
\begin{table}[h!]
\footnotesize
	\centering
	\caption{Simulations conducted in this paper.}
	\label{ALLSIMs}
\resizebox{\columnwidth}{!}{
\begin{tabular}{cccccccccc}
\hline\hline
Distribution $\mathcal{F}$&Changing scaling parameter $\rho$& Changing sparse parameter $p$&Changing $\alpha_{\mathrm{in}}$ and $\alpha_{\mathrm{out}}$\\
\hline
Bernoulli&Simulation 1 (a)&Simulation 1 (b)&Simulations 1 (c) and 1(d)\\
Poisson&Simulation 2 (a)&Simulation 2 (b)&Simulations 2 (c) and 2(d)\\
Binomial&Simulation 3 (a)&Simulation 3 (b)&Simulations 3 (c) and 3(d)\\
Normal&Simulation 4 (a)&Simulation 4 (b)&Simulations 4 (c) and 4(d)\\
Exponential&Simulation 5 (a)&Simulation 5 (b)&Simulations 5 (c) and 5(d)\\
Uniform&Simulation 6 (a)&Simulation 6 (b)&Simulations 6 (c) and 6(d)\\
Logistic&Simulation 7 (a)&Simulation 7 (b)&Simulations 7 (c) and 7(d)\\
Bipartite signed network&Simulation 8 (a)&Simulation 8 (b)&Simulations 8 (c) and 8(d)\\
\hline\hline
\end{tabular}
}
\end{table}

For all simulations in this section, unless specified, the parameters $(n_{r}, n_{c}, K, P,\rho, \Pi_{r}, \Pi_{c})$ and distribution $\mathcal{F}$ under BiMMDF are set as follows. Let $K=2$, each row block own $n_{r,0}$ number of pure nodes where the top $Kn_{r,0}$ row nodes $\{1,2, \ldots, Kn_{r,0}\}$ are pure and the rest row nodes $\{Kn_{r,0}+1, Kn_{r,0}+2,\ldots, n_{r}\}$ are mixed with membership $(1/K,1/K,\ldots,1/K)$. Similarly, let each column block own $n_{c,0}$ number of pure nodes where the top $Kn_{c,0}$ column nodes $\{1,2, \ldots, Kn_{c,0}\}$ are pure and column nodes $\{Kn_{c,0}+1, Kn_{c,0}+2,\ldots, n_{c}\}$ are mixed with membership $(1/K,1/K,\ldots,1/K)$. $n_{r}, n_{c}, \mathrm{scaling~parameter~} \rho, \mathrm{sparsity~parameter~} p$, and distribution $\mathcal{F}$ are set independently for each experiment. For distribution (see, Bernoulli, Poisson, Binomial, Exponential, and Uniform distributions) which needs all elements of $P$ to be nonnegative, we set $P$ as
\[P_{1}=\begin{bmatrix}
    1&0.2\\
    0.3&0.8\\
\end{bmatrix}.\]
For distribution (see, Normal and Logistic  distributions as well as bipartite signed network) which allows $P$ to have negative elements, we set $P$ as
\[P_{2}=\begin{bmatrix}
    1&-0.2\\
    0.3&-0.8\\
\end{bmatrix}.\]
Meanwhile, when we consider the case $n_{r}=n_{c}=n$ for $BiMMDF(n,2,\Pi_{r},\Pi_{c},\alpha_{\mathrm{in}},\alpha_{\mathrm{out}},\mathcal{F})$, we set $\rho P$ as
\[\tilde{P}=\begin{bmatrix}
    \alpha_{\mathrm{in}}&\alpha_{\mathrm{out}}\\
    \alpha_{\mathrm{out}}&\alpha_{\mathrm{in}}\\
\end{bmatrix}\frac{\mathrm{log}(n)}{n},\]
where we aim at changing $\alpha_{\mathrm{in}}$ and $\alpha_{\mathrm{out}}$ to investigate their influences on DiSP's performance and verify Equation (\ref{CondAlphaInOut}) for different distribution $\mathcal{F}$.
\begin{rem}
The only  criteria for choosing the $K\times K$ matrix $P$ is, $P$ should be a full rank asymmetric (or symmetric) matrix, $\mathrm{max}_{k,l\in[K]}|P(k,l)|=1$, and elements of $P$ are positive or nonnegative or can be negative depending on distribution $\mathcal{F}$ as analyzed in Examples \ref{Bernoulli}-\ref{Signed}. The only criteria for setting $\Pi_{r}$ and $\Pi_{c}$ is, they should satisfy conditions in Definition \ref{BiMMDF}, and there exists at least one pure row (and column) node for each row (and column) community.
\end{rem}
To generate a random adjacency matrix $A$ with $K$ row (and column) communities and missing edges from distribution $\mathcal{F}$ under our model BiMMDF, each simulation experiment contains the following steps:

(a) Set $\Omega=\rho\Pi_{r}P\Pi'_{c}$.

(b) Let $A(i,j)$ be a random number generated from distribution $\mathcal{F}$ with expectation $\Omega(i,j)$ for $i\in[n_{r}], j\in[n_{c}]$.

(c) Generate $\mathcal{A}\in\{0,1\}^{n_{r}\times n_{c}}$ from the  bipartite Erd\"os-R\'enyi random graph $G(n_{r}, n_{c},p)$ such that $\mathbb{P}(\mathcal{A}(i,j)=1)=p$ and $\mathbb{P}(\mathcal{A}(i,j)=0)=1-p$ for $i\in[n_{r}], j\in[n_{c}]$.

(d) To generate missing edges in $A$, update $A(i,j)$ by $A(i,j)\mathcal{A}(i,j)$ for $i\in[n_{r}], j\in[n_{c}]$.

(e) Apply an overlapping community detection approach to $A$ with $K$ row (and column) communities. Record Hamming Error and Relative Error.

(f) Repeat (b)-(e) 50 times, and report the averaged Hamming Error and Relative Error over the 50 repetitions.

We consider the following simulation setups.
\subsubsection{Bernoulli distribution}
When $A(i,j)\sim \mathrm{Bernoulli}(\Omega(i,j))$ for $i\in[n_{r}], j\in[n_{c}]$, by Example \ref{Bernoulli} we know that all entries of $P$ should be nonnegative.

\begin{figure}
\centering
\resizebox{\columnwidth}{!}{
\subfigure[Simulation 1(a)]{\includegraphics[width=0.33\textwidth]{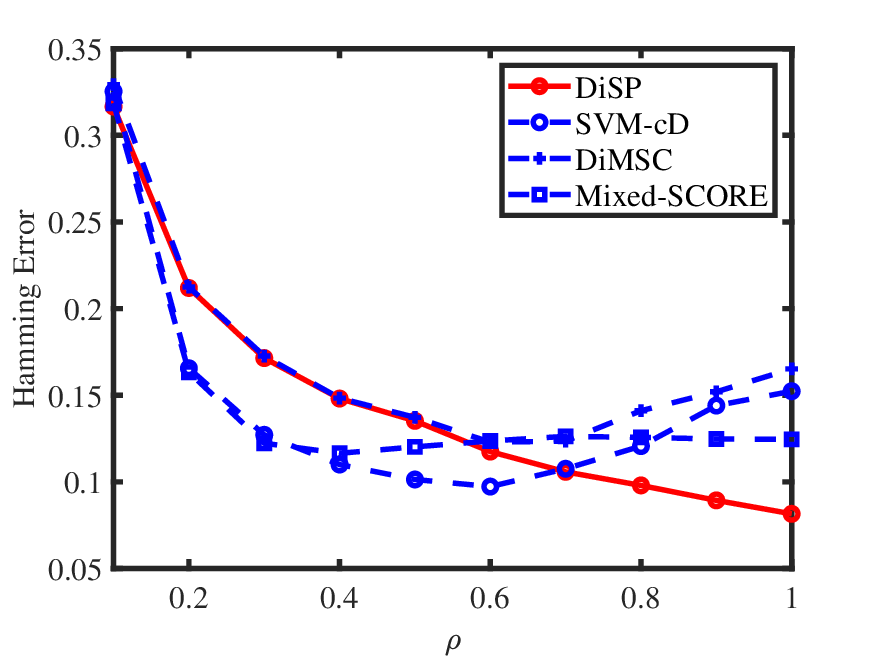}}
\subfigure[Simulation 1(a)]{\includegraphics[width=0.33\textwidth]{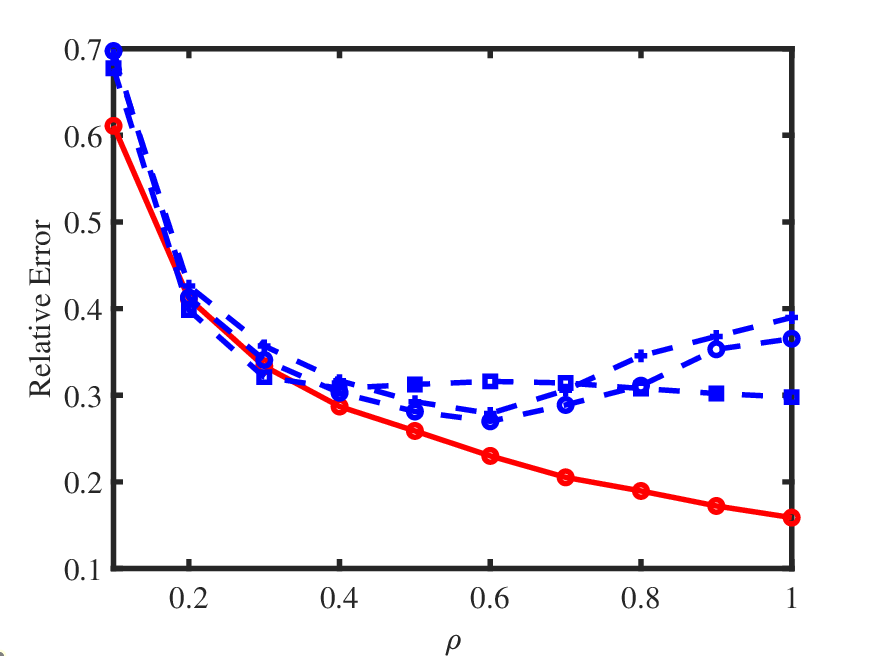}}
\subfigure[Simulation 1(b)]{\includegraphics[width=0.33\textwidth]{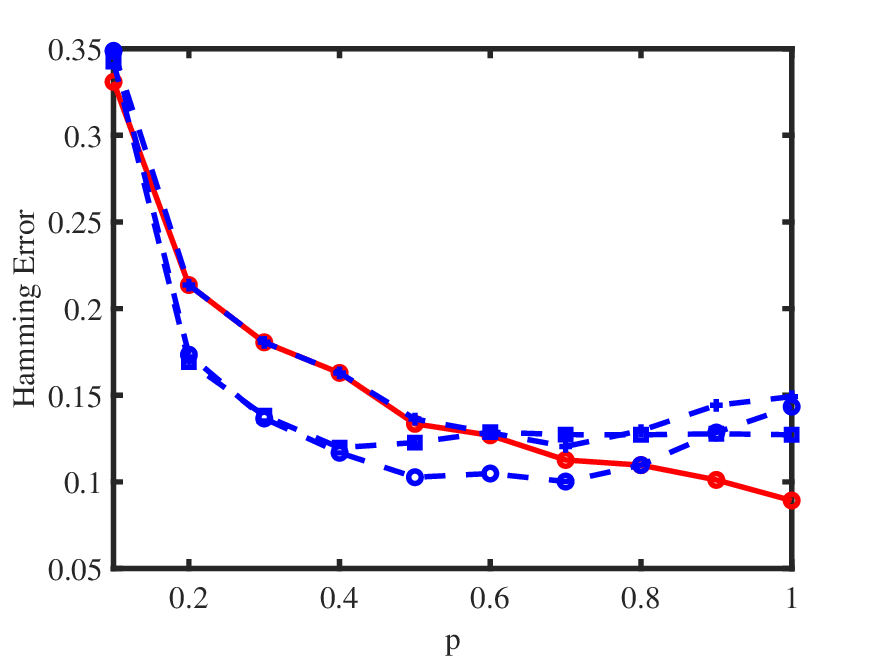}}
}
\resizebox{\columnwidth}{!}{
\subfigure[Simulation 1(b)]{\includegraphics[width=0.33\textwidth]{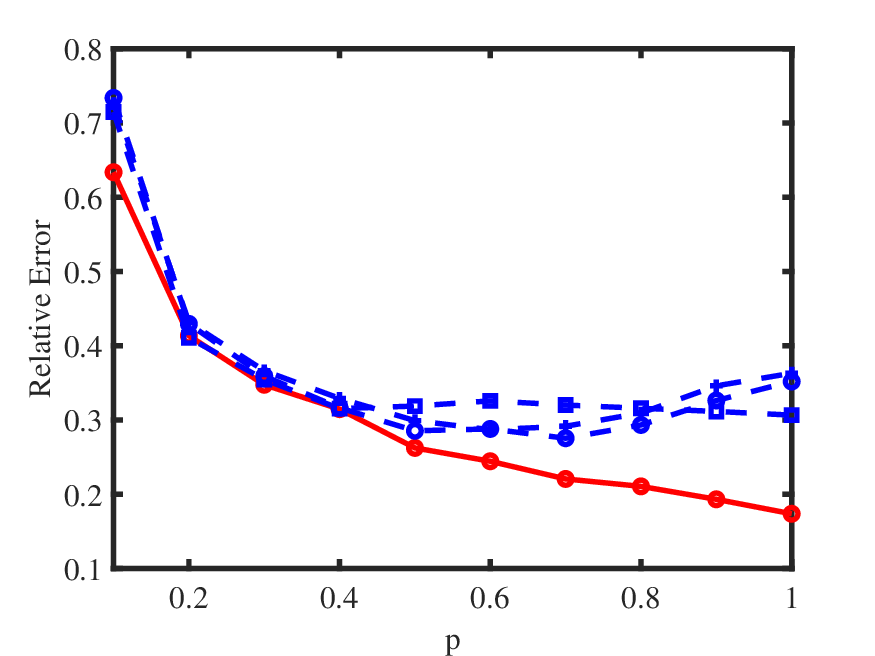}}
\subfigure[Simulation 1(c)]{\includegraphics[width=0.33\textwidth]{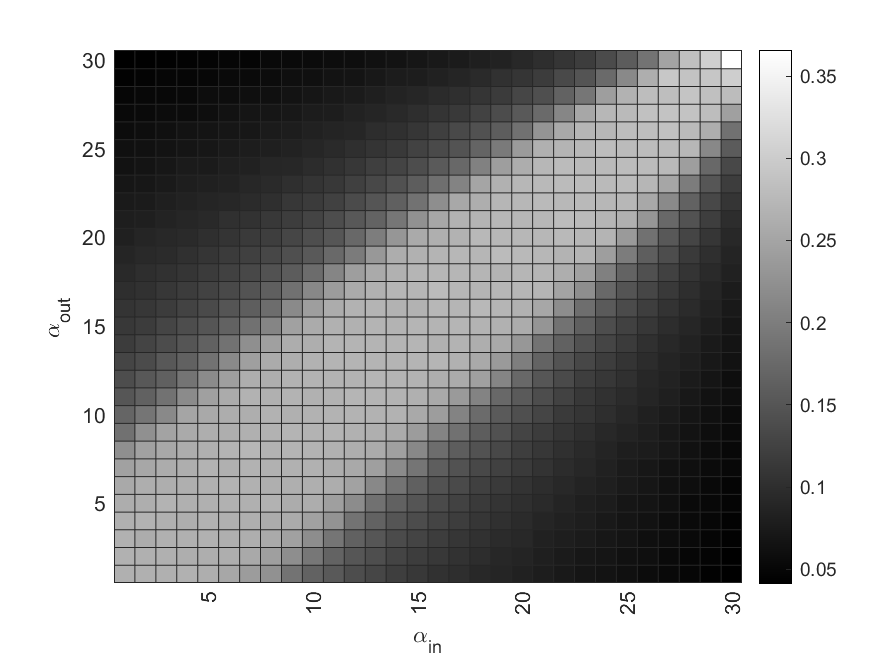}}
\subfigure[Simulation 1(d)]{\includegraphics[width=0.33\textwidth]{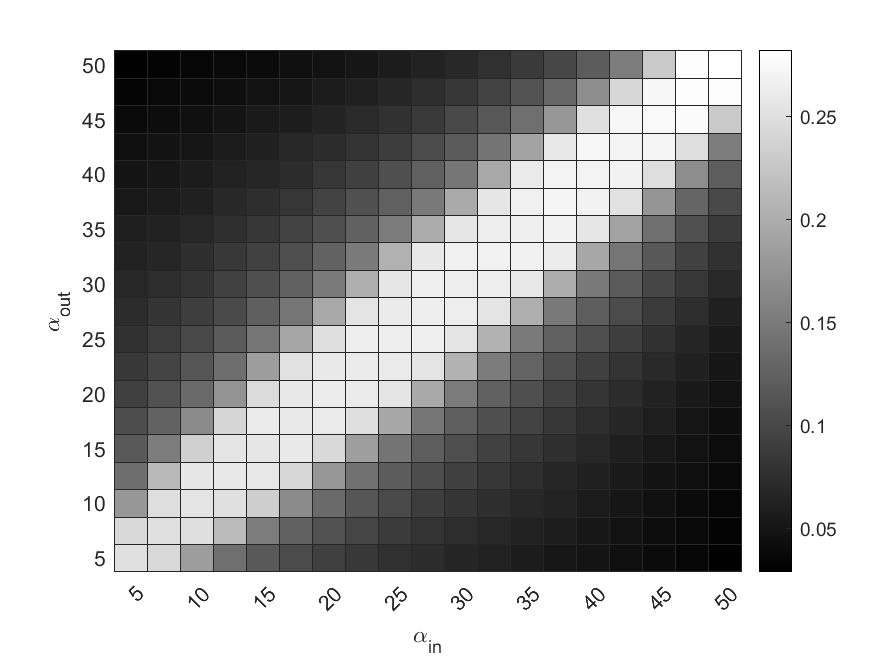}}
}
\caption{Bernoulli distribution. For panels (e) and (f): the darker pixel represents a lower Hamming Error of DiSP.}
\label{S1} 
\end{figure}

\textbf{Simulation 1 (a): changing $\rho$}. Let $n_{r}=200,n_{c}=300, n_{r,0}=50, n_{c,0}=100$, and $P=P_{1}$. Let the sparsity parameter $p$ in the bipartite Erd\"os-R\'enyi random graph $G(n_{r}, n_{c},p)$ be 0.9, i.e., we consider missing edges here. Since $\rho$ should be set no larger than 1 for the Bernoulli distribution, we let $\rho$ range in $\{0.1,0.2,0.3,\ldots,1\}$. The numerical results are displayed in panels (a) and (b) of Figure \ref{S1}. We see that DiSP performs better as $\rho$ increases and this matches analysis in Example \ref{Bernoulli}. Though SVM-cD, DiMSC, and Mixed-SCORE enjoy competitive performances with DiSP when $\rho$ is small, they perform poorer than DiSP when $\rho$ is larger than 0.6.

\textbf{Simulation 1 (b): Changing $p$.} All parameters are set the same as Simulation 1 (a) except we let $\rho=0.8$ and $p$ range in $\{0.1, 0.2, \ldots, 1\}$, i.e., we study the influence of the sparsity parameter $\rho$ on the performances of these four approaches in this simulation. Panels (c) and (d) of Figure \ref{S1} show the results. DiSP's error rates decrease when $p$ increases such that the number of missing edges decreases. When $p$ is small, all methods perform similarly. However, when $p$ is large, DiSP performs best.

\textbf{Simulation 1 (c): changing $\alpha_{\mathrm{in}}$ and $\alpha_{\mathrm{out}}$}. Let $n=n_{r}=n_{c}=300, n_{r,0}=50, n_{c,0}=100, p=1$, and $\rho P=\tilde{P}$. For Bernoulli distribution, $\alpha_{\mathrm{in}}$ and $\alpha_{\mathrm{out}}$ should be set in $(0,\frac{n}{\mathrm{log}(n)}]$ by Example \ref{Bernoulli}. For this simulation, we let $\alpha_{\mathrm{in}}$ and $\alpha_{\mathrm{out}}$ range in $\{1,2,3,\ldots,30\}$, where $|\alpha_{\mathrm{in}}-\alpha_{\mathrm{out}}|\geq1$ when $\alpha_{\mathrm{in}}\neq\alpha_{\mathrm{out}}$. The numerical results are shown in panel (e) of Figure \ref{S1}. We see that DiSP's error rates are large if $\mathrm{max}(\alpha_{\mathrm{in}},\alpha_{\mathrm{out}})$ is too small even when $|\alpha_{\mathrm{in}}-\alpha_{\mathrm{out}}|\gg1$ holds, and DiSP's error rates are small if we increase $\mathrm{max}(\alpha_{\mathrm{in}},\alpha_{\mathrm{out}})$ when $|\alpha_{\mathrm{in}}-\alpha_{\mathrm{out}}|\gg1$ holds. These results are consistent with the separation condition on $\alpha_{\mathrm{in}}$ and $\alpha_{\mathrm{out}}$ provided in Equation (\ref{CondAlphaInOutBernoulli}).

\textbf{Simulation 1 (d): changing $\alpha_{\mathrm{in}}$ and $\alpha_{\mathrm{out}}$}. All parameters are set the same as Simulation 1 (c) except that we let $\alpha_{\mathrm{in}}$ and $\alpha_{\mathrm{out}}$ range in $\{5,7.5,10,12.5,\ldots,50\}$, where $|\alpha_{\mathrm{in}}-\alpha_{\mathrm{out}}|\geq2.5$ when $\alpha_{\mathrm{in}}\neq\alpha_{\mathrm{out}}$ (note that $2.5$ is larger than 1 in Simulation 1 (c)). The numerical results are shown in panel (f) of Figure \ref{S1}. We see that the ``white'' area (i.e., large error rates area of $\alpha_{\mathrm{in}}$ and $\alpha_{\mathrm{out}}$) is narrower than that of the panel (e). This suggests that when the first inequality of Equation (\ref{CondAlphaInOutBernoulli}) holds, increasing $|\alpha_{\mathrm{in}}-\alpha_{\mathrm{out}}|$ decreases DiSP's error rates. This phenomenon holds naturally because the community detection problem is easier as $|p_{\mathrm{in}}-p_{\mathrm{out}}|=|\alpha_{\mathrm{in}}-\alpha_{\mathrm{out}}|\frac{\mathrm{log}(n)}{n}$ increases.
\subsubsection{Poisson distribution}
When $A(i,j)\sim\mathrm{Poisson}(\Omega(i,j))$ for $i\in[n_{r}], j\in[n_{c}]$, by Example \ref{Poisson}, all elements of $P$ should be positive.

\begin{figure}
\centering
\resizebox{\columnwidth}{!}{
\subfigure[Simulation 2(a)]{\includegraphics[width=0.33\textwidth]{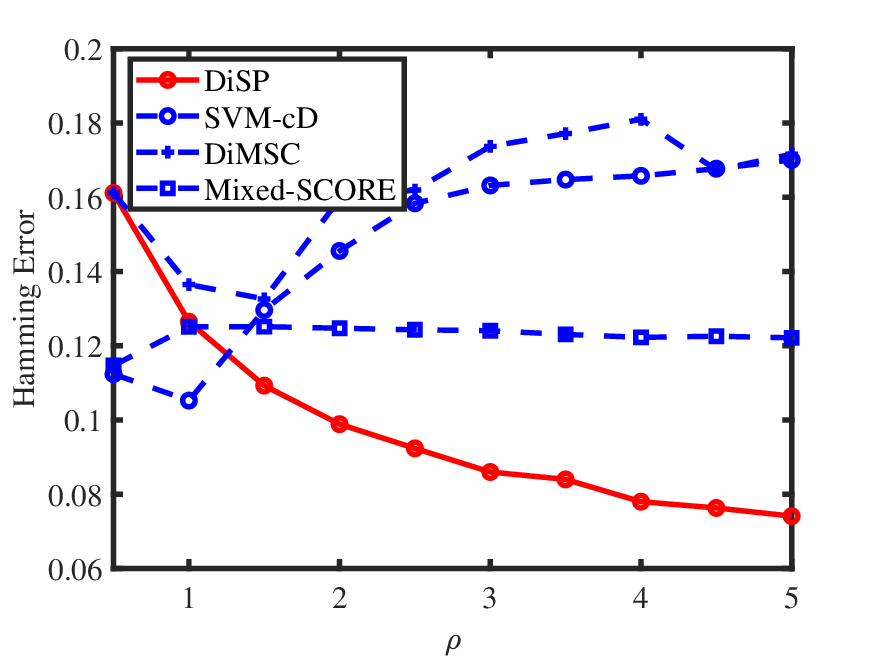}}
\subfigure[Simulation 2(a)]{\includegraphics[width=0.33\textwidth]{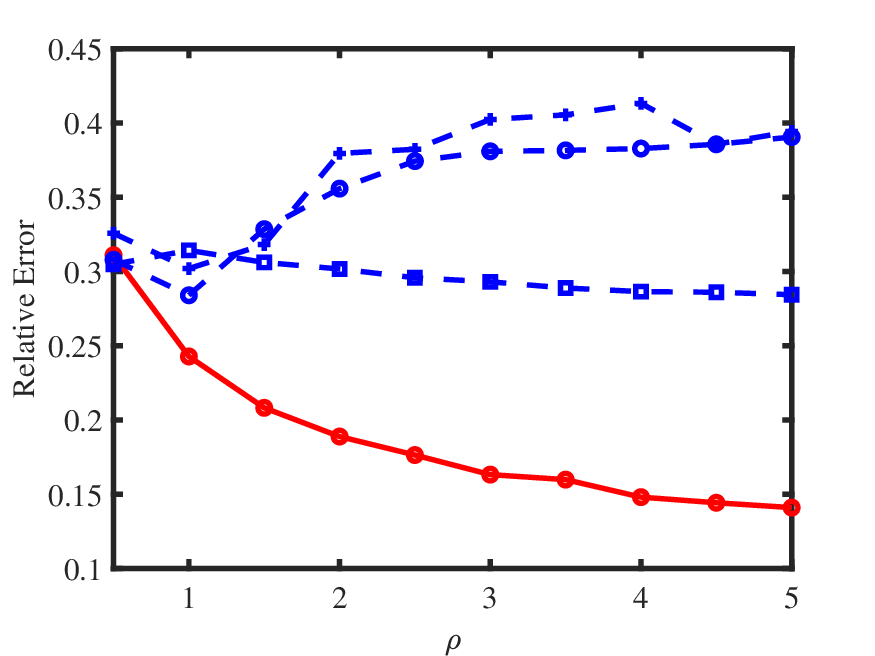}}
\subfigure[Simulation 2(b)]{\includegraphics[width=0.33\textwidth]{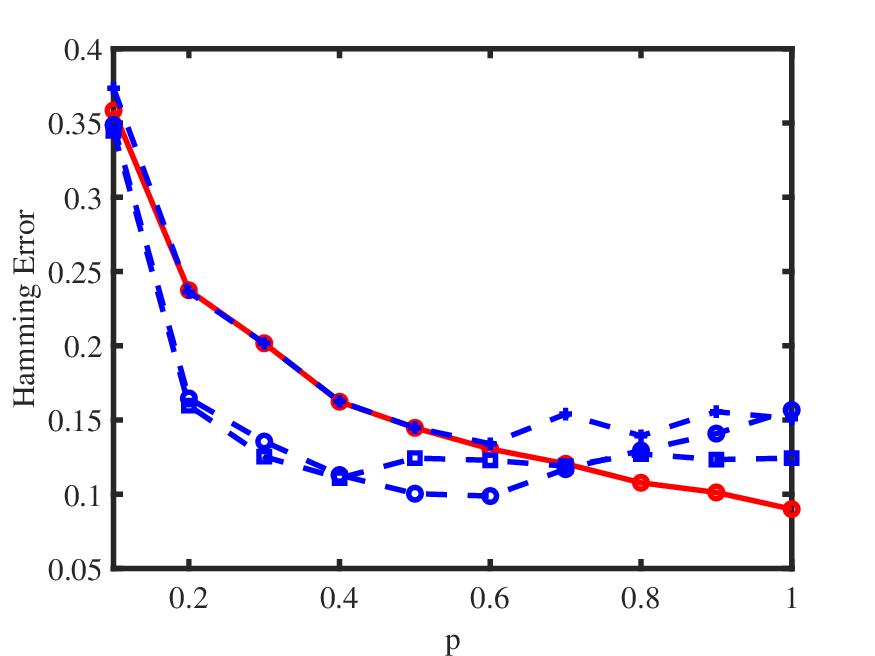}}
}
\resizebox{\columnwidth}{!}{
\subfigure[Simulation 2(b)]{\includegraphics[width=0.33\textwidth]{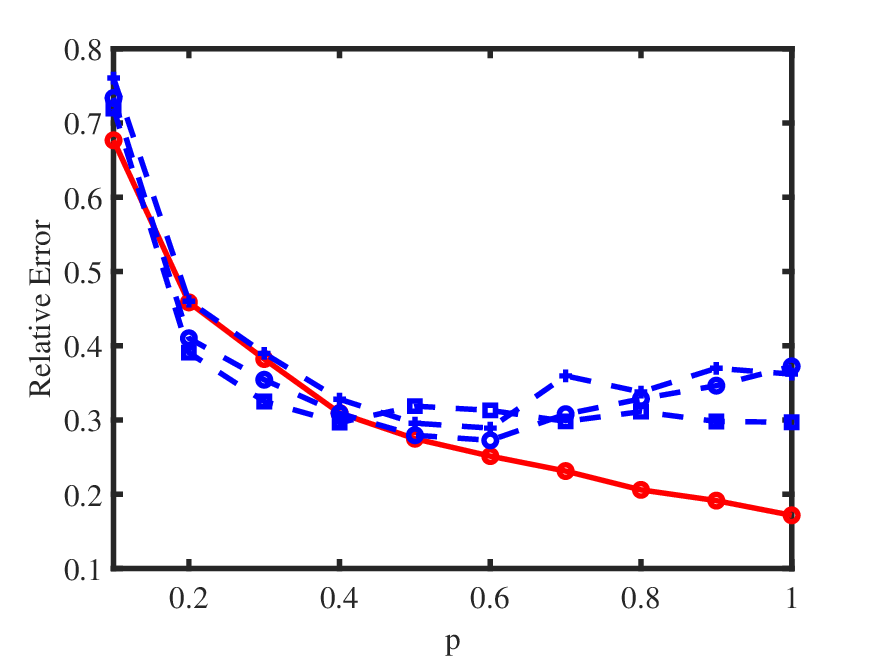}}
\subfigure[Simulation 2(c)]{\includegraphics[width=0.33\textwidth]{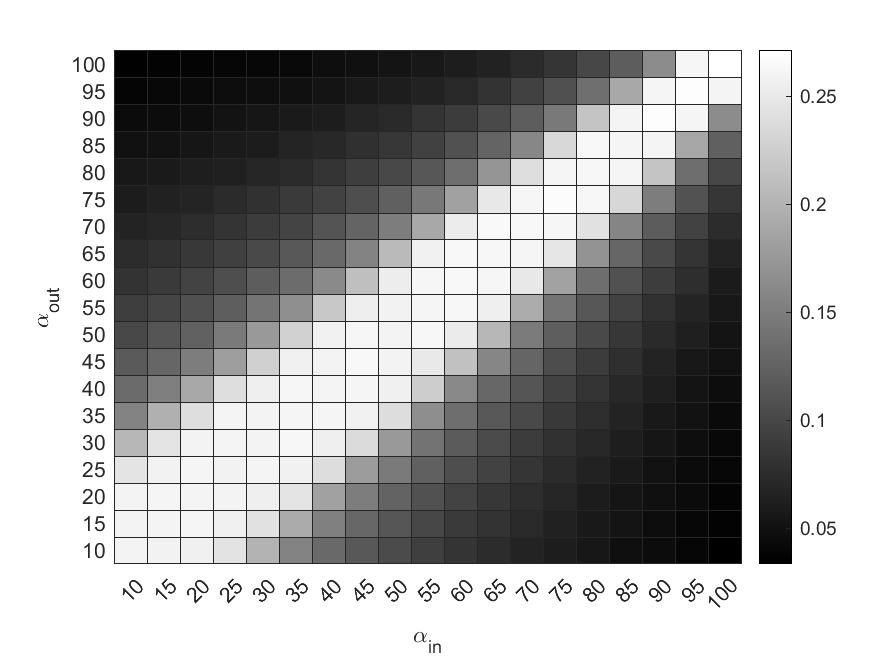}}
\subfigure[Simulation 2(d)]{\includegraphics[width=0.33\textwidth]{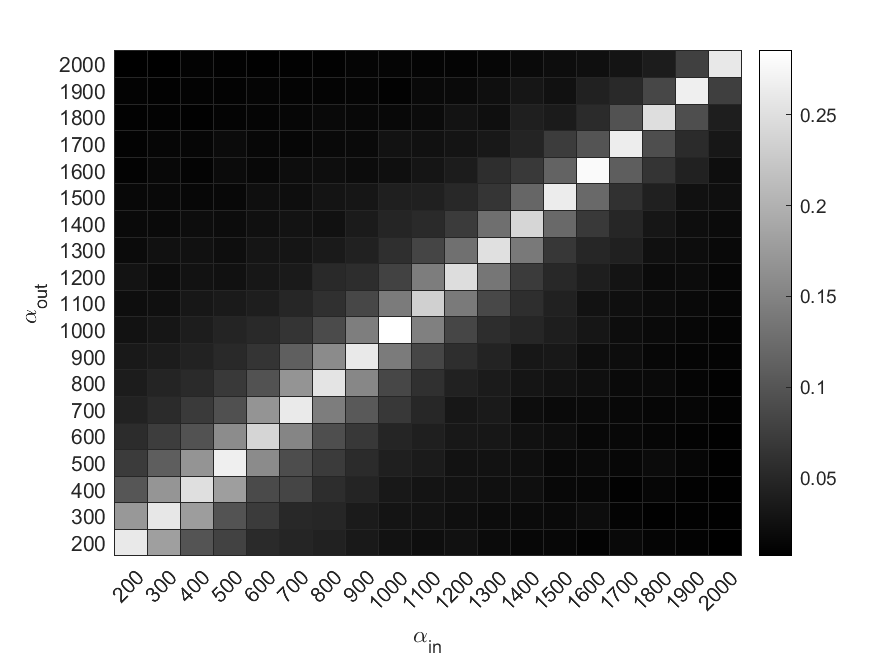}}
}
\caption{Poisson distribution. For panels (e) and (f): the darker pixel represents a lower Hamming Error.}
\label{S2} 
\end{figure}

\textbf{Simulation 2 (a): changing $\rho$}. Let $n_{r}=200,n_{c}=300, n_{r,0}=50, n_{c,0}=100, p=0.9$, and $P=P_{1}$. Since $\rho$ can be set in $(0,+\infty)$ for Poisson distribution, we let $\rho$ range in $\{0.5,1,1.5,\ldots,5\}$. The results are displayed in panels (a) and (b) of Figure \ref{S2}. We see that increasing $\rho$ decreases DiSP's error rates, and this is consistent with findings in Example \ref{Poisson}. In this setting, DiSP has the best performance among all 4 procedures.

\textbf{Simulation 2 (b): Changing $p$.} All parameters are set the same as Simulation 2 (a) except we let $\rho=2$ and $p$ range in $\{0.1, 0.2, \ldots, 1\}$. The results are reported in panels (c) and (d) of Figure \ref{S2}, which suggest that DiSP performs better when there are lesser missing edges and DiSP outperforms its competitors when $p$ is larger than 0.6.

\textbf{Simulation 2 (c): changing $\alpha_{\mathrm{in}}$ and $\alpha_{\mathrm{out}}$}. Let $n=n_{r}=n_{c}=300, n_{r,0}=50, n_{c,0}=100, p=1,$ and $\rho P=\tilde{P}$. $\alpha_{\mathrm{out}}$ should be set in $(0, +\infty)$ by Example \ref{Poisson}.  Here, we let $\alpha_{\mathrm{in}}$ and $\alpha_{\mathrm{out}}$ be in the range of $\{10,15,20,\ldots,100\}$, where $|\alpha_{\mathrm{in}}-\alpha_{\mathrm{out}}|\geq5$ when $\alpha_{\mathrm{in}}\neq\alpha_{\mathrm{out}}$. The numerical results are shown in panel (e) of Figure \ref{S2}. The analysis is similar to Simulation 1 (e), and we omit it here.

\textbf{Simulation 2 (d): changing $\alpha_{\mathrm{in}}$ and $\alpha_{\mathrm{out}}$}. All parameters are set the same as Simulation 2 (c) except that we let $\alpha_{\mathrm{in}}$ and $\alpha_{\mathrm{out}}$ be in the range of $\{200,300,400,\ldots,2000\}$ for this simulation, where $|\alpha_{\mathrm{in}}-\alpha_{\mathrm{out}}|\geq100$ when $\alpha_{\mathrm{in}}\neq\alpha_{\mathrm{out}}$. Panel (f) of Figure \ref{S2} shows the results. The analysis is similar to Simulation 1 (f), and we omit it here.
\subsubsection{Binomial distribution}
\begin{figure}
\centering
\resizebox{\columnwidth}{!}{
\subfigure[Simulation 3(a)]{\includegraphics[width=0.33\textwidth]{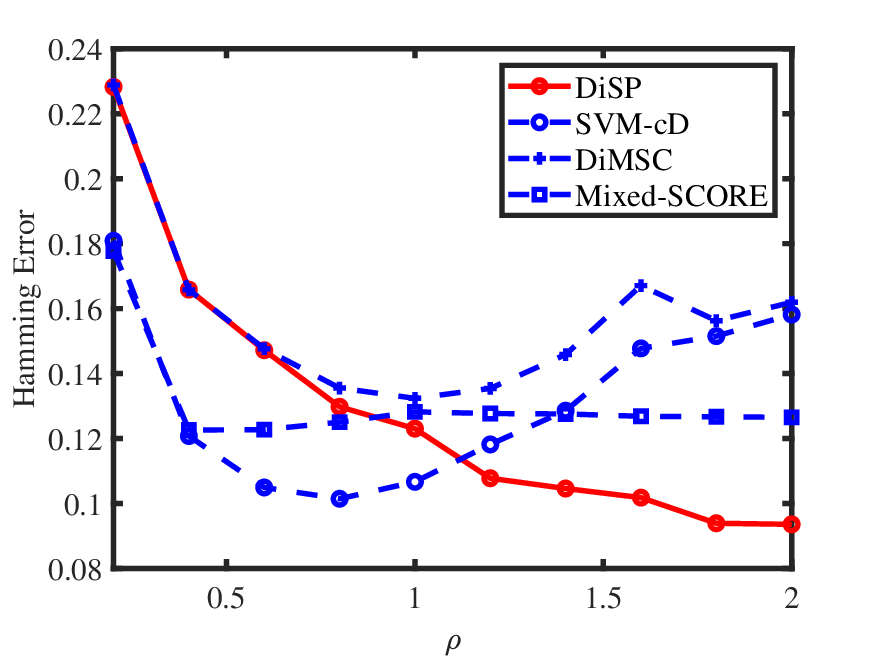}}
\subfigure[Simulation 3(a)]{\includegraphics[width=0.33\textwidth]{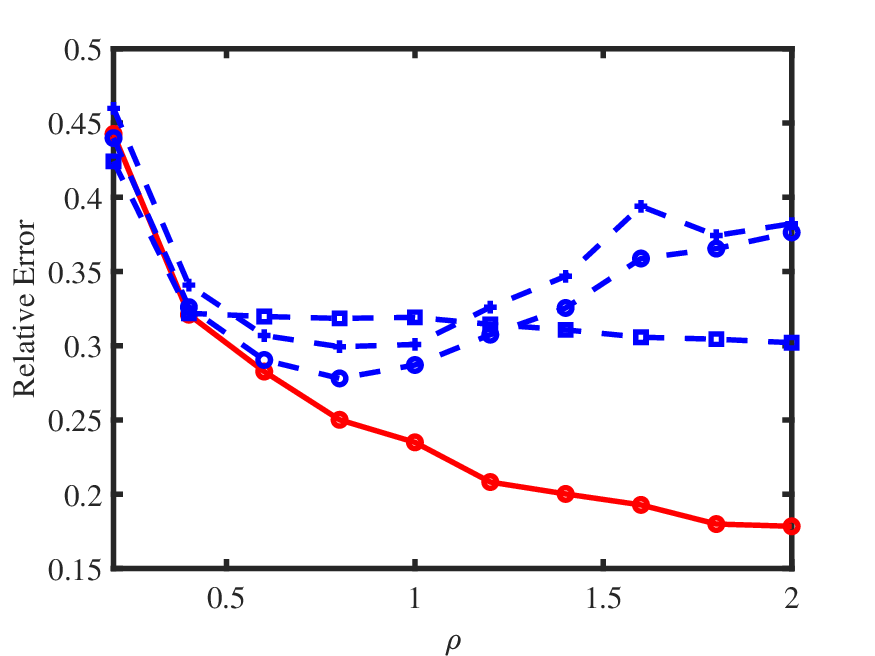}}
\subfigure[Simulation 3(b)]{\includegraphics[width=0.33\textwidth]{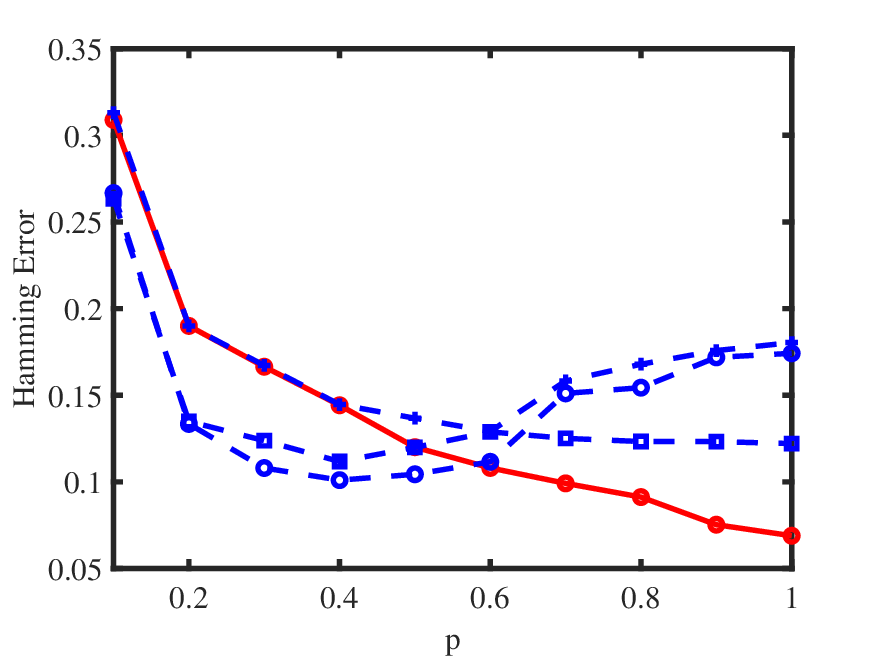}}
}
\resizebox{\columnwidth}{!}{
\subfigure[Simulation 3(b)]{\includegraphics[width=0.33\textwidth]{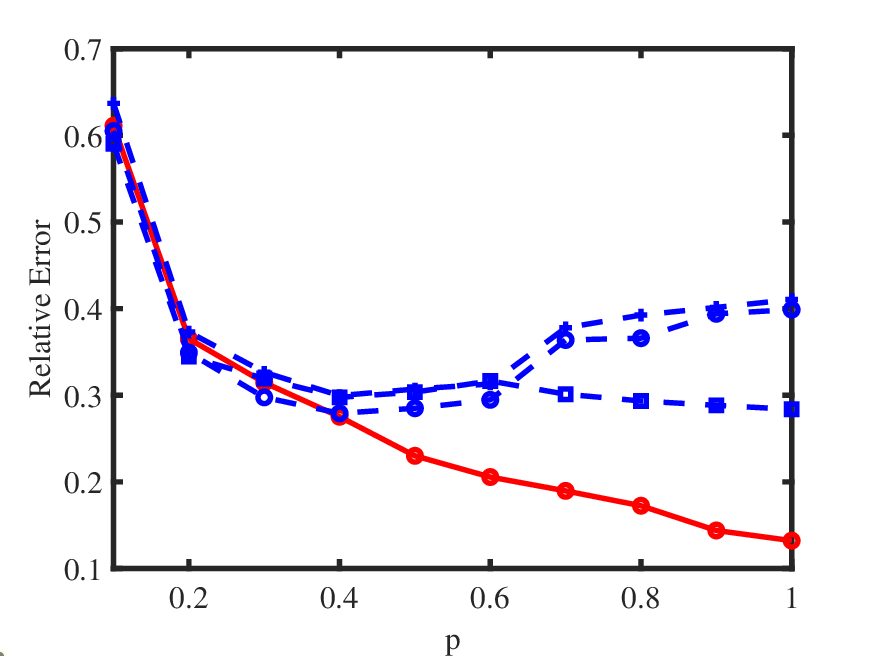}}
\subfigure[Simulation 3(c)]{\includegraphics[width=0.33\textwidth]{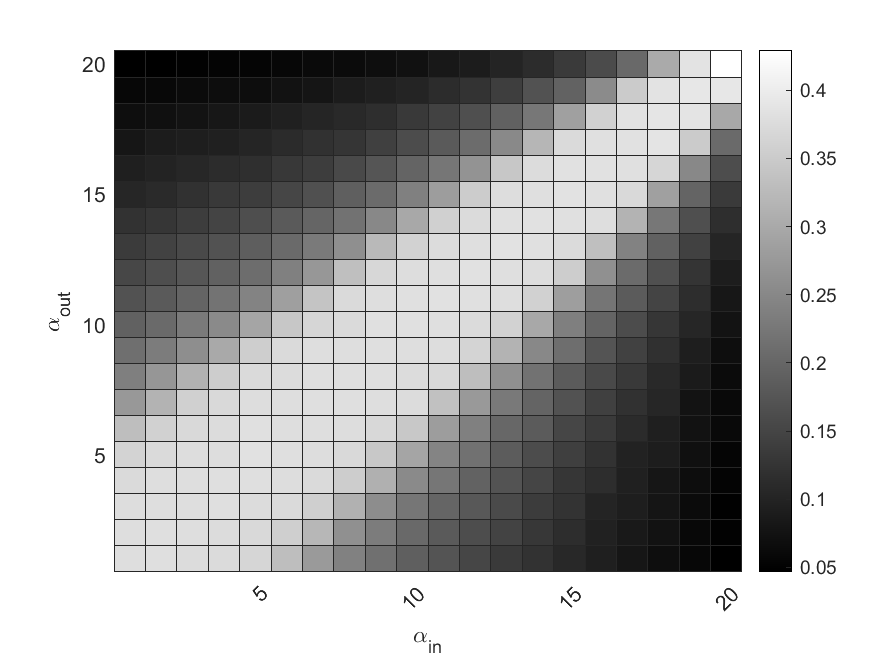}}
\subfigure[Simulation 3(d)]{\includegraphics[width=0.33\textwidth]{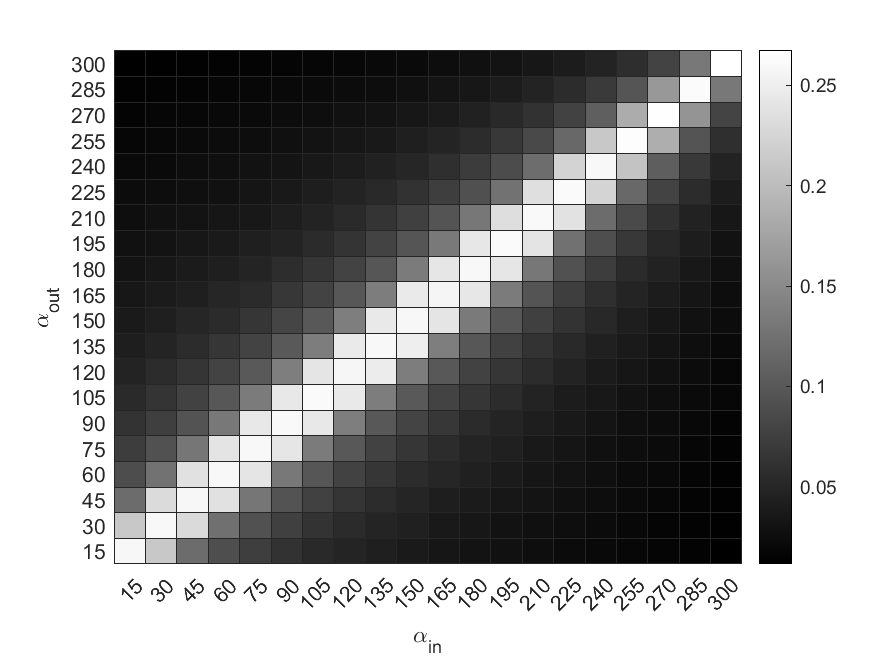}}
}
\caption{Binomial distribution. For panels (e) and (f): the darker pixel represents a lower Hamming Error.}
\label{S3} 
\end{figure}
When $A(i,j)\sim\mathrm{Binomial}(m,\frac{\Omega(i,j)}{m})$ for any positive integer $m$ for $i\in[n_{r}], j\in[n_{c}]$, by Example \ref{Binomial}, all elements of $P$ should be nonnegative and $\rho$ should be set less than $m$.

\textbf{Simulation 3 (a): changing $\rho$}. Let $n_{r}=200,n_{c}=300,n_{r,0}=50, n_{c,0}=100, p=0.9$, $P=P_{1}$, and $m=7$. Let $\rho$ range in $\{0.2,0.4,0.6,\ldots,2\}$. Panels (a) and (b) of Figure \ref{S3} display the results, and we see that DiSP's error rates decrease when increasing $\rho$, which is consistent with the analysis in Example \ref{Binomial}. In this experiment, DiSP and its competitors have very similar error rates when $\rho$ is small while DiSP outperforms its competitors when $\rho$ is larger than 1.

\textbf{Simulation 3 (b): Changing $p$.} All parameters are set the same as Simulation 3 (a) except we let $\rho=3$ and $p$ range in $\{0.1, 0.2, \ldots, 1\}$. The results are shown in panels (c) and (d) of Figure \ref{S3}. The analysis is similar to Simulation 2 (b), and we omit it here.

\textbf{Simulation 3 (c): changing $\alpha_{\mathrm{in}}$ and $\alpha_{\mathrm{out}}$}. Let $n=n_{r}=n_{c}=300, n_{r,0}=50, n_{c,0}=100, m=7, p=1$, and $\rho P=\tilde{P}$. For Poisson distribution, $\alpha_{\mathrm{in}}$ and $\alpha_{\mathrm{out}}$ should be set in $(0,\frac{mn}{\mathrm{log}(n)}]$ by Example \ref{Binomial}. For this simulation, we let $\alpha_{\mathrm{in}}$ and $\alpha_{\mathrm{out}}$ be in the range of $\{1,2,3,\ldots,20\}$, where $|\alpha_{\mathrm{in}}-\alpha_{\mathrm{out}}|\geq1$ when $\alpha_{\mathrm{in}}\neq\alpha_{\mathrm{out}}$. Panel (e) of Figure \ref{S3} shows the results. The analysis is similar to Simulation 1 (e), and we omit it here.

\textbf{Simulation 3 (d): changing $\alpha_{\mathrm{in}}$ and $\alpha_{\mathrm{out}}$}. All parameters are set the same as Simulation 3(c) except that we let $\alpha_{\mathrm{in}}$ and $\alpha_{\mathrm{out}}$ be in the range of $\{15,30,45,\ldots,300\}$ for this simulation, where $|\alpha_{\mathrm{in}}-\alpha_{\mathrm{out}}|\geq15$ when $\alpha_{\mathrm{in}}\neq\alpha_{\mathrm{out}}$. Panel (f) of Figure \ref{S3} shows the results. The analysis is similar to Simulation 1 (f), and we omit it here.
\subsubsection{Normal distribution}
When $A(i,j)\sim\mathrm{Normal}(\Omega(i,j),\sigma^{2}_{A})$ for some $\sigma^{2}_{A}>0$ for $i\in[n_{r}], j\in[n_{c}]$, by Example \ref{Normal}, all elements of $P$ are real values and $\rho$ can be set in $(0,+\infty)$.

\begin{figure}
\centering
\resizebox{\columnwidth}{!}{
\subfigure[Simulation 4(a)]{\includegraphics[width=0.33\textwidth]{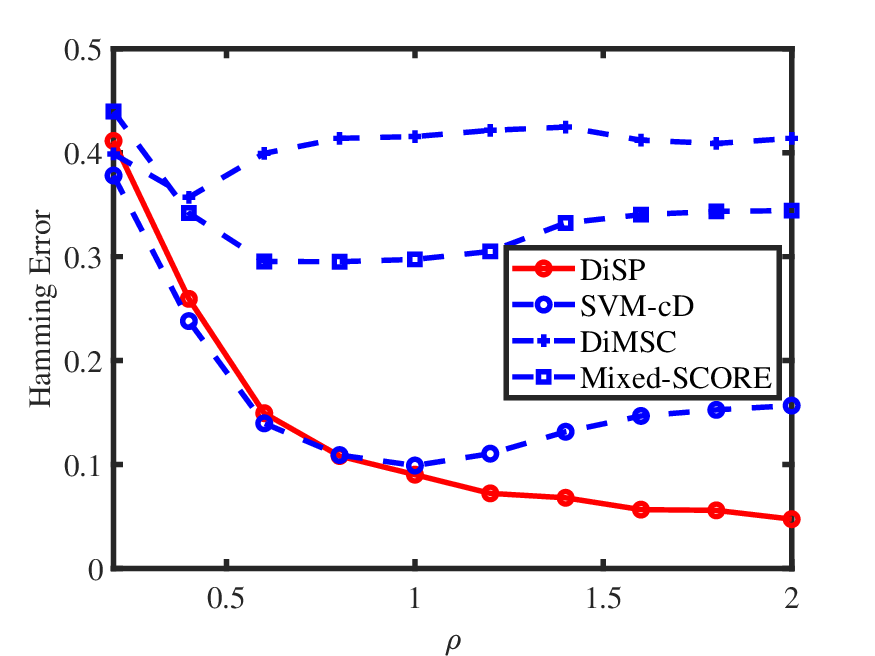}}
\subfigure[Simulation 4(a)]{\includegraphics[width=0.33\textwidth]{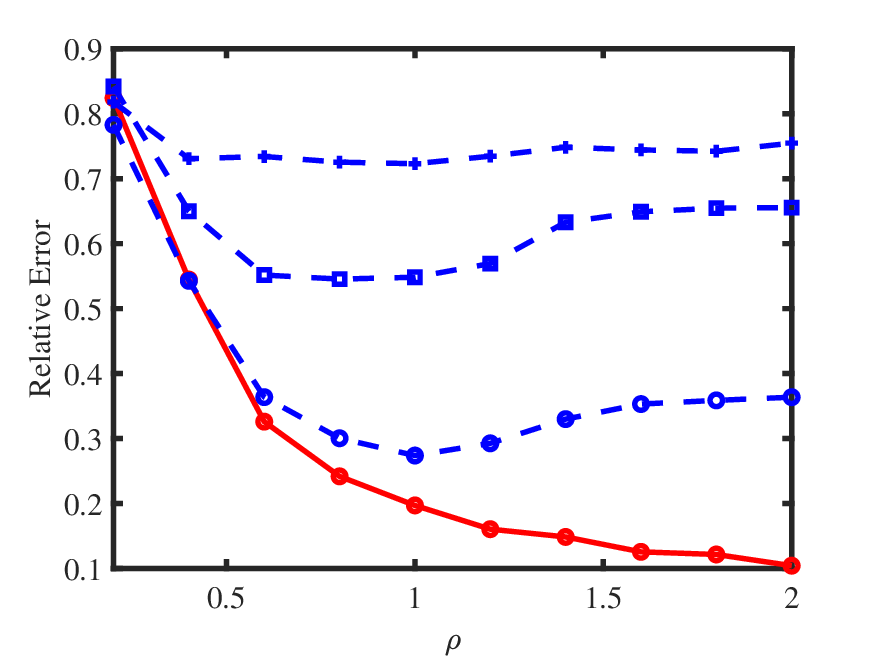}}
\subfigure[Simulation 4(b)]{\includegraphics[width=0.33\textwidth]{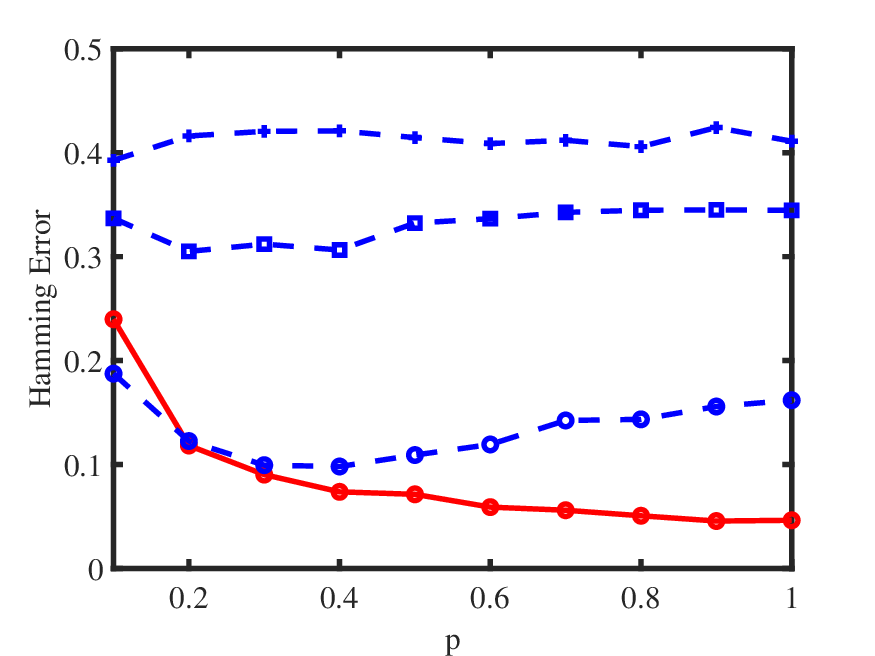}}
}
\resizebox{\columnwidth}{!}{
\subfigure[Simulation 4(b)]{\includegraphics[width=0.33\textwidth]{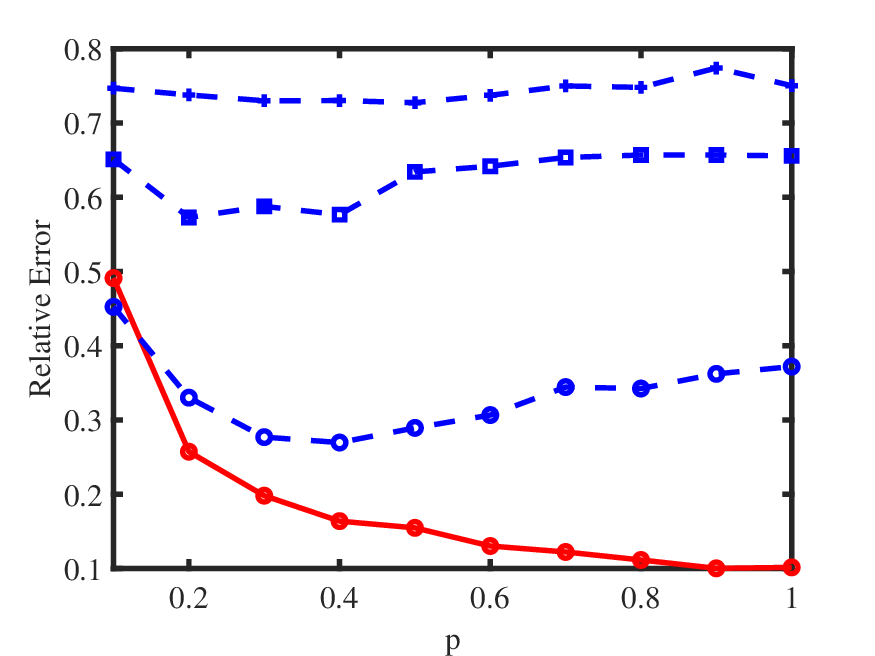}}
\subfigure[Simulation 4(c)]{\includegraphics[width=0.33\textwidth]{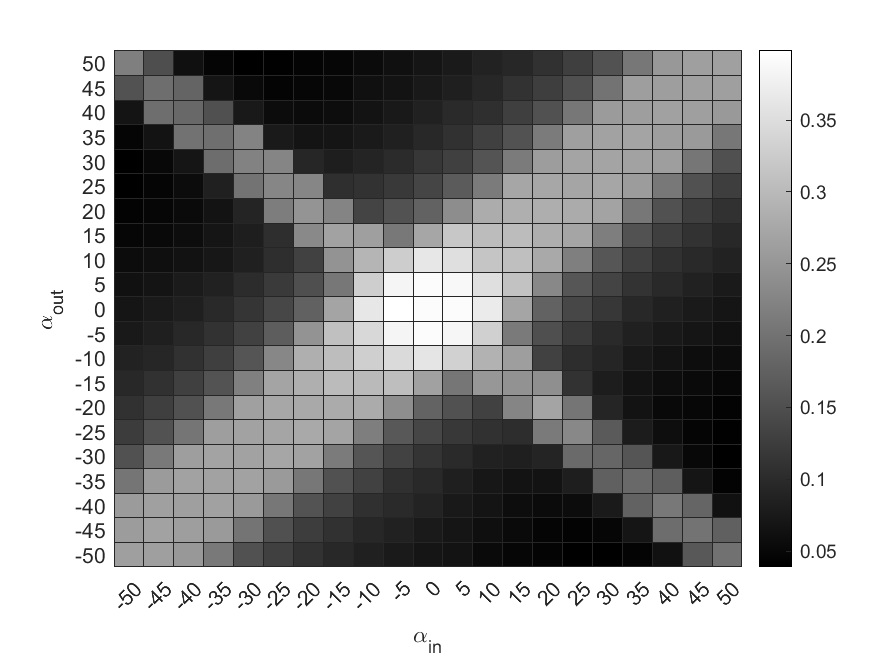}}
\subfigure[Simulation 4(d)]{\includegraphics[width=0.33\textwidth]{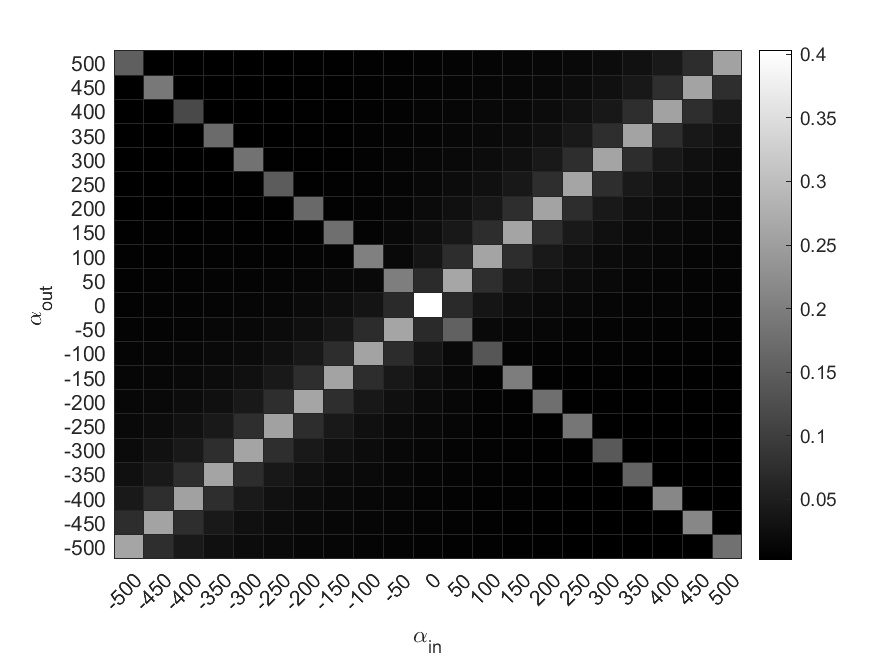}}
}
\caption{Normal distribution. For panels (e) and (f): the darker pixel represents a lower Hamming Error.}
\label{S4} 
\end{figure}

\textbf{Simulation 4 (a): changing $\rho$}. Let $n_{r}=200,n_{c}=300, n_{r,0}=50, n_{c,0}=100, p=0.9$, $P=P_{2}$, and $\sigma^{2}_{A}=1$. Let $\rho$ range in $\{0.2,0.4,0.6,\ldots,2\}$. The results are shown in panels (a) and (b) of Figure \ref{S4}. We see that DiSP's error rates decrease when $\rho$ increases and this is consistent with findings in Example \ref{Normal}. When $\rho$ is less than 1, DiSP and SVM-cD have similar error rates, which are smaller than those of DiMSC and Mixed-SCORE. When $\rho$ is larger than 1, DiSP has the best performance among all approaches.

\textbf{Simulation 4 (b): changing $p$}. All parameters are set the same as Simulation 4 (a) except we let $\rho=2$ and $p$ range in $\{0.1, 0.2, \ldots, 1\}$. The results are displayed in panels (c) and (d) of Figure \ref{S4}. It suggests that DiSP performs better when $p$ increases and DiSP significantly outperforms its competitors.

\textbf{Simulation 4 (c): changing $\alpha_{\mathrm{in}}$ and $\alpha_{\mathrm{out}}$}. Let $n=n_{r}=n_{c}=300, n_{r,0}=50, n_{c,0}=100, \sigma^{2}_{A}=1, p=1$, and $\rho P=\tilde{P}$. For Normal distribution, $\alpha_{\mathrm{in}}$ and $\alpha_{\mathrm{out}}$ can be set in $(-\infty,+\infty)$ by Example \ref{Normal}. For this simulation, we let $\alpha_{\mathrm{in}}$ and $\alpha_{\mathrm{out}}$ be in the range of $\{-50,-45,-40,\ldots,50\}$, where $||\alpha_{\mathrm{in}}|-|\alpha_{\mathrm{out}}||\geq15$ when $|\alpha_{\mathrm{in}}|\neq|\alpha_{\mathrm{out}}|$. Panel (e) of Figure \ref{S4} shows the results. Because the first inequality of Equation (\ref{CondAlphaInOutNormal}) does not add a constraint on $\mathrm{max}(|\alpha_{\mathrm{in}}|,|\alpha_{\mathrm{out}}|)$, DiSP's error rates are small as long as the second inequality holds. We see that results of Simulation 4 (c) support Equation (\ref{CondAlphaInOutNormal}) because the ``white'' area in panel (e) of Figure \ref{S4} enjoys a symmetric structure while the ``white'' areas in panel (e) of Figure \ref{S1}, panel (e) of Figure \ref{S2} and panel (e) of Figure \ref{S3} have an asymmetric structure because the first inequality of Equations (\ref{CondAlphaInOutBernoulli})-(\ref{CondAlphaInOutBinomial}) has requirement on $\mathrm{max}(\alpha_{\mathrm{in}},\alpha_{\mathrm{out}})$.

\textbf{Simulation 4 (d): changing $\alpha_{\mathrm{in}}$ and $\alpha_{\mathrm{out}}$}. All parameters are set the same as Simulation 4(c) except that we let $\alpha_{\mathrm{in}}$ and $\alpha_{\mathrm{out}}$ be in the range of $\{-500,-450,-400,\ldots,500\}$ for this simulation, where $||\alpha_{\mathrm{in}}|-|\alpha_{\mathrm{out}}||\geq150$ when $|\alpha_{\mathrm{in}}|\neq|\alpha_{\mathrm{out}}|$ (note that 150 is much larger than 15 in Simulation 4 (c)). The results are displayed in panel (f) of Figure \ref{S4}. Error rates of Simulation 4 (d) are much smaller than that of Simulation 4 (c) because we increase $||\alpha_{\mathrm{in}}|-|\alpha_{\mathrm{out}}||$ when $|\alpha_{\mathrm{in}}|\neq|\alpha_{\mathrm{out}}|$. So, the results of this simulation also support Equation (\ref{CondAlphaInOutNormal}).
\subsubsection{Exponential distribution}
When $A(i,j)\sim \mathrm{Exponential}(\frac{1}{\Omega(i,j)})$ for $i\in[n_{r}], j\in[n_{c}]$, by Example \ref{Exponential}, $P$ should have positive entries. and $\rho$ can be set in $(0,+\infty)$.

\begin{figure}
\centering
\resizebox{\columnwidth}{!}{
\subfigure[Simulation 5(a)]{\includegraphics[width=0.33\textwidth]{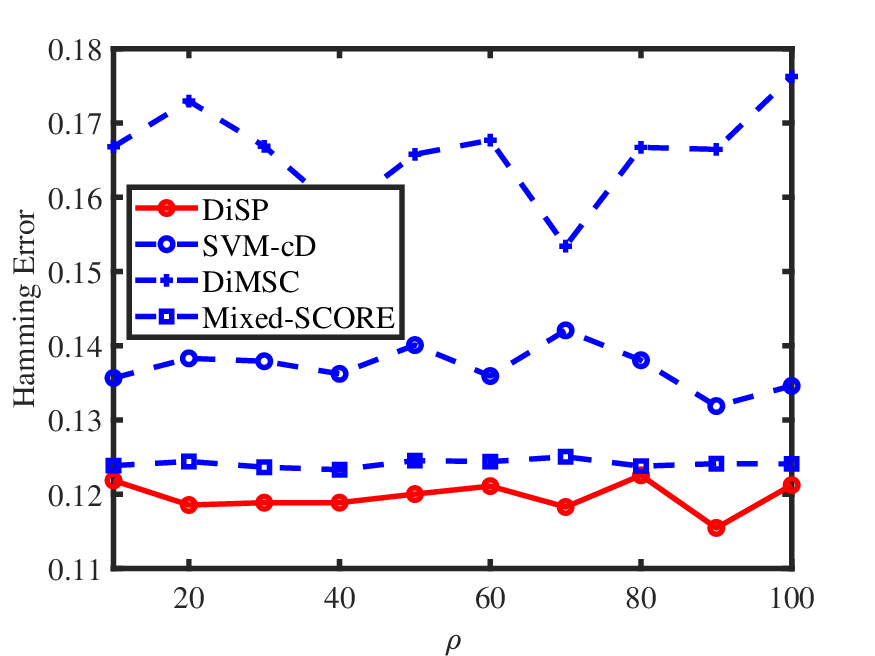}}
\subfigure[Simulation 5(a)]{\includegraphics[width=0.33\textwidth]{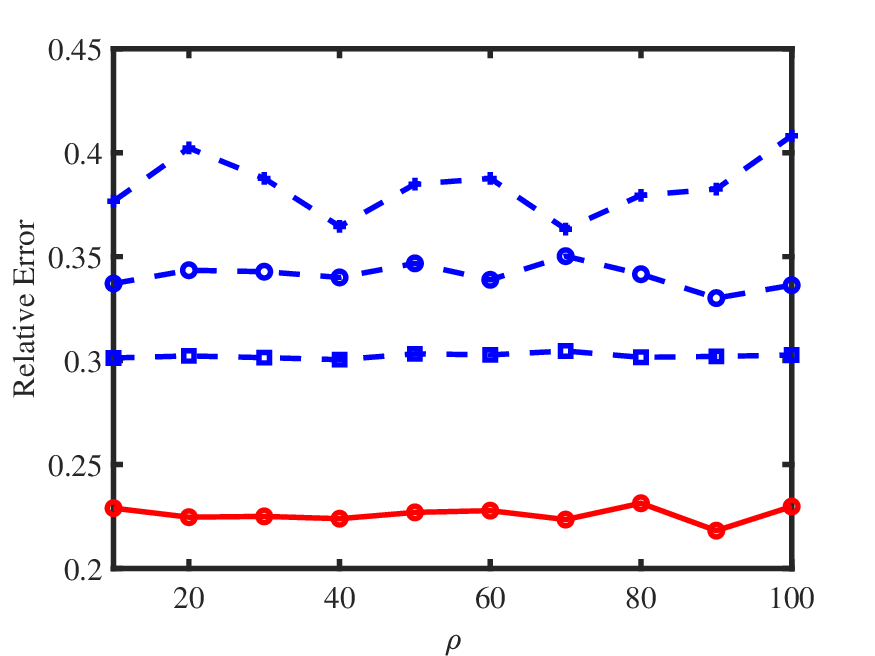}}
\subfigure[Simulation 5(b)]{\includegraphics[width=0.33\textwidth]{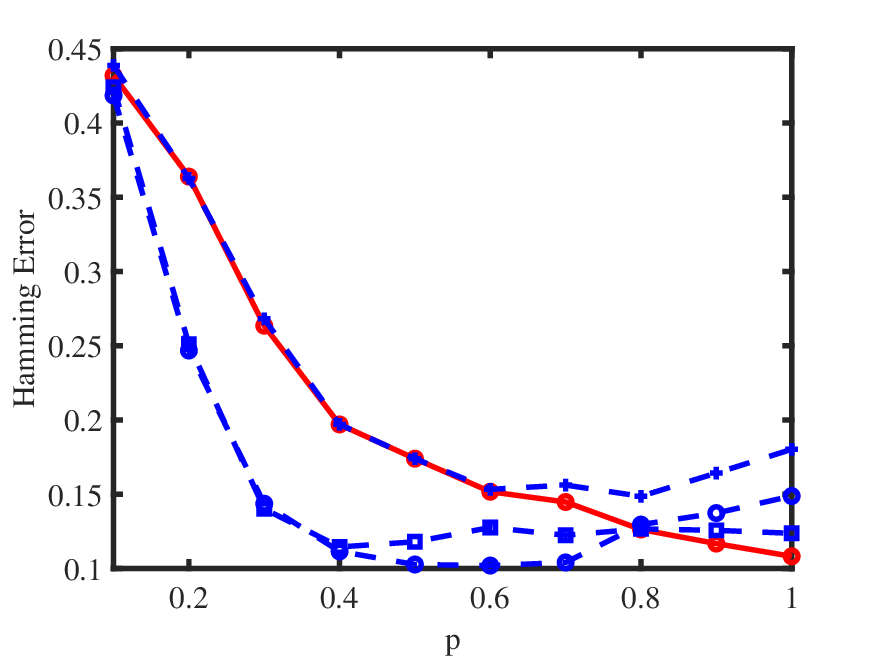}}
}
\resizebox{\columnwidth}{!}{
\subfigure[Simulation 5(b)]{\includegraphics[width=0.33\textwidth]{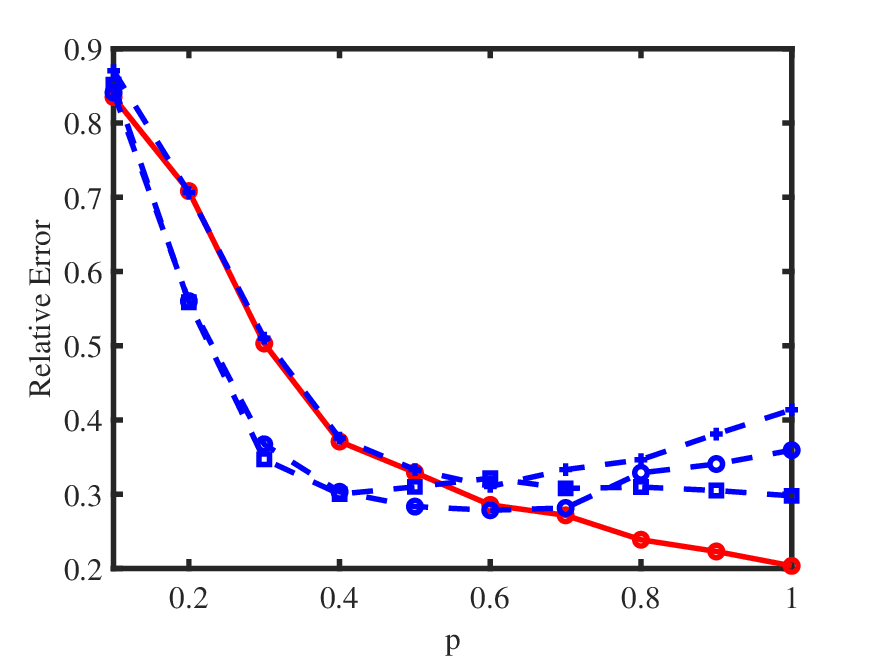}}
\subfigure[Simulation 5(c)]{\includegraphics[width=0.33\textwidth]{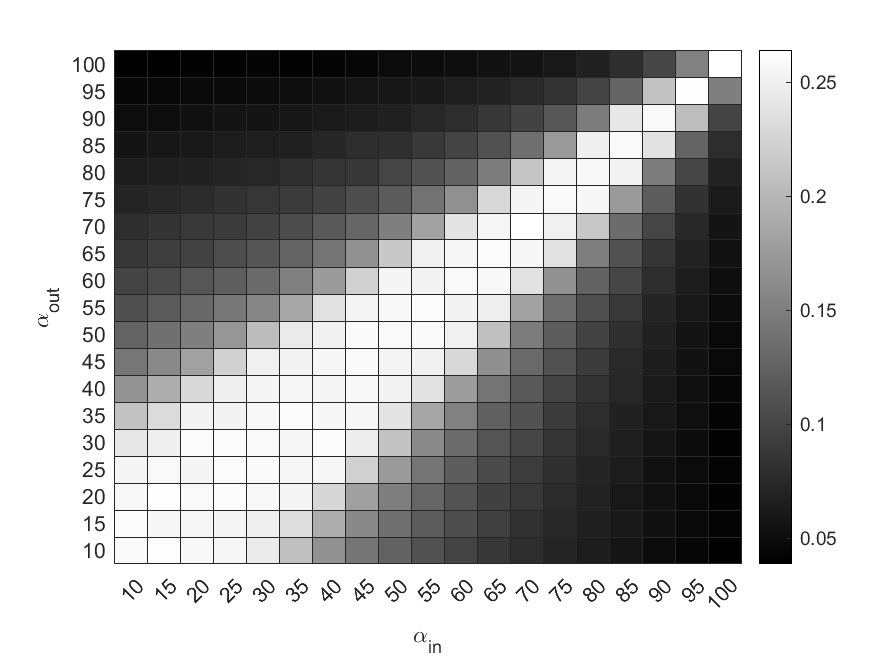}}
\subfigure[Simulation 5(d)]{\includegraphics[width=0.33\textwidth]{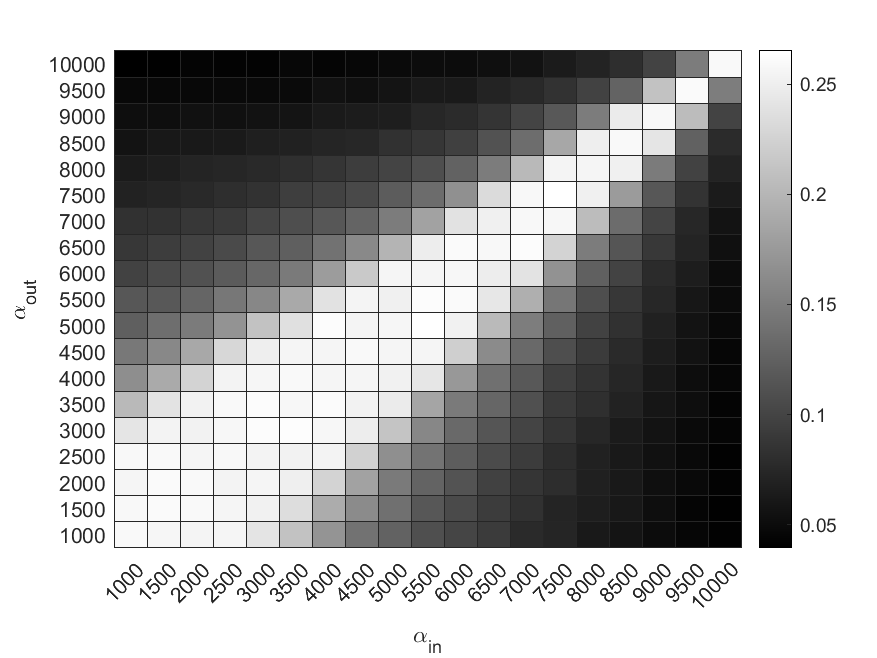}}
}
\caption{Exponential distribution. For panels (e) and (f): the darker pixel represents a lower Hamming Error.}
\label{S5} 
\end{figure}

\textbf{Simulation 5 (a): changing $\rho$}. Let $n_{r}=200,n_{c}=300, n_{r,0}=50, n_{c,0}=100, p=0.9$, $P=P_{1}$. Let $\rho$ range in $\{10,20,30,\ldots,100\}$. In the plot of the result (Figure \ref{S5} (a) and (b)), we see that increasing $\rho$ has no influence on DiSP's performance, and this phenomenon matches our findings in Example \ref{Exponential} because $\rho$ vanishes in the theoretical upper bounds of error rates when setting $\gamma$ as $\rho$ for Exponential distribution. In this experiment, the error rates of DiSP are smaller than that of the best-performing algorithm among the others.

\textbf{Simulation 5 (b): changing $p$}. All parameters are set the same as Simulation 5 (a) except we let $\rho=10$ and $p$ range in $\{0.1, 0.2, \ldots, 1\}$. The results are shown in panels (c) and (d) of Figure \ref{S5}. The analysis is similar to Simulation 2 (b), and we omit it here.

\textbf{Simulation 5 (c): changing $\alpha_{\mathrm{in}}$ and $\alpha_{\mathrm{out}}$}. Let $n=n_{r}=n_{c}=300, n_{r,0}=50, n_{c,0}=100, p=1$, and $\rho P=\tilde{P}$. For Exponential distribution, $\alpha_{\mathrm{in}}$ and $\alpha_{\mathrm{out}}$ can be set in $(0,+\infty)$ by Example \ref{Exponential}. Here, we let $\alpha_{\mathrm{in}}$ and $\alpha_{\mathrm{out}}$ be in the range of $\{10,15,20,\ldots,100\}$, where $|\alpha_{\mathrm{in}}-\alpha_{\mathrm{out}}|\geq5$ when $\alpha_{\mathrm{in}}\neq\alpha_{\mathrm{out}}$. The numerical results are shown in panel (e) of Figure \ref{S5}. We see that the ``white'' area of panel (e) has an asymmetric structure, and this phenomenon occurs because $\mathrm{max}(\alpha_{\mathrm{in}}, \alpha_{\mathrm{out}})$ should be sufficiently large to make DiSP's error rates small even when the second inequality of Equation (\ref{CondAlphaInOutExponential}) holds. We also see that DiSP performs better when increasing $|\alpha_{\mathrm{in}}-\alpha_{\mathrm{out}}|$.

\textbf{Simulation 5 (d): changing $\alpha_{\mathrm{in}}$ and $\alpha_{\mathrm{out}}$}. All parameters are set the same as Simulation 5(c) except that we let $\alpha_{\mathrm{in}}$ and $\alpha_{\mathrm{out}}$ be in the range of $\{1000,1500,2000,\ldots,10000\}$ for this simulation, where $|\alpha_{\mathrm{in}}-\alpha_{\mathrm{out}}|\geq500$ when $\alpha_{\mathrm{in}}\neq\alpha_{\mathrm{out}}$ (note that 500 is much larger than 5 in Simulation 5 (c)). The results are displayed in panel (f) of Figure \ref{S5}. Unlike Simulations 1-4, the ``white'' area in panel (f) of Figure \ref{S5} still has an asymmetric structure even though 500 is much larger than 5. This phenomenon occurs because the first inequality of Equation (\ref{CondAlphaInOutExponential}) is $\mathrm{max}(\alpha^{2}_{\mathrm{in}},\alpha^{2}_{\mathrm{out}})\frac{\mathrm{log}(n)}{n}\geq\tau^{2}+o(1)$. The $\frac{\mathrm{log}(n)}{n}$ term makes that to make Equation (\ref{CondAlphaInOutExponential}) hold, $\mathrm{max}(\alpha_{\mathrm{in}}, \alpha_{\mathrm{out}})$ must be large enough even when the second inequality of Equation (\ref{CondAlphaInOutExponential}) holds. Therefore, the asymmetric structures of ``white'' areas in panels (e) and (f) of Figure \ref{S5} support our findings in Equation (\ref{CondAlphaInOutExponential}) for Exponential distribution.
\subsubsection{Uniform distribution}
\begin{figure}
\centering
\resizebox{\columnwidth}{!}{
\subfigure[Simulation 6(a)]{\includegraphics[width=0.33\textwidth]{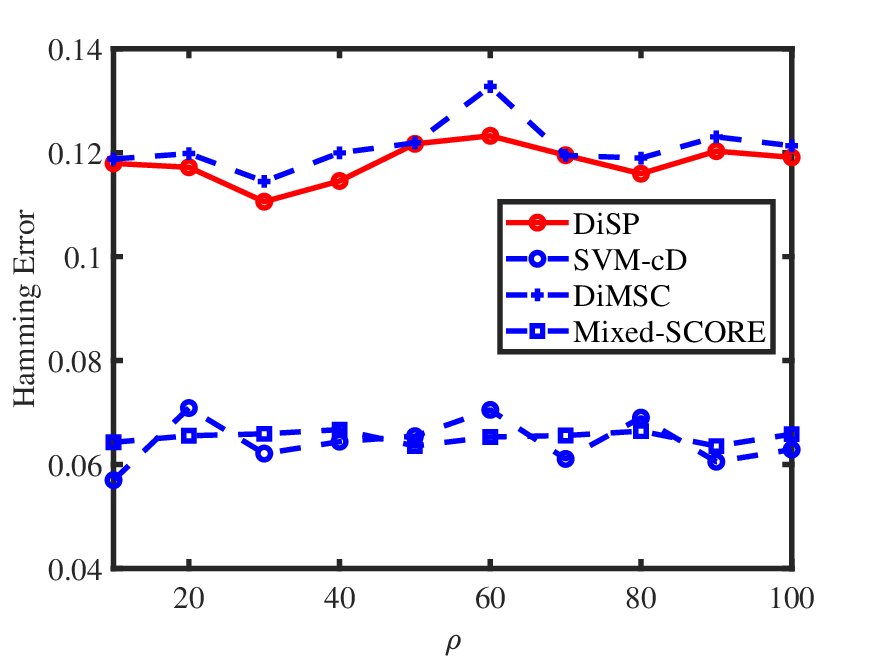}}
\subfigure[Simulation 6(a)]{\includegraphics[width=0.33\textwidth]{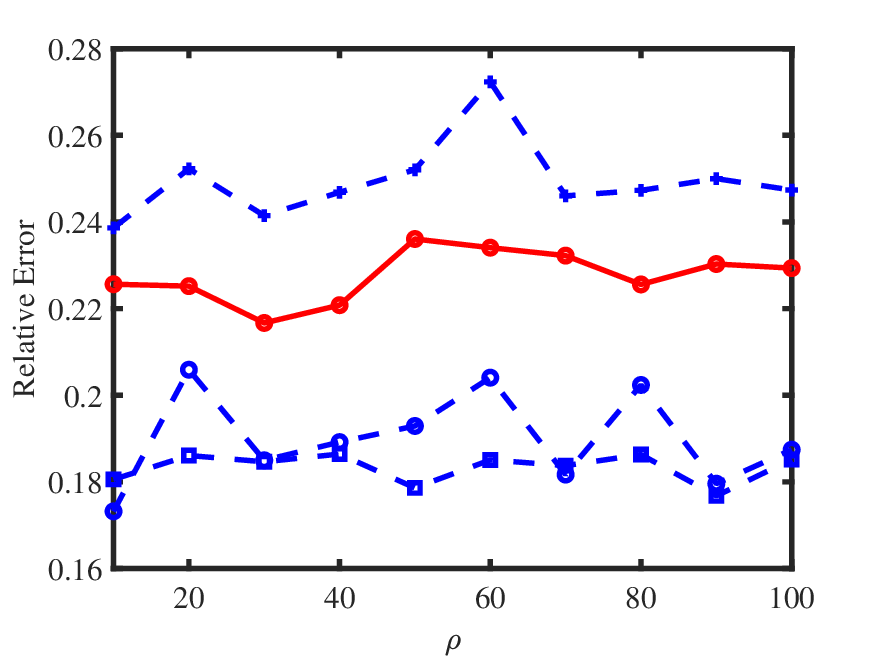}}
\subfigure[Simulation 6(b)]{\includegraphics[width=0.33\textwidth]{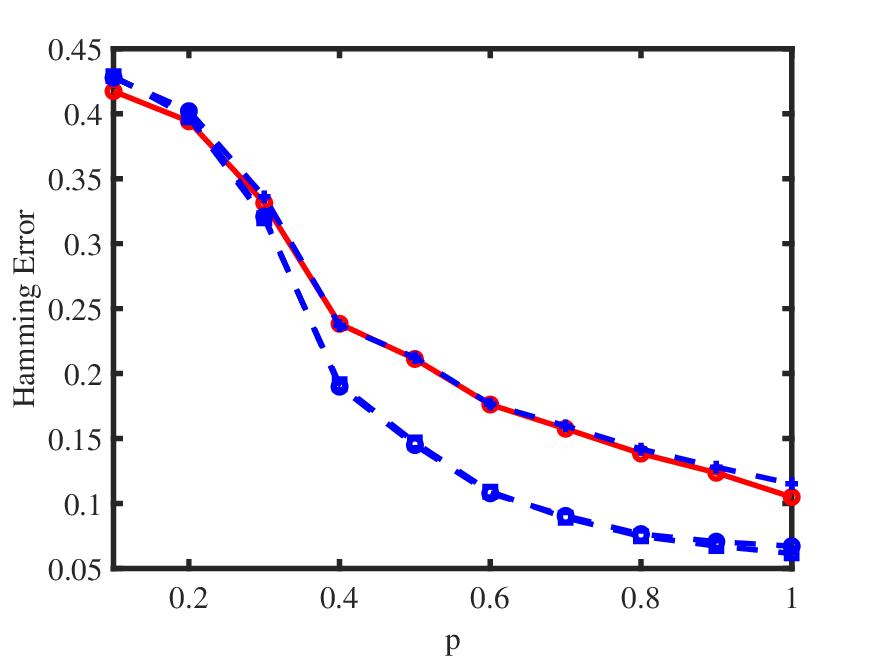}}
}
\resizebox{\columnwidth}{!}{
\subfigure[Simulation 6(b)]{\includegraphics[width=0.33\textwidth]{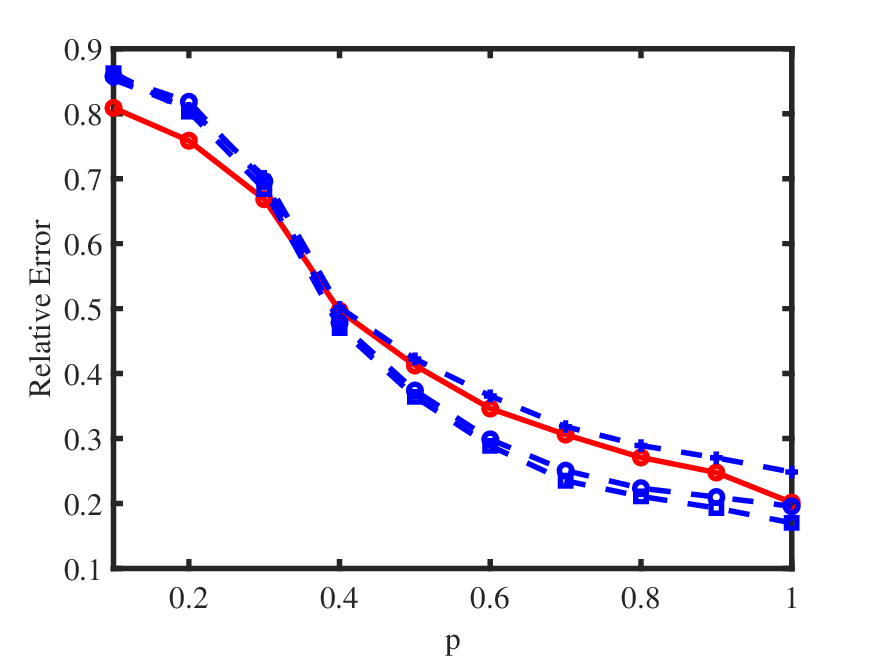}}
\subfigure[Simulation 6(c)]{\includegraphics[width=0.33\textwidth]{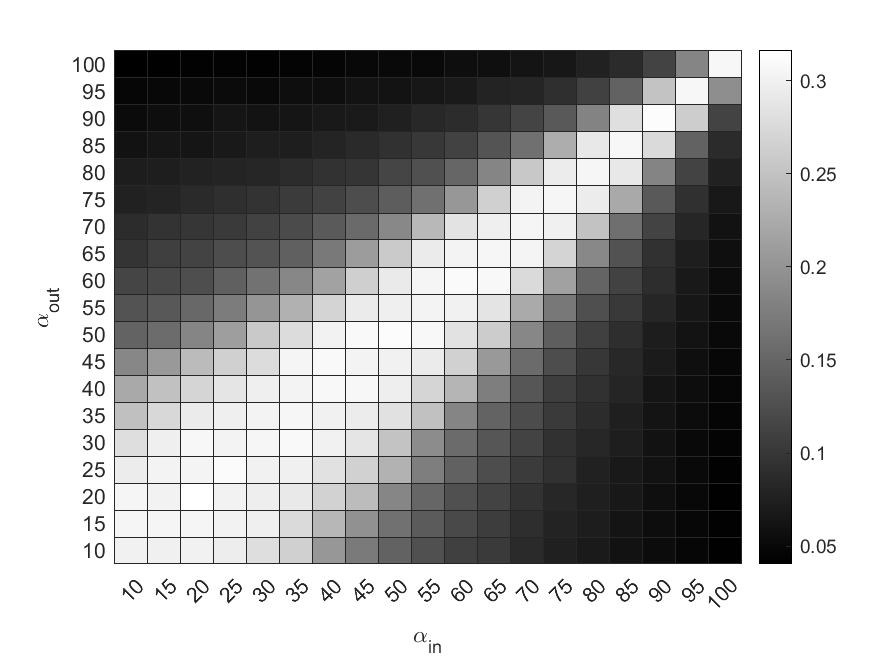}}
\subfigure[Simulation 6(d)]{\includegraphics[width=0.33\textwidth]{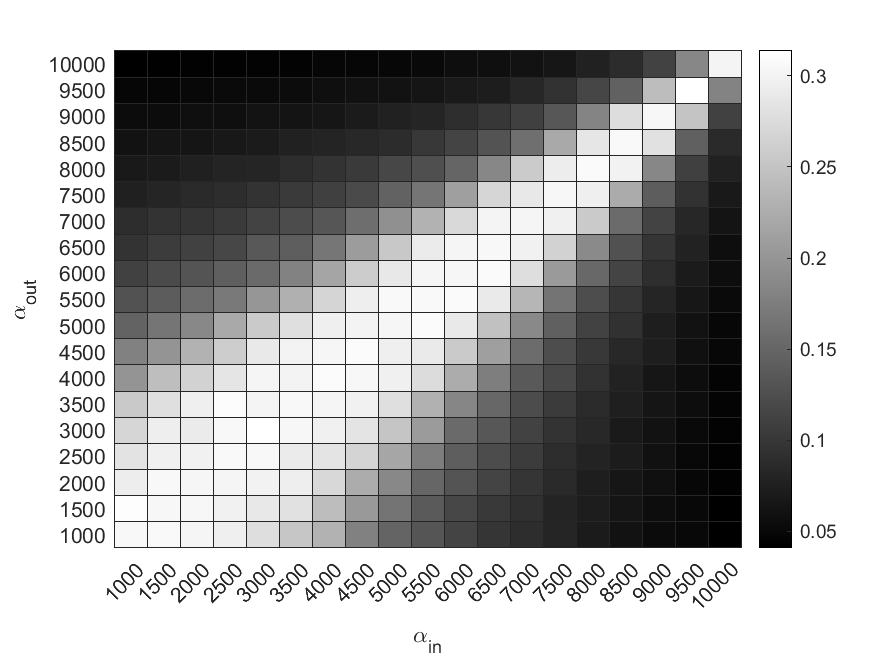}}
}
\caption{Uniform distribution. For panels (e) and (f): the darker pixel represents a lower Hamming Error.}
\label{S6} 
\end{figure}
When  $A(i,j)\sim\mathrm{Uniform}(0,2\Omega(i,j))$ for $i\in[n_{r}], j\in[n_{c}]$, by Example \ref{Uniform}, all entries of $P$ should be nonnegative and $\rho$ can be set in $(0,+\infty)$.

\textbf{Simulation 6 (a): changing $\rho$}. Let $n_{r}=30,n_{c}=50, n_{r,0}=10, n_{c,0}=20, p=0.9$, $P=P_{1}$. Let $\rho$ range in $\{10,20,30,\ldots,100\}$. Panels (a) and (b) of Figure \ref{S6} show the results. We see that $\rho$ has no significant influence on the performances for all 4 algorithms for Uniform distribution, and this verifies our findings in Example \ref{Uniform}. In this experiment, though DiSP outperforms DiMSC, it performs slightly poorer than SVM-cD and Mixed-SCORE.

\textbf{Simulation 6 (b): changing $p$}. All parameters are set the same as Simulation 6 (a) except we let $\rho=10$ and $p$ range in $\{0.1, 0.2, \ldots, 1\}$. The results are displayed in panels (c) and (d) of Figure \ref{S6}. We see that all procedures have similar error rates and they perform better when there are lesser missing edges as $p$ increases.

\textbf{Simulation 6 (c): changing $\alpha_{\mathrm{in}}$ and $\alpha_{\mathrm{out}}$}. Let $n=n_{r}=n_{c}=50, n_{r,0}=10, n_{c,0}=20, p=1$, and $\rho P=\tilde{P}$. For Uniform distribution, $\alpha_{\mathrm{in}}$ and $\alpha_{\mathrm{out}}$ can be set in $(0,+\infty)$ by Example \ref{Uniform}. Here, we let $\alpha_{\mathrm{in}}$ and $\alpha_{\mathrm{out}}$ be in the range of $\{10,15,20,\ldots,100\}$. The numerical results are shown in panel (e) of Figure \ref{S6}. The analysis for this simulation is similar to that of Simulation 5 (e), and we omit it here.

\textbf{Simulation 6 (d): changing $\alpha_{\mathrm{in}}$ and $\alpha_{\mathrm{out}}$}. All parameters are set the same as Simulation 6(c) except that we let $\alpha_{\mathrm{in}}$ and $\alpha_{\mathrm{out}}$ be in the range of $\{1000,1500,2000,\ldots,10000\}$ for this simulation. The numerical results are displayed in the last panel of Figure \ref{S6}. The analysis is similar to that of Simulation 5 (f), and we omit it here.
\subsubsection{Logistic distribution}
When $A(i,j)\sim\mathrm{Logistic}(\Omega(i,j),\beta)$ for $\beta>0$ for $i\in[n_{r}], j\in[n_{c}]$, by Example \ref{Logistic}, all entries of $P$ are real values and $\rho$ can be set in $(0,+\infty)$.

\textbf{Simulation 7(a): changing $\rho$}. Let $n_{r}=30,n_{c}=50, n_{r,0}=10, n_{c,0}=20, \beta=1, p=0.9$, and $P=P_{2}$. Let $\rho$ range in $\{0.2,0.4,0.6,\ldots,2\}$.
\begin{figure}
\centering
\resizebox{\columnwidth}{!}{
\subfigure[Simulation 7(a)]{\includegraphics[width=0.33\textwidth]{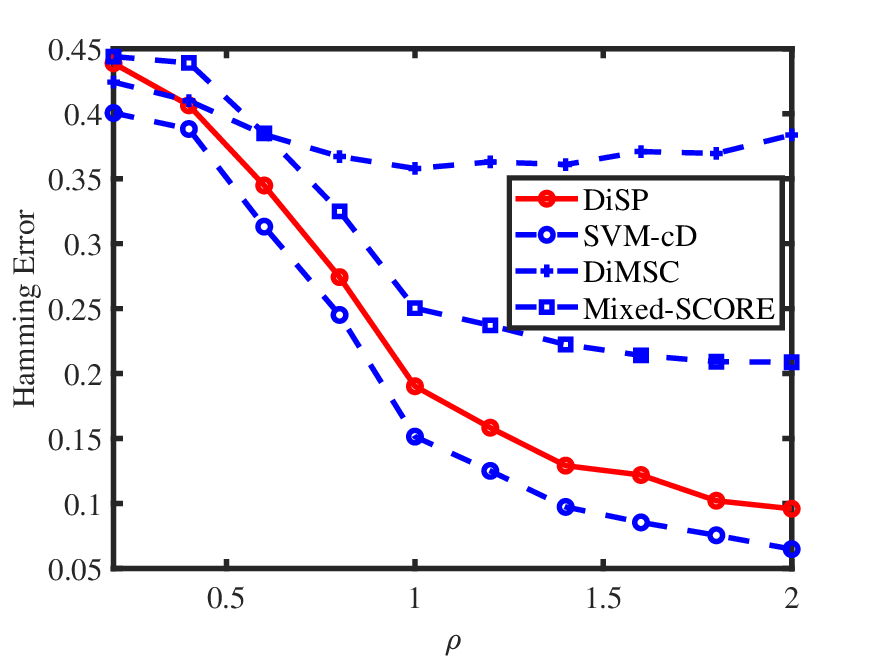}}
\subfigure[Simulation 7(a)]{\includegraphics[width=0.33\textwidth]{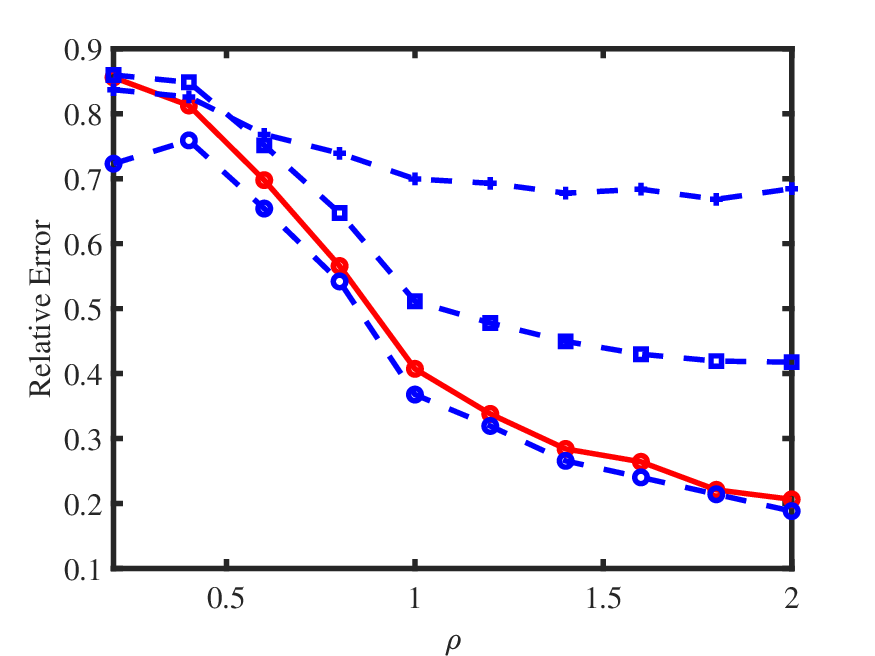}}
\subfigure[Simulation 7(b)]{\includegraphics[width=0.33\textwidth]{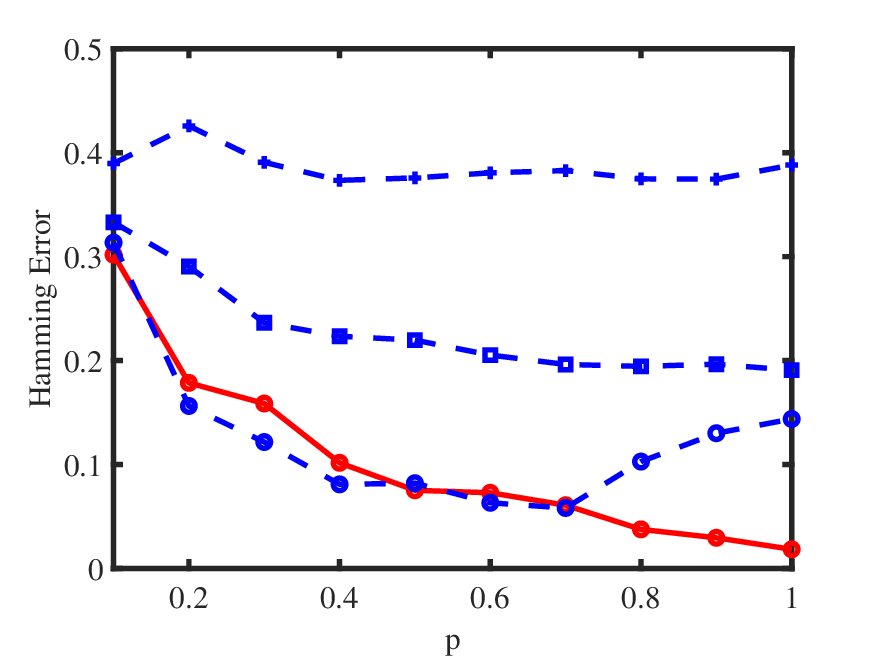}}
}
\resizebox{\columnwidth}{!}{
\subfigure[Simulation 7(b)]{\includegraphics[width=0.33\textwidth]{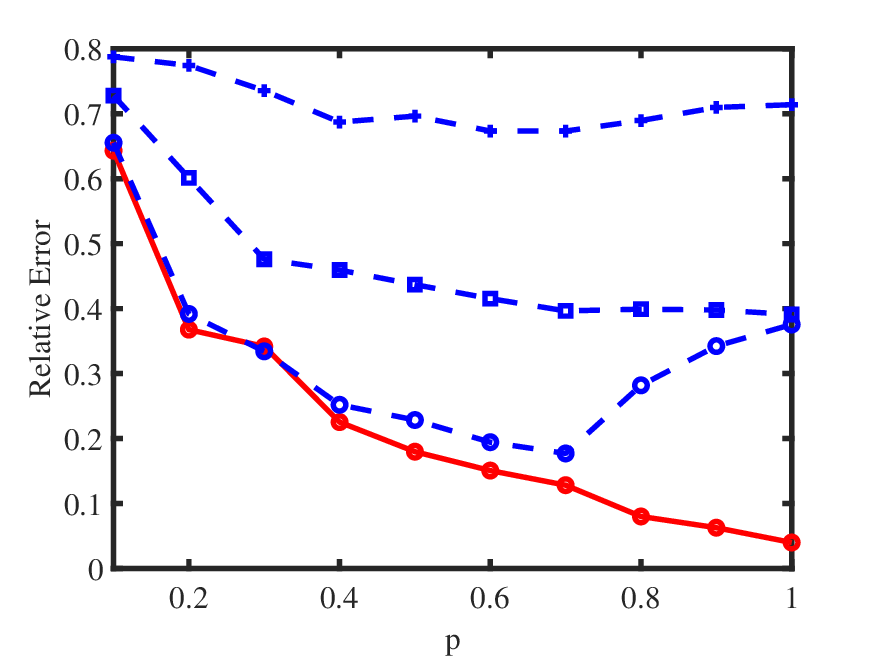}}
\subfigure[Simulation 7(c)]{\includegraphics[width=0.33\textwidth]{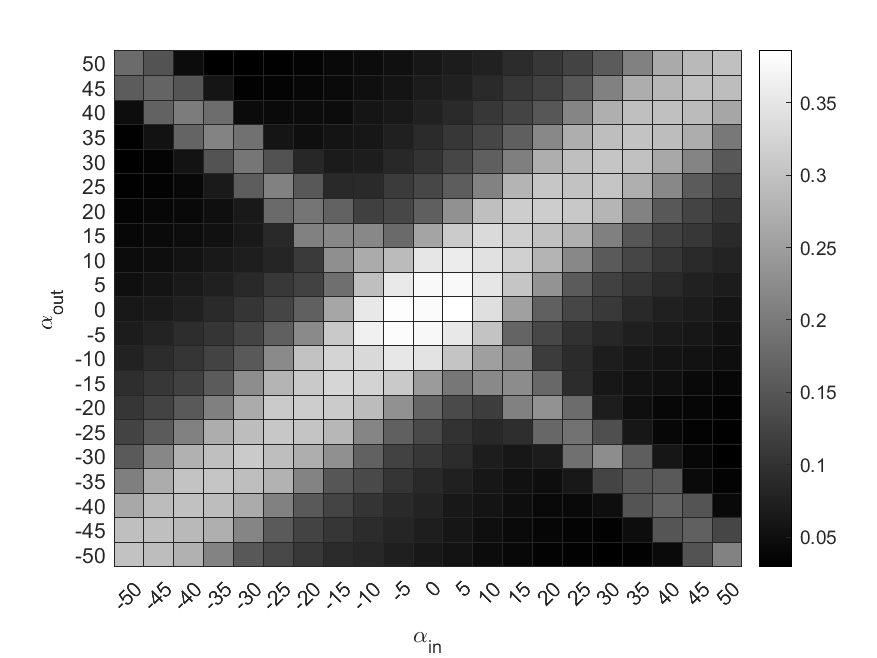}}
\subfigure[Simulation 7(d)]{\includegraphics[width=0.33\textwidth]{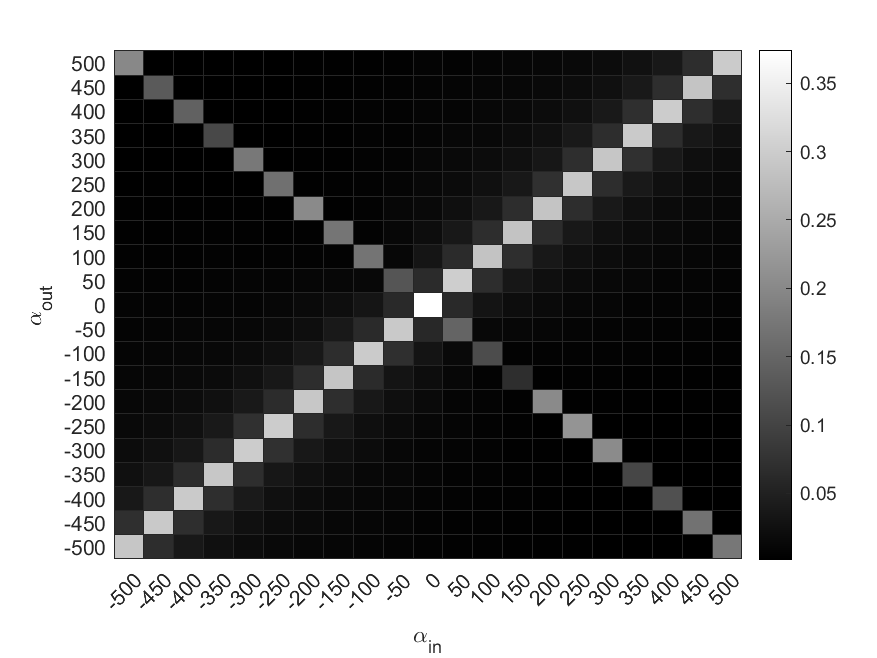}}
}
\caption{Logistic distribution. For panels (e) and (f): the darker pixel represents a lower Hamming Error.}
\label{S7} 
\end{figure}

\textbf{Simulation 7 (b): changing $p$}. All parameters are set the same as Simulation 7 (a) except we let $\rho=2$ and $p$ range in $\{0.1, 0.2, \ldots, 1\}$.

\textbf{Simulation 7 (c): changing $\alpha_{\mathrm{in}}$ and $\alpha_{\mathrm{out}}$}. Let $n=n_{r}=n_{c}=50, n_{r,0}=10, n_{c,0}=20, \beta=1, p=1$, and $\rho P=\tilde{P}$. For Logistic distribution, $\alpha_{\mathrm{in}}$ and $\alpha_{\mathrm{out}}$ can be set in $(-\infty,+\infty)$ by Example \ref{Logistic}. For this simulation, we let $\alpha_{\mathrm{in}}$ and $\alpha_{\mathrm{out}}$ be in the range of $\{-50,-45,-40,\ldots,50\}$.

\textbf{Simulation 7 (d): changing $\alpha_{\mathrm{in}}$ and $\alpha_{\mathrm{out}}$}. All parameters are set the same as Simulation 7(c) except that we let $\alpha_{\mathrm{in}}$ and $\alpha_{\mathrm{out}}$ be in the range of $\{-500,-450,-400,\ldots,500\}$ for this simulation.

Figure \ref{S7} displays the results of Simulation 7. The analysis is similar to that of Simulation 4, and we omit it here.
\subsubsection{Bipartite signed network}
\begin{figure}
\centering
\resizebox{\columnwidth}{!}{
\subfigure[Simulation 8(a)]{\includegraphics[width=0.33\textwidth]{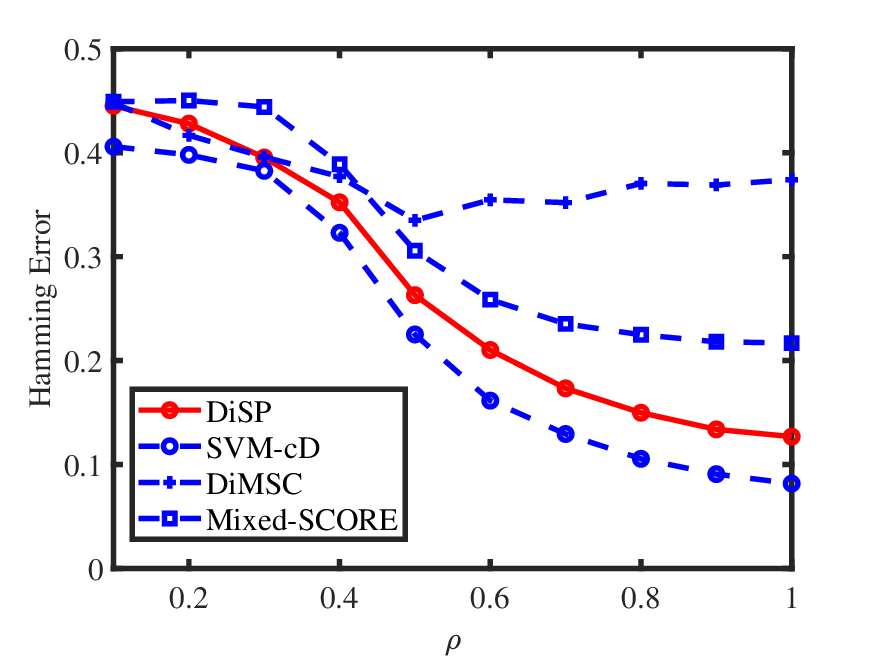}}
\subfigure[Simulation 8(a)]{\includegraphics[width=0.33\textwidth]{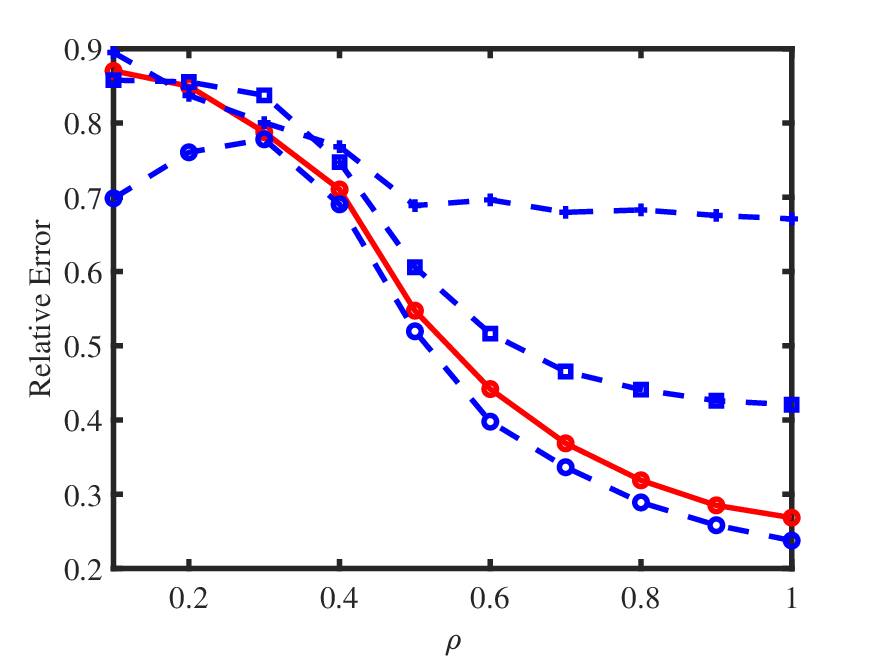}}
\subfigure[Simulation 8(b)]{\includegraphics[width=0.33\textwidth]{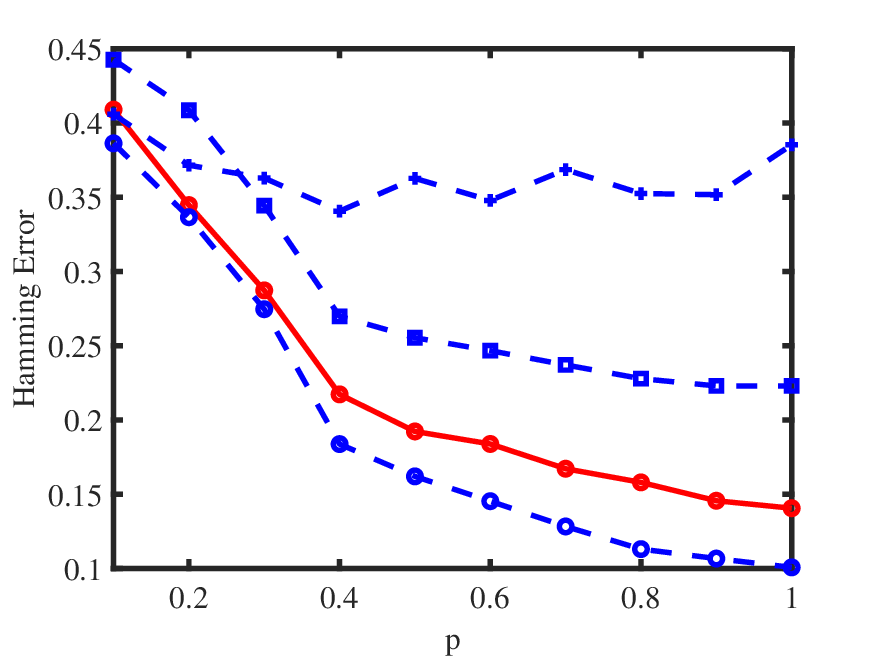}}
}
\resizebox{\columnwidth}{!}{
\subfigure[Simulation 8(b)]{\includegraphics[width=0.33\textwidth]{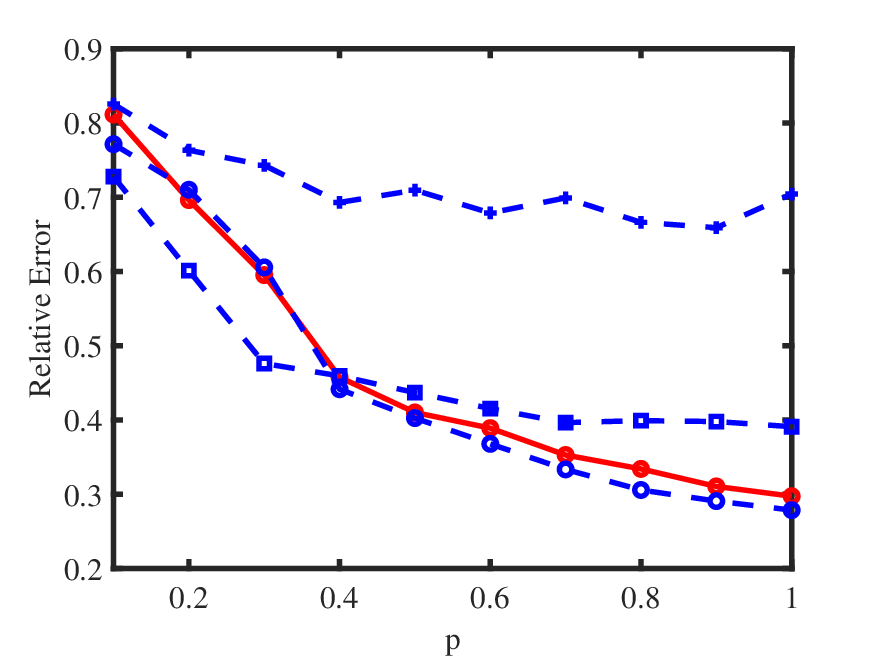}}
\subfigure[Simulation 8(c)]{\includegraphics[width=0.33\textwidth]{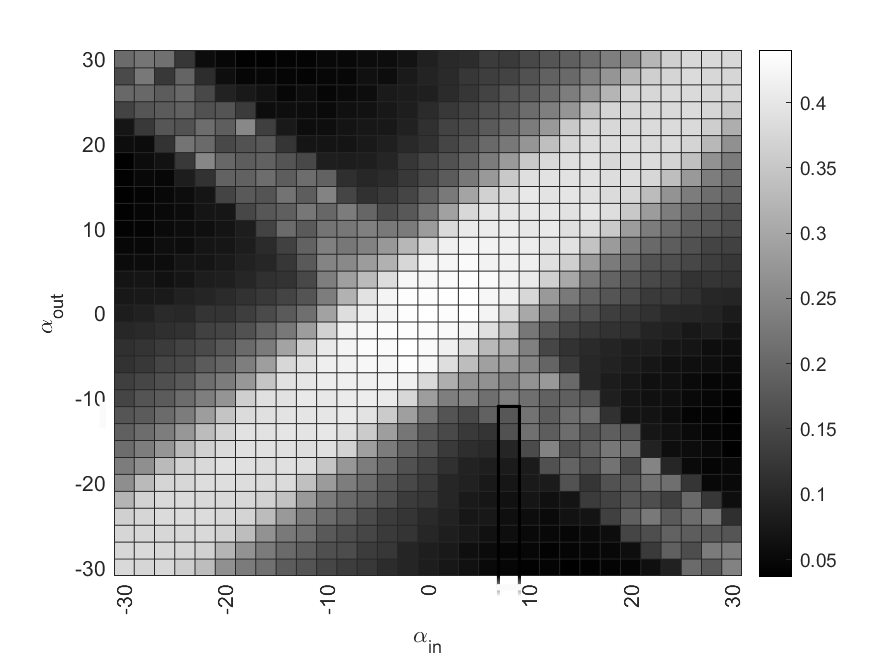}}
\subfigure[Simulation 8(d)]{\includegraphics[width=0.33\textwidth]{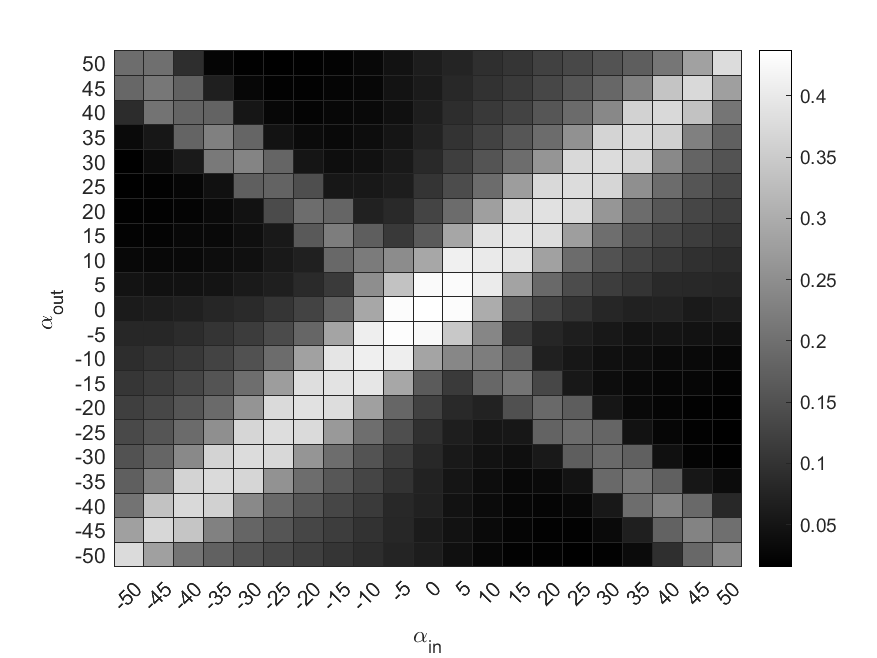}}
}
\caption{Bipartite signed network. For panels (e) and (f): the darker pixel represents a lower Hamming Error.}
\label{S8} 
\end{figure}
For bipartite signed network when $\mathbb{P}(A(i,j)=1)=\frac{1+\Omega(i,j)}{2}$ and $\mathbb{P}(A(i,j)=-1)=\frac{1-\Omega(i,j)}{2}$ for $i\in[n_{r}], j\in[n_{c}]$, by Example \ref{Signed}, $P$'s entries are real values and $\rho$ should be set in $(0,1)$.

\textbf{Simulation 8 (a): changing $\rho$}. Let $n_{r}=100,n_{c}=150, n_{r,0}=30, n_{c,0}=60, p=0.9$, and $P=P_{2}$. Let $\rho$ range in $\{0.1,0.2,0.3,\ldots,1\}$.

\textbf{Simulation 8 (b): changing $p$}. All parameters are set the same as Simulation 8 (a) except we let $\rho=0.8$ and $p$ range in $\{0.1, 0.2, \ldots, 1\}$.

\textbf{Simulation 8 (c): changing $\alpha_{\mathrm{in}}$ and $\alpha_{\mathrm{out}}$}.  Let $n=n_{r}=n_{c}=300, n_{r,0}=100, n_{c,0}=120, p=1$, and $\rho P=\tilde{P}$. $\alpha_{\mathrm{in}}$ and $\alpha_{\mathrm{out}}$ can be set in $(-\frac{n}{\mathrm{log}(n)},\frac{n}{\mathrm{log}(n)})$ by Example \ref{Signed}.  For this simulation, we let $\alpha_{\mathrm{in}}$ and $\alpha_{\mathrm{out}}$ be in the range of $\{-30,-28,-26,\ldots,30\}$.

\textbf{Simulation 8 (d): changing $\alpha_{\mathrm{in}}$ and $\alpha_{\mathrm{out}}$}.  All parameters are set the same as Simulation 8(c) except that we let $\alpha_{\mathrm{in}}$ and $\alpha_{\mathrm{out}}$ be in the range of $\{-50,-45,-40,\ldots,50\}$.

Figure \ref{S8} displays the results of Simulation 8. The analysis is similar to that of Simulation 4, and we omit it here.

\subsubsection{Adjacency matrices with missing edges under different distributions}
\textbf{Simulation 9:} For visuality, we plot adjacency matrices of overlapping bipartite weighted networks generated under BiMMDF for different distribution $\mathcal{F}$. For $P$, we set it as
\begin{flalign*}
P_{a}&=\begin{bmatrix}
    1&0.2\\
    0.1&0.9
\end{bmatrix} \mathrm{~or~}
P_{b}=\begin{bmatrix}
    1&-0.2\\
    0.1&-0.9
\end{bmatrix}.
\end{flalign*}
Under different $\mathcal{F}$, $\rho$ should be set in the interval obtained in Examples \ref{Bernoulli}-\ref{Signed}.  We consider below eight settings.

\begin{figure*}
\centering
\subfigure[$A$ of Set-up 1]{\includegraphics[width=0.4\textwidth]{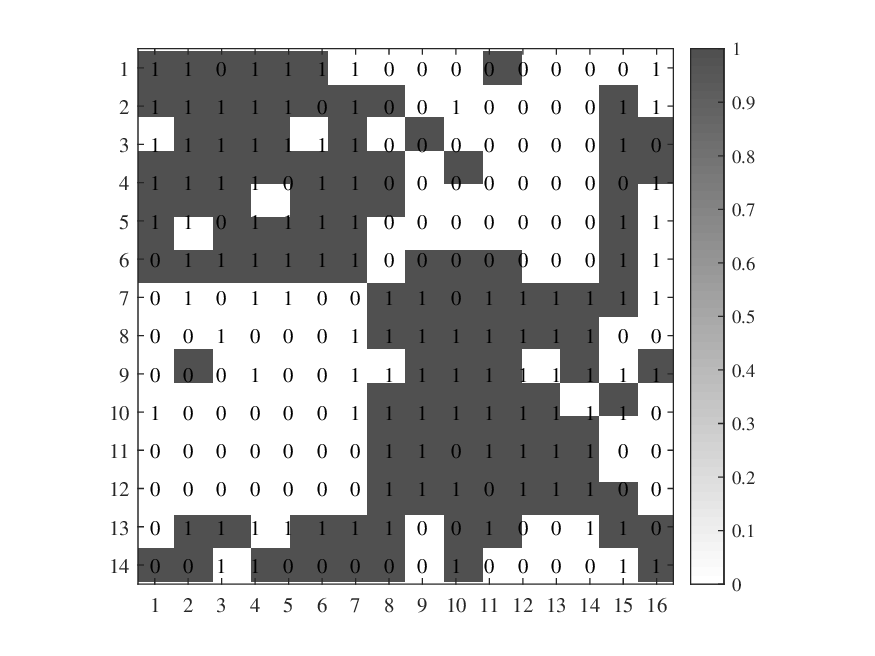}}
\subfigure[$A$ of Set-up 2]{\includegraphics[width=0.4\textwidth]{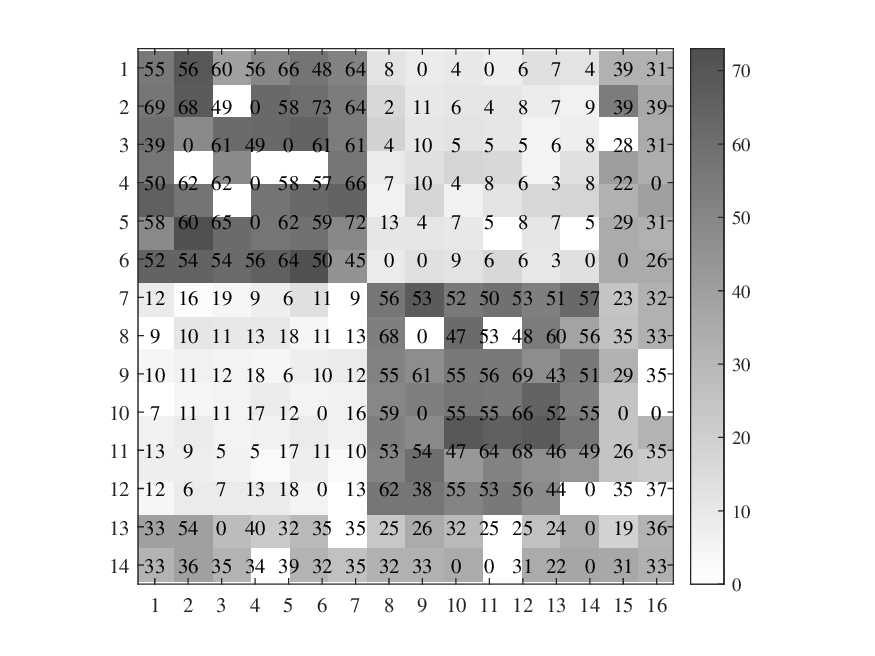}}
\subfigure[$A$ of Set-up 3]{\includegraphics[width=0.4\textwidth]{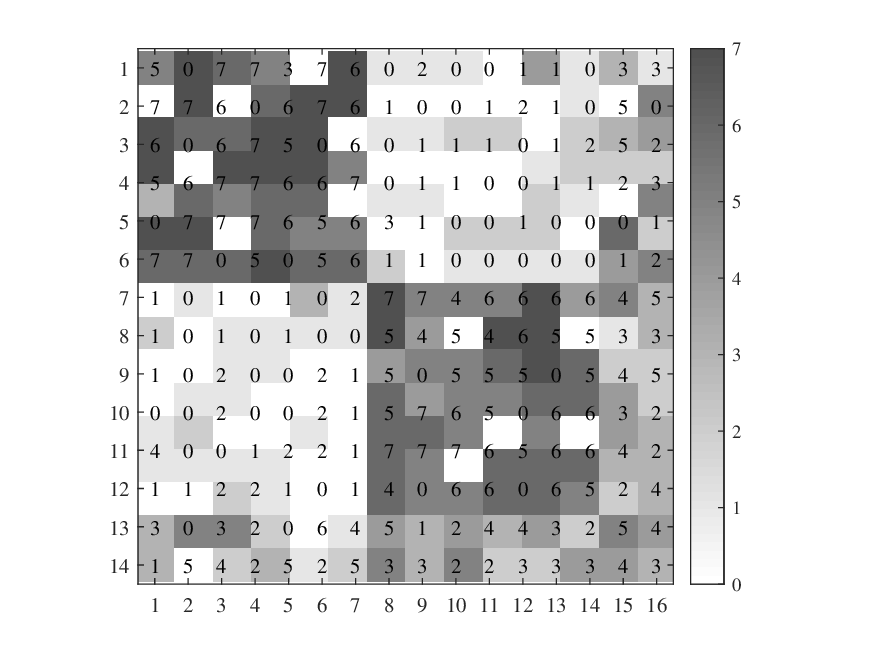}}
\subfigure[$A$ of Set-up 4]{\includegraphics[width=0.4\textwidth]{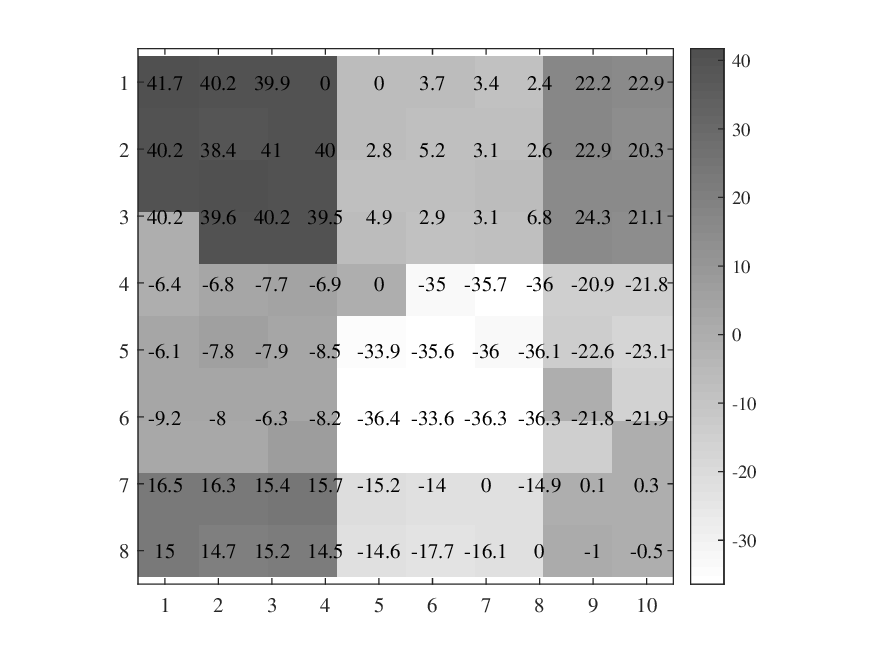}}
\subfigure[$A$ of Set-up 5]{\includegraphics[width=0.4\textwidth]{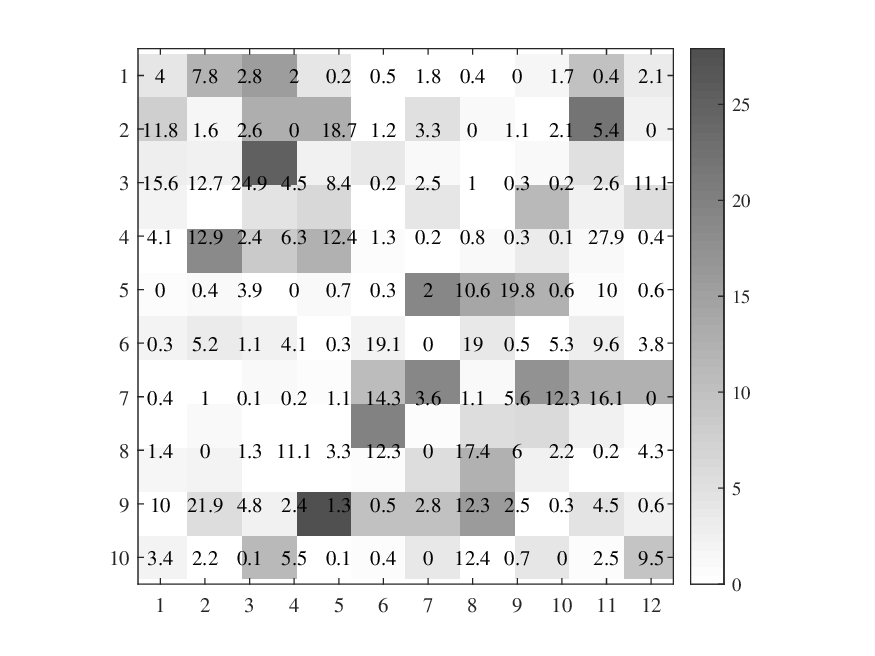}}
\subfigure[$A$ of Set-up 6]{\includegraphics[width=0.4\textwidth]{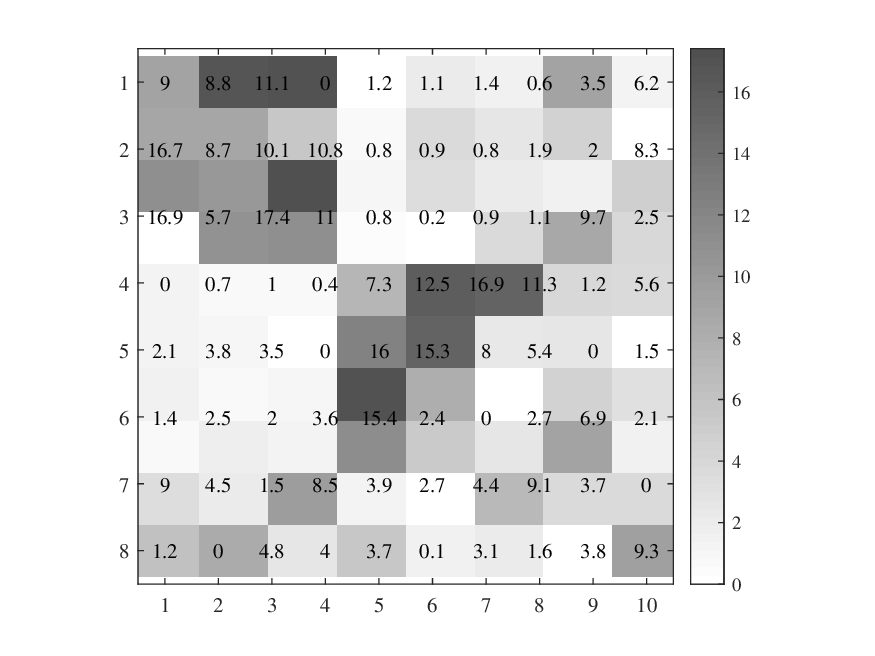}}
\subfigure[$A$ of Set-up 7]{\includegraphics[width=0.4\textwidth]{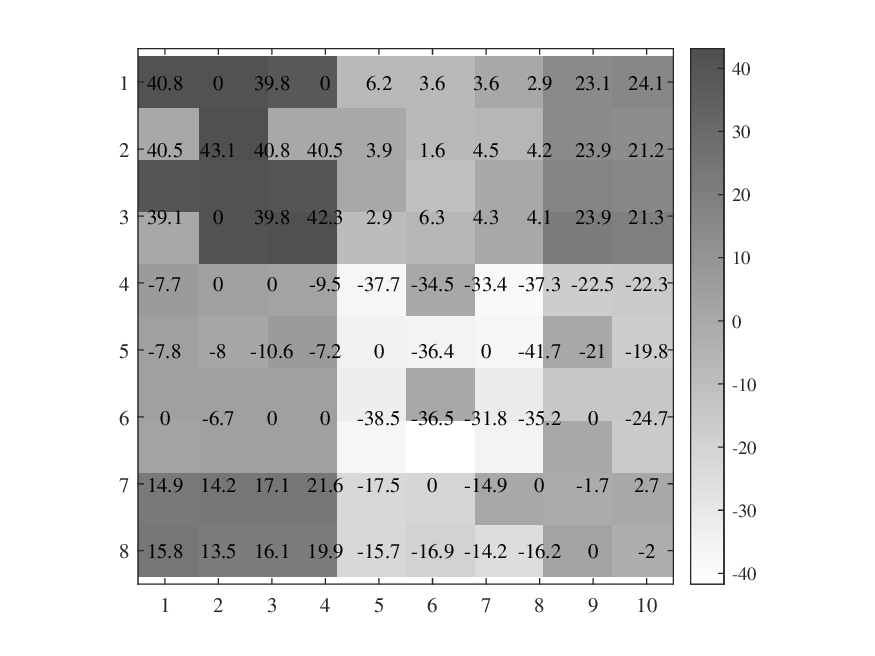}}
\subfigure[$A$ of Set-up 8]{\includegraphics[width=0.4\textwidth]{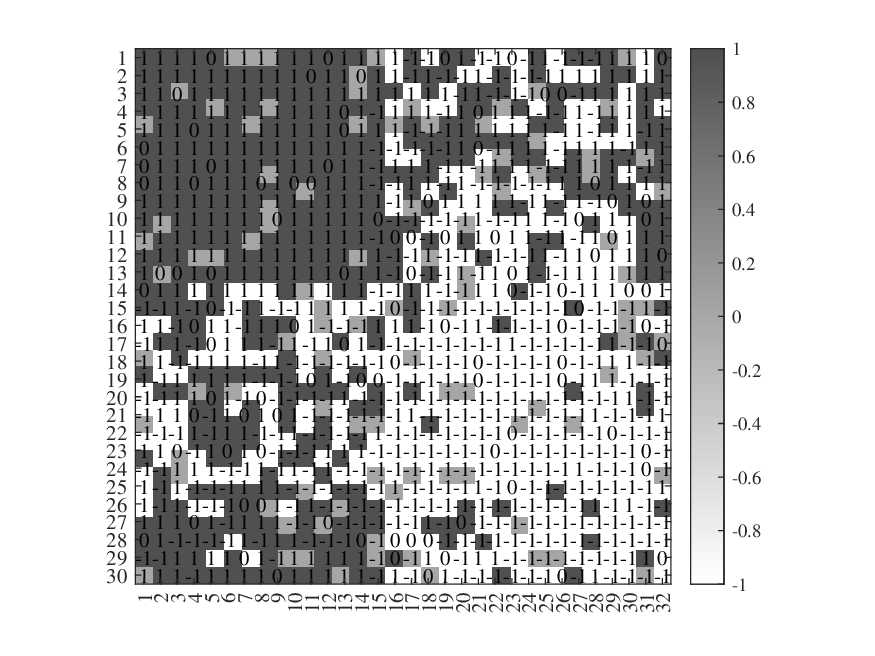}}
\caption{Illustration for bipartite weighted networks' adjacency matrices generated from BiMMDF. In panels (d)-(g), we keep $A$'s elements in one decimal for visualization beauty. In all panels, some elements of $A$ are zero, suggesting that there are missing edges in $A$.}
\label{ABi}
\end{figure*}

\emph{Set-up 1}: When $A(i,j)\sim \mathrm{Bernoulli}(\Omega(i,j))$ for $i\in[n_{r}], j\in[n_{c}]$, set $n_{r}=16, n_{r,0}=7, n_{c}=14, n_{c,0}=6, \rho=0.9, p=1$, and $P=P_{a}$. For this set-up, a bipartite un-weighted network with 16 row nodes and 14 column nodes is generated from BiMMDF. Panel (a) of Figure \ref{ABi} shows an adjacency matrix $A$ generated from BiMMDF for Set-up 1.

\emph{Set-up 2}: When $A(i,j)\sim \mathrm{Poisson}(\Omega(i,j))$ for $i\in[n_{r}], j\in[n_{c}]$, set $n_{r}=16, n_{r,0}=7, n_{c}=14, n_{c,0}=6, \rho=60, p=0.9$, and $P=P_{a}$. Panel (b) of Figure \ref{ABi} shows an $A$ generated from BiMMDF for this set-up.

\emph{Set-up 3}: When $A(i,j)\sim \mathrm{Binomial}(m,\Omega(i,j)/m)$ for $i\in[n_{r}], j\in[n_{c}]$, set $n_{r}=16, n_{r,0}=7, n_{c}=14, n_{c,0}=6, m=7, \rho=6, p=0.9$, and $P=P_{a}$. Panel (c) of Figure \ref{ABi} shows an $A$ generated from BiMMDF for this set-up.

\emph{Set-up 4}: When $A(i,j)\sim \mathrm{Normal}(\Omega(i,j),\sigma^{2}_{A})$ for $i\in[n_{r}], j\in[n_{c}]$, set $n_{r}=10, n_{r,0}=4, n_{c}=8, n_{c,0}=3, \sigma^{2}_{A}=1, \rho=40, p=0.9$, and $P=P_{b}$. Panel (d) of Figure \ref{ABi} shows an $A$ generated from BiMMDF for this set-up.

\emph{Set-up 5}: When $A(i,j)\sim \mathrm{Exponential}(\frac{1}{\Omega(i,j)})$ for $i\in[n_{r}], j\in[n_{c}]$, set $n_{r}=12, n_{r,0}=5, n_{c}=10, n_{c,0}=4,\rho=10, p=0.9$, and $P=P_{a}$. Panel (e) of Figure \ref{ABi} shows an $A$ generated from BiMMDF for this set-up.

\emph{Set-up 6}: When $A(i,j)\sim \mathrm{Uniform}(0,2\Omega(i,j))$ for $i\in[n_{r}], j\in[n_{c}]$, set $n_{r}=10, n_{r,0}=4, n_{c}=8, n_{c,0}=3, \rho=10, p=0.9$, and $P=P_{a}$. Panel (f) of Figure \ref{ABi} shows an $A$ generated from BiMMDF for this set-up.

\emph{Set-up 7}: When $A(i,j)\sim \mathrm{Logistic}(\Omega(i,j),\beta)$ for $i\in[n_{r}], j\in[n_{c}]$, set $n_{r}=10, n_{r,0}=4, n_{c}=8, n_{c,0}=3, \beta=1, \rho=40, p=0.9$, and $P=P_{b}$. Panel (g) of Figure \ref{ABi} shows an $A$ generated from BiMMDF for this set-up.

\emph{Set-up 8}: For bipartite signed network when $\mathbb{P}(A(i,j)=1)=\frac{1+\Omega(i,j)}{2}$ and $\mathbb{P}(A(i,j)=-1)=\frac{1-\Omega(i,j)}{2}$ for $i\in[n_{r}], j\in[n_{c}]$, set $n_{r}=32, n_{r,0}=14, n_{c}=30, n_{c,0}=14, \rho=1, p=0.9$, and $P=P_{b}$. Panel (h) of Figure \ref{ABi} shows an $A$ generated from BiMMDF for this set-up.

Tables \ref{HESetUp} and \ref{RESetUp} record Hamming Error and Relative Error of all four approaches for adjacency matrices generated from set-ups 1-8, respectively. The results show that, for set-ups 1, 2, 5, 6, and 7, DiSP outperforms its competitors; for the other three set-ups, DiSP performs similarly to SVM-cD and both methods outperform DiMSC and Mixed-SCORE. With given $A$ and known memberships $\Pi_{r}$ and $\Pi_{c}$ for these set-ups, readers can apply DiSP to $A$ to check its effectiveness.
\begin{table}[h!]
\footnotesize
	\centering
	\caption{Hamming Error of methods used in this paper for adjacency matrices generated by Set-ups 1-8.}
	\label{HESetUp}
	\begin{tabular}{cccccccccccc}
\hline\hline&Set-up 1&Set-up 2&Set-up 3&Set-up 4&Set-up 5&Set-up 6&Set-up 7&Set-up 8\\
\hline
DiSP&0.0681&0.0295&0.0365&0.0197&0.1490&0.0804&0.0332&0.0774\\
SVM-cD&0.0710&0.0321&0.0311&0.0169&0.1505&0.0860&0.0358&0.0733\\
DiMSC&0.1442&0.0704&0.0365&0.3747&0.3379&0.1148&0.3151&0.4010\\
Mixed-SCORE&0.0734&0.0313&0.0402&0.1906&0.1555&0.0744&0.2213&0.1482\\
\hline\hline
\end{tabular}
\end{table}

\begin{table}[h!]
\footnotesize
	\centering
	\caption{Relative Error of methods used in this paper for adjacency matrices generated by Set-ups 1-8.}
	\label{RESetUp}
	\begin{tabular}{cccccccccccc}
\hline\hline&Set-up 1&Set-up 2&Set-up 3&Set-up 4&Set-up 5&Set-up 6&Set-up 7&Set-up 8\\
\hline
DiSP&0.2091&0.1030&0.1001&0.0441&0.4632&0.1810&0.0943&0.2225\\
SVM-cD&0.2142&0.1093&0.0988&0.0507&0.4633&0.2683&0.1060&0.1772\\
DiMSC&0.4275&0.2310&0.0860&0.7112&0.6743&0.3228&0.5788&0.7298\\
Mixed-SCORE&0.2038&0.1051&0.1079&0.4076&0.4501&0.2279&0.4327&0.4237\\
\hline\hline
\end{tabular}
\end{table}
\subsection{Real data applications}\label{RealDataSection}
In addition to the synthetic datasets, we also use DiSP to find community memberships in several real-world networks. Table \ref{realdata} presents basic information and summarized statistics of real-world networks used in this article. Among these networks, Crisis in a Cloister, Highschool, and Facebook-like Social Network are directed weighted networks whose row nodes are the same as column nodes while the other networks are bipartite. Facebook-like Social Network can be downloaded from \url{https://toreopsahl.com/datasets/#online_social_network} (accessed on 9 June 2023) while the other networks can be downloaded from \url{http://konect.cc/networks} (accessed on 9 June 2023). Since it is meaningless to detect community memberships for isolated nodes which have no connection with any other nodes, we need to remove these isolated nodes before processing data. For Facebook-like Social Network, the original data has 1899 nodes, after removing isolated nodes from both row and column sides, it has 1302 nodes. Unicode languages originally has 614 languages and we remove 86 languages that have not been spoken by the 254 countries. For Marvel, the original data has 19428 works and we remove 6486 works that have no connection with the 6486 characters.

\begin{figure}
\centering
\resizebox{\columnwidth}{!}{
\subfigure[Highschool]{\includegraphics[width=0.15\textwidth]{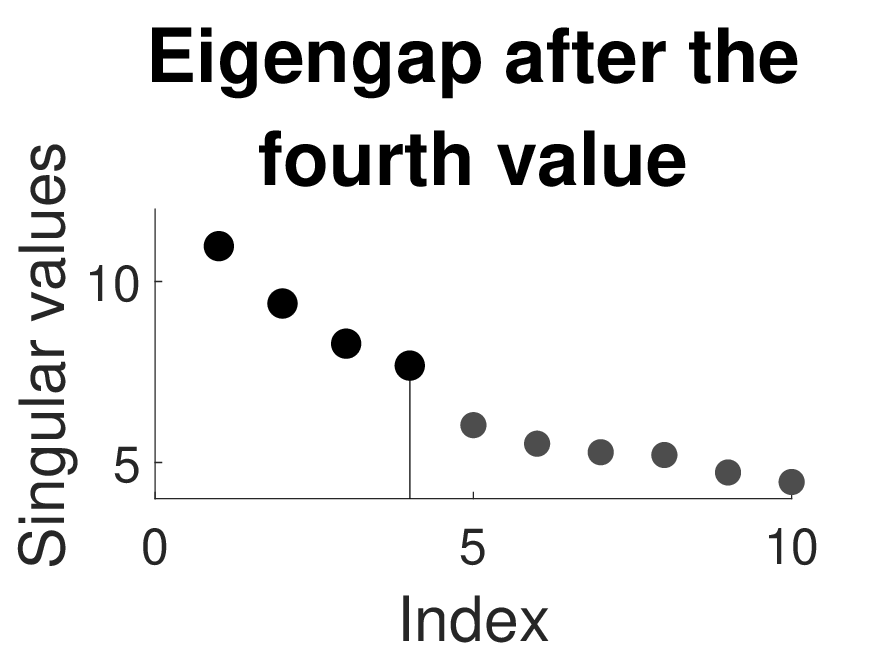}}
\subfigure[Facebook-like Social Network]{\includegraphics[width=0.15\textwidth]{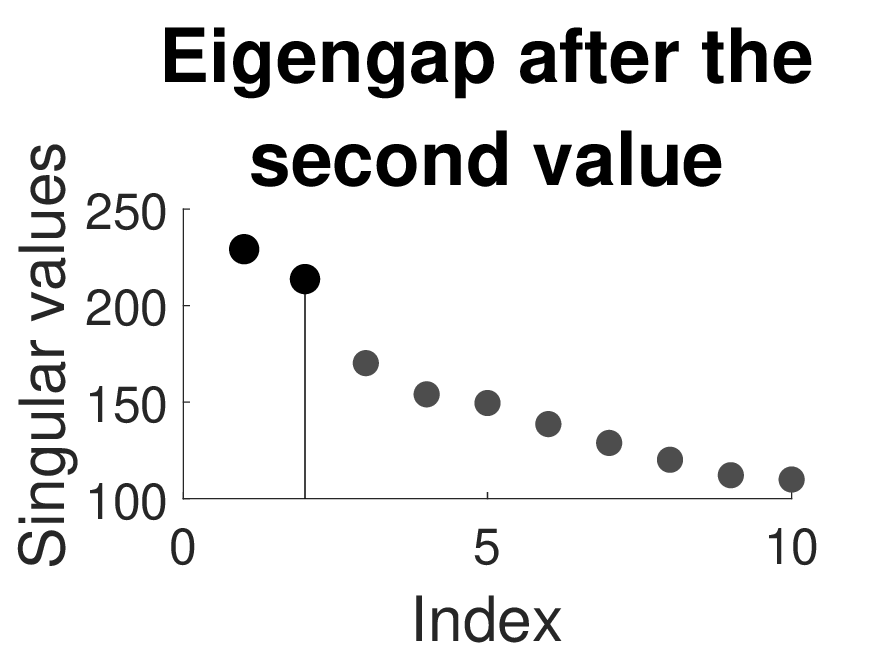}}
\subfigure[Unicode languages]{\includegraphics[width=0.15\textwidth]{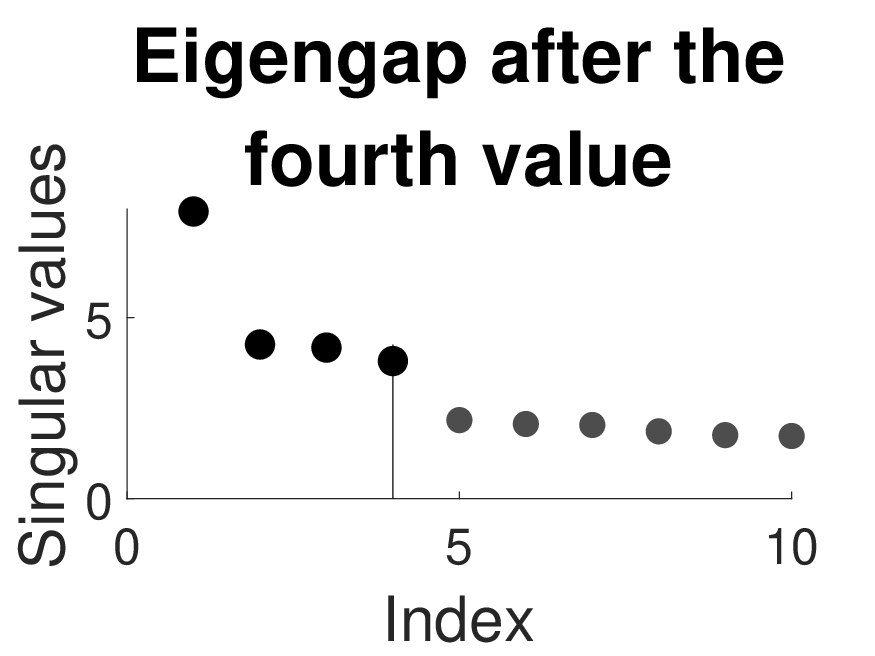}}
\subfigure[Marvel]{\includegraphics[width=0.15\textwidth]{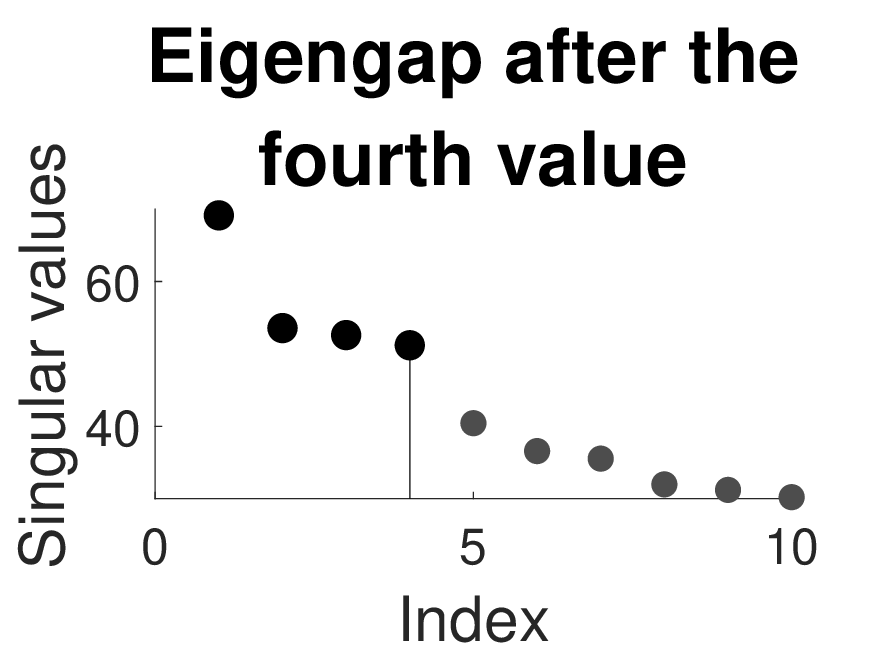}}
\subfigure[Amazon (Wang)]{\includegraphics[width=0.15\textwidth]{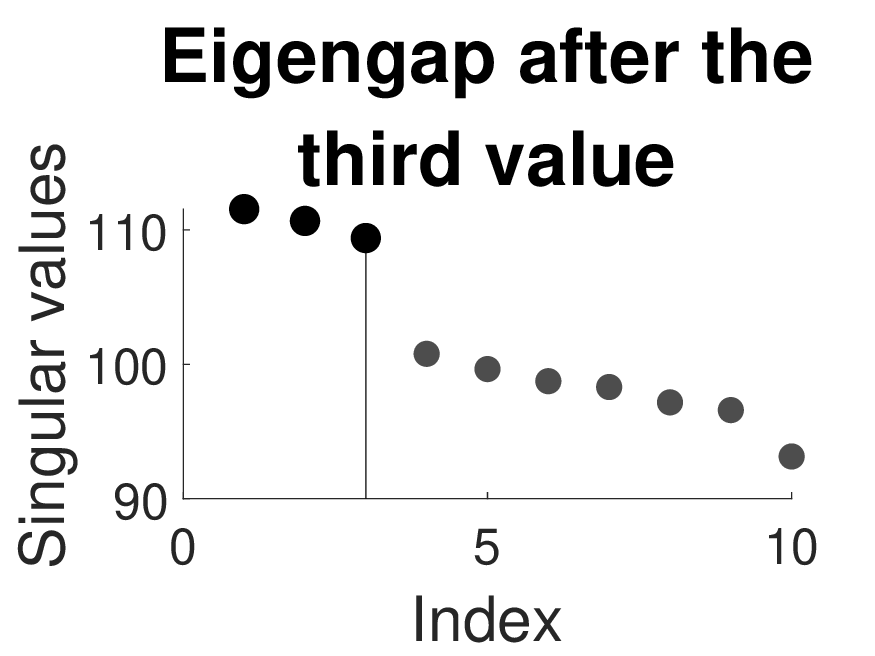}}
\subfigure[CiaoDVD movie ratings]{\includegraphics[width=0.15\textwidth]{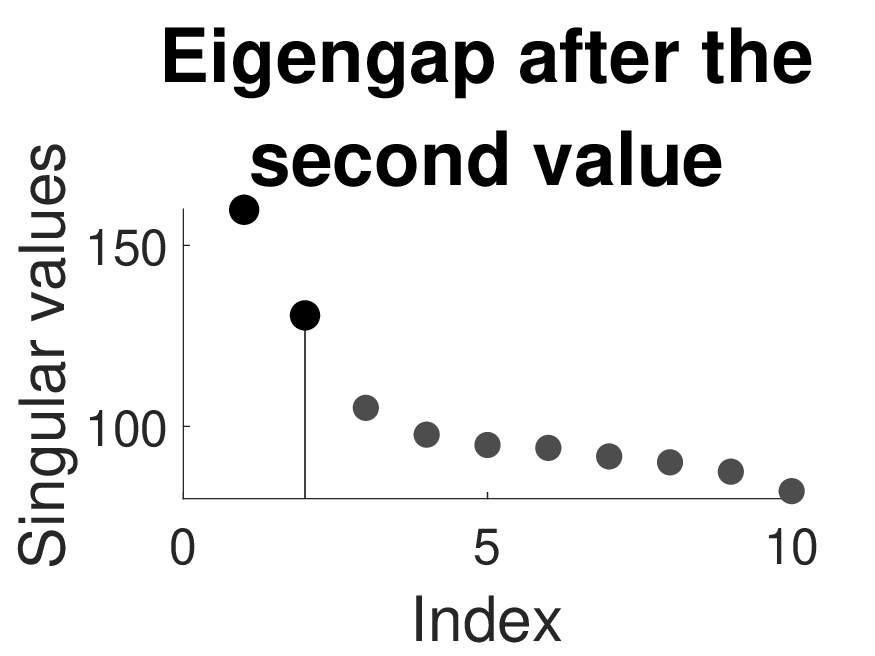}}
\subfigure[arXiv cond-mat]{\includegraphics[width=0.15\textwidth]{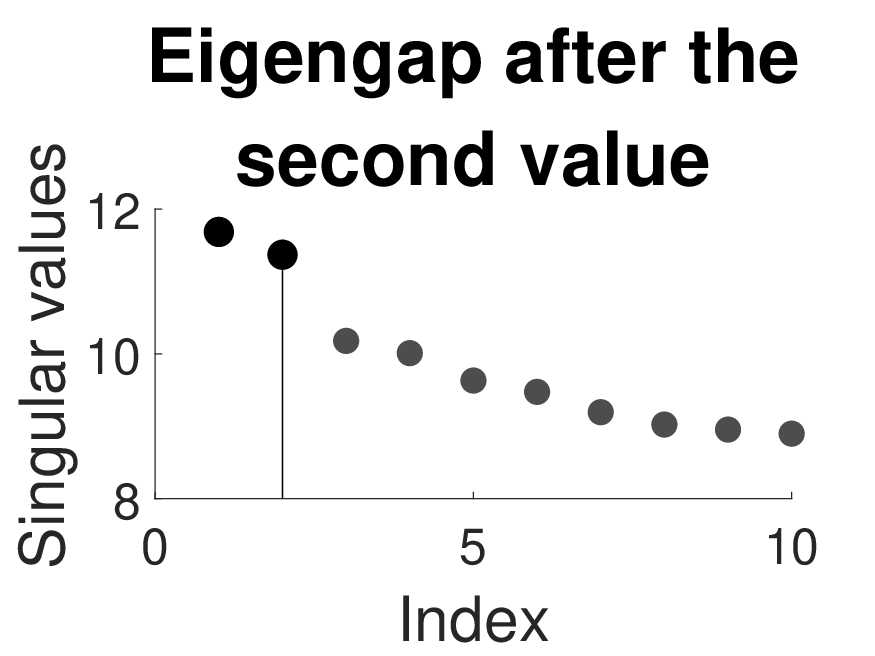}}
}
\caption{Top 10 singular values of $A$ for seven real-world networks.}
\label{SVDS10} 
\end{figure}

For these networks, their community memberships are unknown and we aim at applying DiSP to have a better understanding of their community structure. For Crisis in a Cloister, $K$ is 3 identified by \citep{sampson1969crisis,handcock2007model,MMSB}. For networks with unknown $K$, eigengap is used to estimate it \citep{DISIM}. Thus, we plot the top 10 singular values of $A$ in Figure \ref{SVDS10} to determine $K$. For Highschool, Unicode languages, and Marvel, the eigengap suggests $K=4$. For Facebook-like Social Network, CiaoDVD movie ratings, and arXiv cond-mat, the eigengap suggests $K=2$. For Amazon (Wang), the eigengap suggests $K=3$.

\begin{table}[h!]
\footnotesize
	\centering
	\caption{Basic information and summarized statistics of real-world networks studied in this paper.}
	\label{realdata}
	\resizebox{\columnwidth}{!}{
	\begin{tabular}{cccccccccccc}
\hline\hline&Row node meaning&Column node meaning&Edge meaning&True memberships&$n_{r}$&$n_{c}$&$K$&$\mathrm{max}_{i,j}A(i,j)$&$\mathrm{min}_{i,j}A(i,j)$&\#Edges\\
\hline
Crisis in a Cloister \cite{sampson1969crisis}&Monk&Monk&Ratings&Unknown&18&18&3&1&-1&184\\
Highschool \cite{coleman1964introduction}&Boy&Boy&Friendship&Unknown&70&70&Unknown&2&0&366\\
Facebook-like Social Network \cite{opsahl2009clustering}&User&User&Messages&Unknown&1302&1302&Unknown&98&0&19044\\
Unicode languages \cite{kunegis2013konect}&Country&Language&Hosts&Unknown&254&528&Unknown&1&0&1106\\
Marvel \cite{alberich2002marvel}&Character&Work&Appearance&Unknown&6486&12942&Unknown&1&0&96662\\
Amazon (Wang) \cite{wang2010latent}&User&Item&Rating&Unknown&26112&799&Unknown&5&0&28901\\
CiaoDVD movie ratings \cite{guo2014etaf}&User&Movie&Rating&Unknonw&17615&16121&Unknown&5&0&72345\\
arXiv cond-mat \cite{newman2001structure}&Author&Paper&Authorship&Unknown&16726&22015&Unknown&1&0&58595\\
\hline\hline
\end{tabular}}
\end{table}

To explore and understand the community structure of a real-world bipartite (and directed) network, we introduce the following items.
\begin{itemize}
  \item Let $\hat{\mathcal{C}}_{r}$ be a vector  whose $i$-th element is $\hat{\mathcal{C}}_{r}(i)=\mathrm{argmax}_{k\in[K]}\hat{\Pi}_{r}(i,k)$ for $i\in[n_{r}]$, and we call $\hat{\mathcal{C}}_{r}(i)$ the home base row community of row node $i$. Define $\hat{\mathcal{C}}_{c}$ by letting $\hat{\mathcal{C}}_{c}(j)=\mathrm{argmax}_{k\in[K]}\hat{\Pi}_{c}(j,k)$ for $j\in[n_{j}]$ and call it home base column community of column node $j$.
  \item For row node $i$, we call it highly mixed row node if $\mathrm{max}_{k\in[K]}\hat{\Pi}_{r}(i,k)\leq0.6$ and call it highly pure row node if $\mathrm{max}_{k\in[K]}\hat{\Pi}_{r}(i,k)\geq0.9$. A highly mixed (and pure) column node is defined similarly. Note that for row node $i$ whose membership satisfies $0.6<\mathrm{max}_{k\in[K]}\hat{\Pi}_{r}(i,k)<0.9$, it is neither highly mixed nor highly pure.
  \item Let $\eta_{r}=\frac{|\{i:\mathrm{max}_{k\in[K]}\hat{\Pi}_{r}(i,k)\leq0.6\}|}{n_{r}}$ be the proportion of highly mixed row nodes and $\zeta_{r}=\frac{|\{i:\mathrm{max}_{k\in[K]}\hat{\Pi}_{r}(i,k)\geq0.9\}|}{n_{r}}$ be the proportion of highly pure row nodes. $\eta_{c}$ and $\zeta_{c}$ are defined similarly for column nodes.
  \item For directed networks, we have $n_{r}=n_{c}=n$. Since row nodes are the same as column nodes, to measure the asymmetric structure between row communities and column communities, we set $\mathrm{Hamm}_{rc}=\frac{\mathrm{min}_{\mathcal{P}\in S}\|\hat{\Pi}_{r}\mathcal{P}-\hat{\Pi}_{c}\|_{1}}{n}$. Large $\mathrm{Hamm}_{rc}$ suggests heavy asymmetric between row and column communities, and vice versa. For un-directed networks, $\mathrm{Hamm}_{rc}$ is 0, suggesting that $\mathrm{Hamm}_{rc}$ is a good way to discover asymmetries in directed networks. Sure, $\mathrm{Hamm}_{rc}$ is inapplicable for bipartite networks.
\end{itemize}

To estimate community memberships of real-world networks in Table \ref{realdata}, we apply DiSP to $A$ with $K$ row (and column) communities, where $K$ is 3 for Crisis in a Cloister, and $K$ used for the other networks is suggested by eigengap in Figure  \ref{SVDS10}. We report $\eta_{r}, \eta_{c}, \zeta_{r}, \zeta_{c}, \mathrm{Hamm}_{rc}$, and DiSP's runtime in Table \ref{RealDataResults}, where runtime is the average of 10 independent repetitions. From the results on real-world networks, we draw the following conclusions.
\begin{itemize}
  \item $\mathrm{Hamm}_{rc}$ for Crisis in a Cloister and Facebook-like Social Network is larger than that of Highschool. This indicates that the asymmetry between row and column communities for Crisis in a Cloister and Facebook-like Social Network is heavier than that of Highschool.
  \item Large $\eta_{r}$ and $\eta_{c}$ for Crisis in a Cloister, Facebook-like Social Network, Marvel, and Amazon (Wang) indicate that there exist large proportions of highly mixed nodes in both row and column communities. For comparison, the other four networks have lesser highly mixed nodes. Meanwhile, Unicode languages, CiaoDVD movie ratings, and arXiv cond-mat have larger proportions of highly pure nodes in both row and column communities than the other networks.
  \item For Crisis in a Cloister, $18\times(1-\eta_{r}-\zeta_{r})\approx4$ indicates that 4 monks in the row nodes side are neither highly mixed nor highly pure; $18\times(1-\eta_{c}-\zeta_{c})\approx6$ indicates that 6 monks in the column nodes side are neither highly mixed nor highly pure.
  \item For Highschool, $70\times(1-\eta_{r}-\zeta_{r})\approx31$, i.e., 31 boys in the row nodes side are neither highly mixed nor highly pure; $70\times(1-\eta_{c}-\zeta_{c})\approx33$, i.e., 33 boys in the column nodes side are neither highly mixed nor highly pure.
  \item For Facebook-like Social Network, most users are neither highly mixed nor highly pure because $1-\eta_{r}-\zeta_{r}=0.4938$ and $1-\eta_{c}-\zeta_{c}=0.6252$.
  \item For Unicode languages, since $\eta_{r}+\zeta_{r}=1$, all countries are either highly mixed or highly pure. Meanwhile, since $\zeta_{r}=0.9449$, we see that $94.99\%$ of countries are highly pure, suggesting that nearly $94.99\%$ of countries only belong to one of the four clusters. Since $528\times\eta_{c}\approx43$, we see that only 43 languages are highly mixed while 390 languages are highly pure because $528\times\zeta_{c}\approx390$.
  \item For Marvel, $81.88\%$ of characters are highly pure and $57.72\%$ of works are neither highly mixed nor highly pure since $1-\eta_{c}-\zeta_{c}=0.5772$. A similar analysis holds for Amazon (Wang).
  \item For CiaoDVD movie ratings, all users are either highly mixed or highly pure since $\eta_{r}+\zeta_{r}=1$, $94.43\%$ of users are highly pure, and less than $5\%$ of movies are highly mixed. A similar analysis holds for arXiv cond-mat.
\end{itemize}

\begin{table}[h!]
\footnotesize
	\centering
	\caption{$\eta_{r}, \eta_{c}, \zeta_{r}, \zeta_{c}, \mathrm{Hamm}_{rc}$, and runtime when applying DiSP to real-world networks used in this paper.}
	\label{RealDataResults}
\begin{tabular}{cccccccccc}
\hline\hline
data&$\eta_{r}$&$\eta_{c}$&$\zeta_{r}$&$\zeta_{c}$&$\mathrm{Hamm_{rc}}$&Runtime\\
\hline
Crisis in a Cloister&0.2222&0.2778&0.5556&0.3889&0.3692&0.0032s\\
Highschool&0.1429&0.1000&0.4143&0.4286&0.1340&0.0046s\\
Facebook-like Social Network&0.1951&0.1928&0.3111&0.1820&0.3084&0.0422s\\
Unicode languages&0.0551&0.0814&0.9449&0.7386&-&0.0117s\\
Marvel&0.1812&0.2186&0.8188&0.2042&-&3.4237s\\
Amazon (Wang)&0.3104&0.5156&0.6896&0.2741&-&1.8655s\\
CiaoDVD movie ratings&0.0557&0.0496&0.9443&0.7153&-&11.0493s\\
arXiv cond-mat&0.0584&0.0551&0.9416&0.6899&-&20.0398s\\
\hline\hline
\end{tabular}
\end{table}

\begin{figure}
\centering
\resizebox{\columnwidth}{!}{
\subfigure[$\hat{\Pi}_{r}$ for row nodes side]{\includegraphics[width=0.5\textwidth]{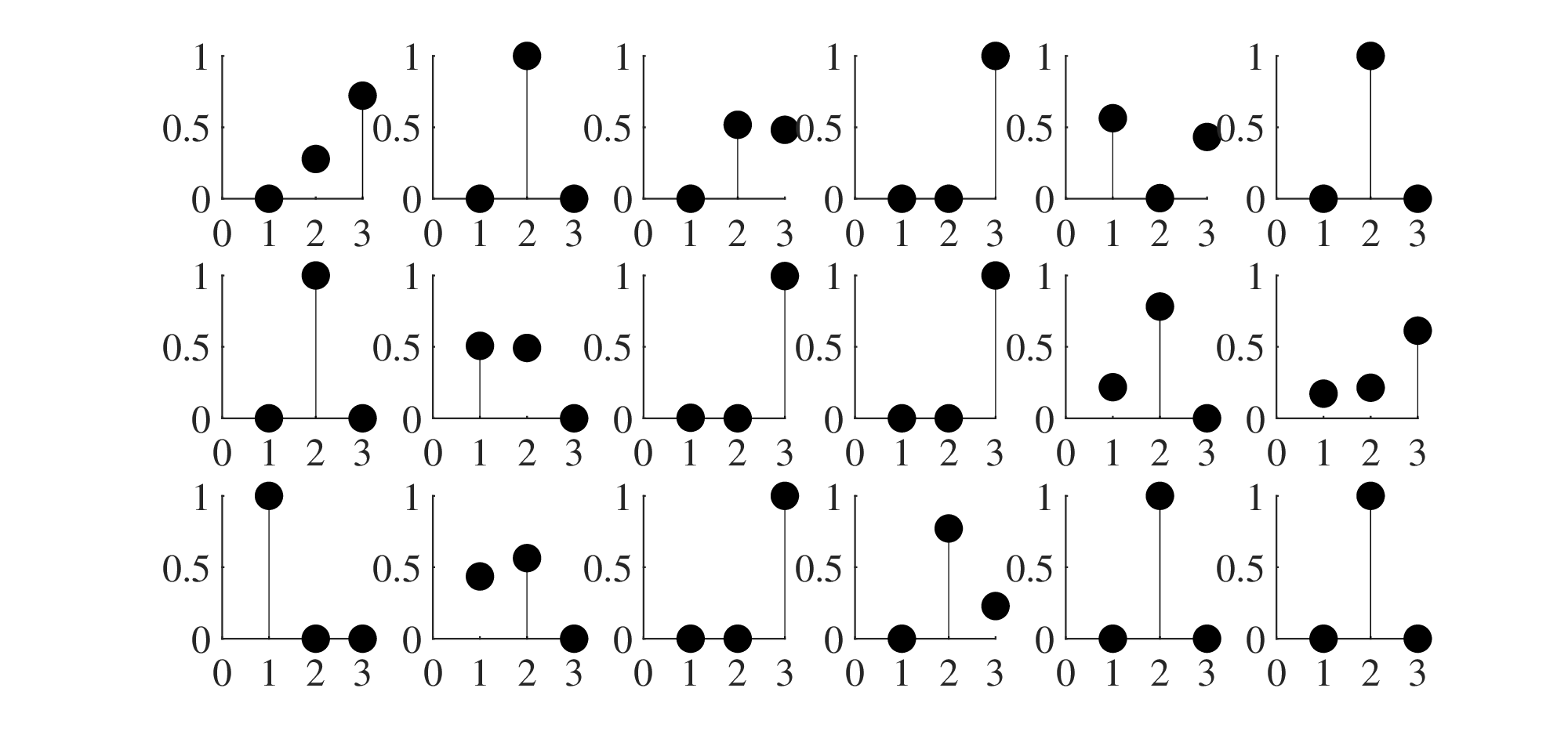}}
\subfigure[$\hat{\Pi}_{c}$ for column nodes side]{\includegraphics[width=0.5\textwidth]{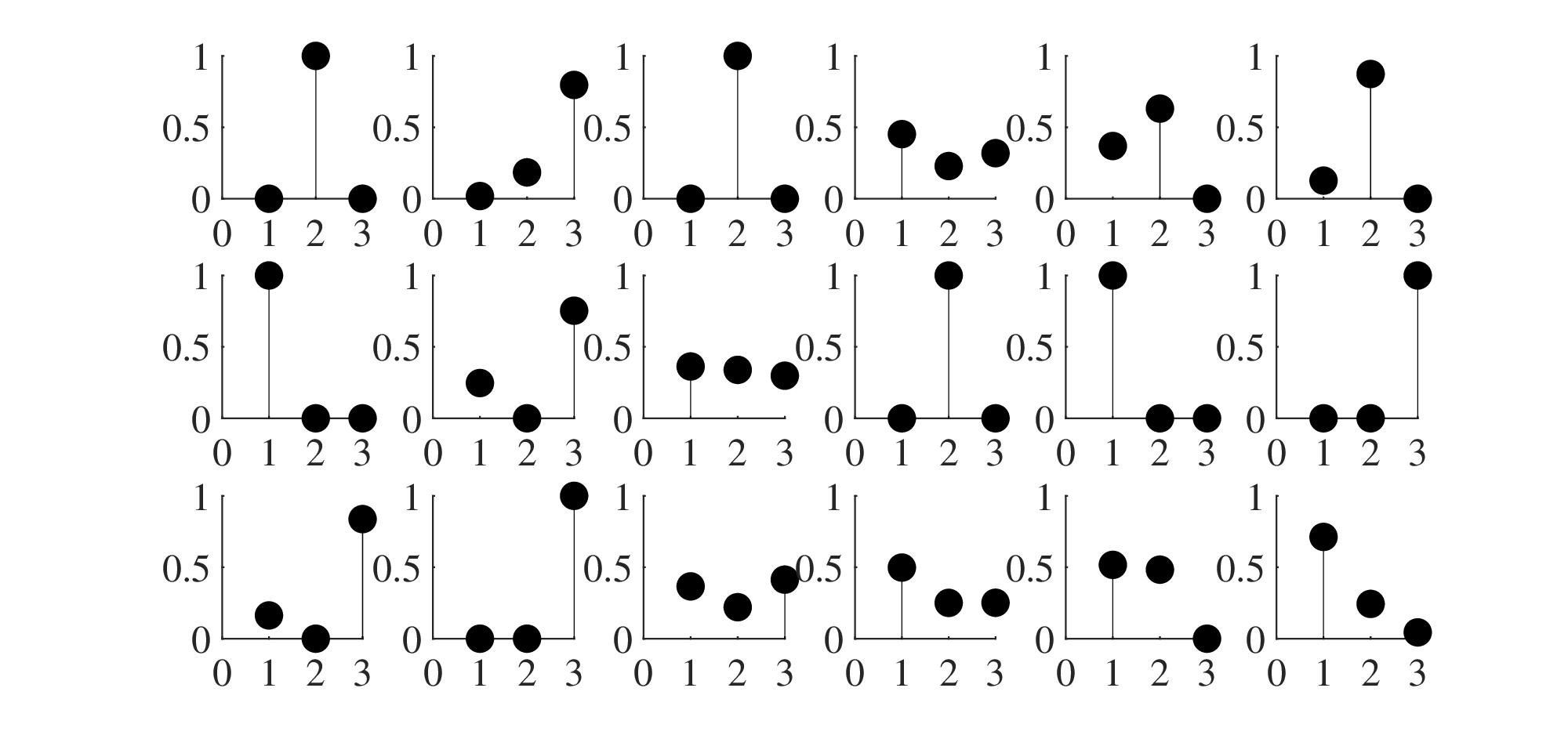}}
}
\caption{The estimated mixed membership matrices $\hat{\Pi}_{r}$ and $\hat{\Pi}_{c}$ detected by DiSP, for the 18 monks in Crisis in a Cloister network. Each sub-panel denotes a membership vector of a monk; we order the communities 1 to 3 on the X axis, and the monk's membership belonging to each cluster is on the Y axis.}
\label{crisisPirc} 
\end{figure}

\begin{figure}
\centering
\resizebox{\columnwidth}{!}{
\subfigure[$\hat{\Pi}_{r}$ for row nodes side]{\includegraphics[width=1\textwidth]{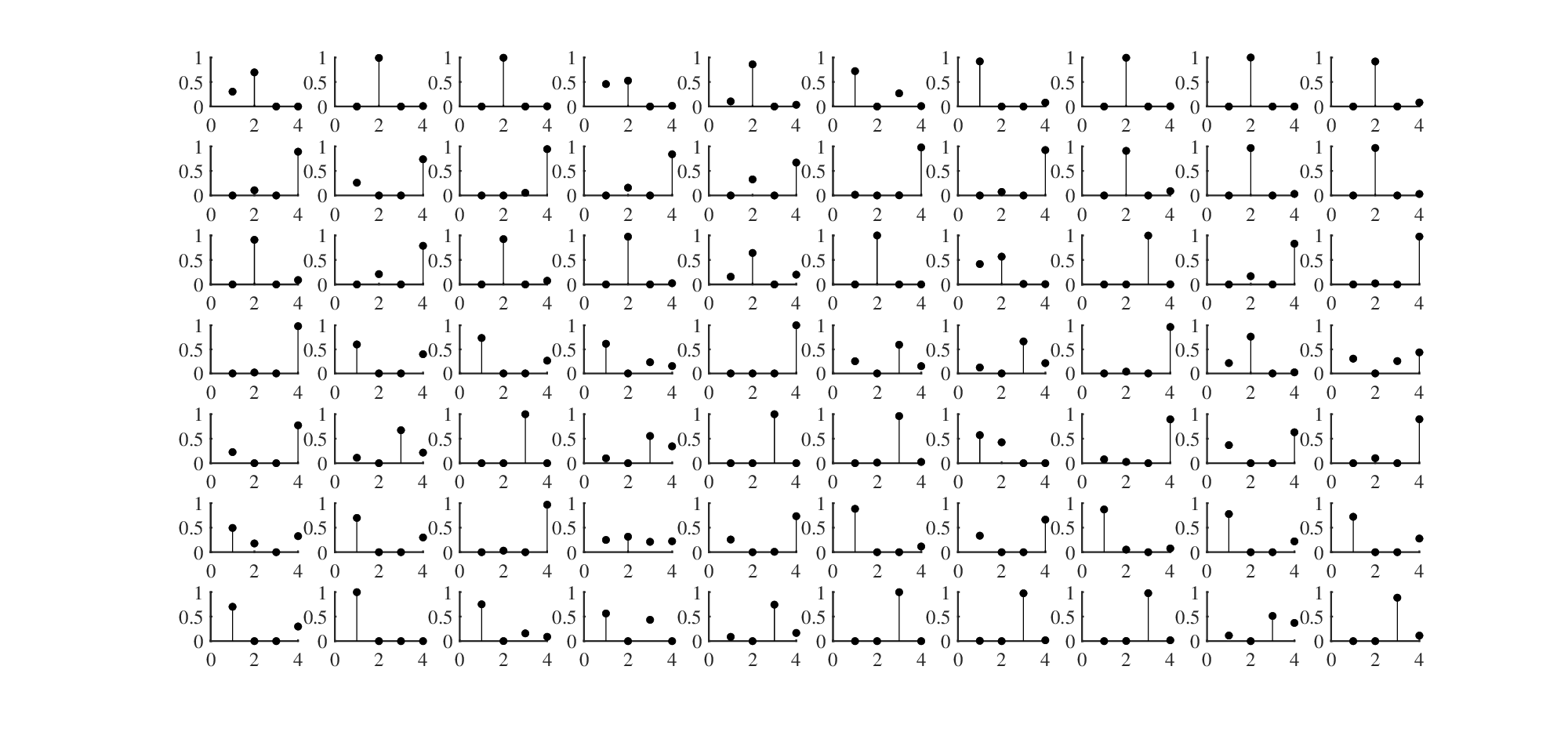}}
}
\resizebox{\columnwidth}{!}{
\subfigure[$\hat{\Pi}_{c}$ for column nodes side]{\includegraphics[width=1\textwidth]{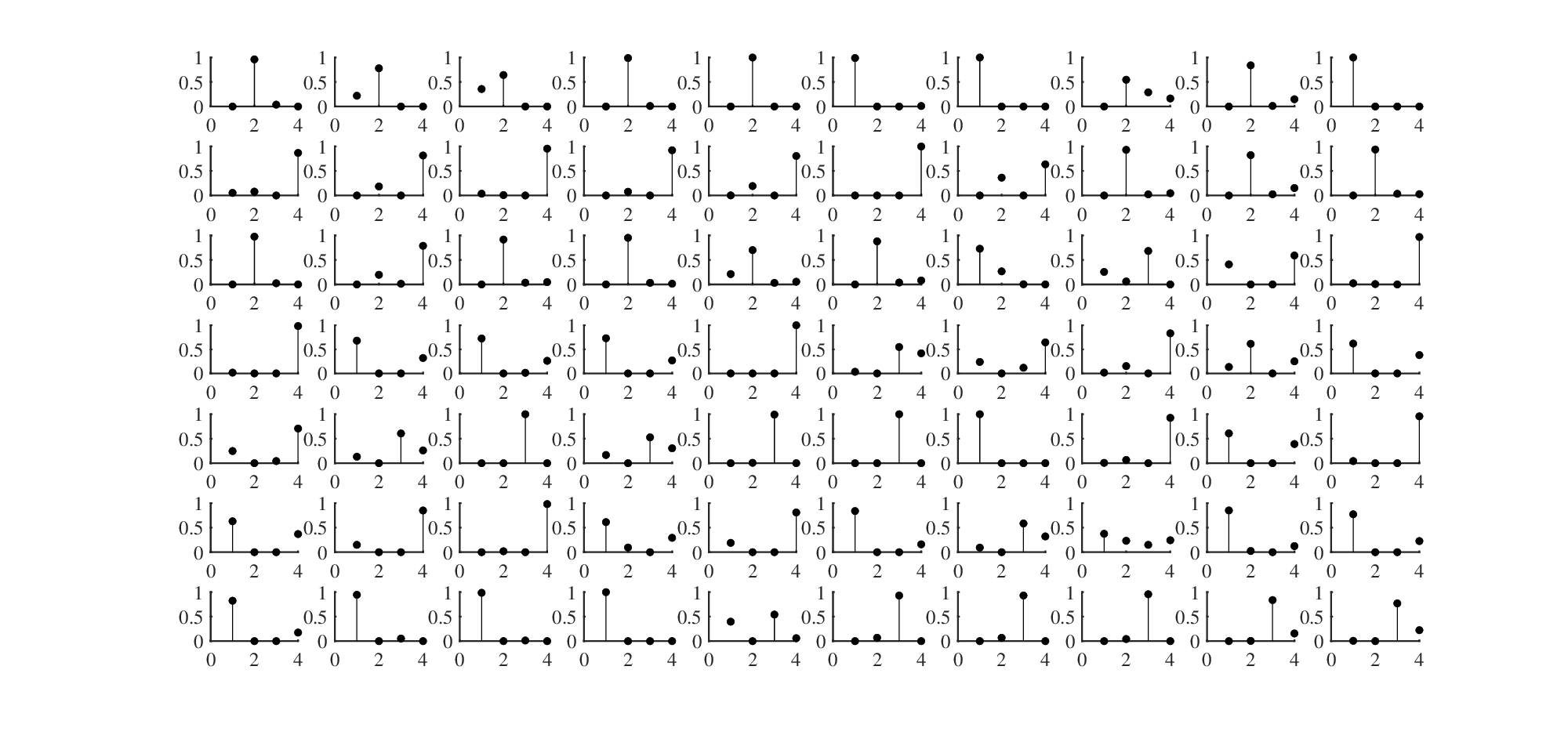}}
}
\caption{The estimated mixed membership matrices $\hat{\Pi}_{r}$ and $\hat{\Pi}_{c}$ detected by DiSP, for the 70 boys in Highschool network. Each sub-panel denotes a membership vector of a boy; we order the communities 1 to 4 on the X axis, and the boy's membership belonging to each cluster is on the Y axis.}
\label{highschoolPirc} 
\end{figure}

\begin{figure}
\centering
\resizebox{\columnwidth}{!}{
\subfigure[$\hat{\Pi}_{r}$ for row nodes side]{\includegraphics[width=1\textwidth]{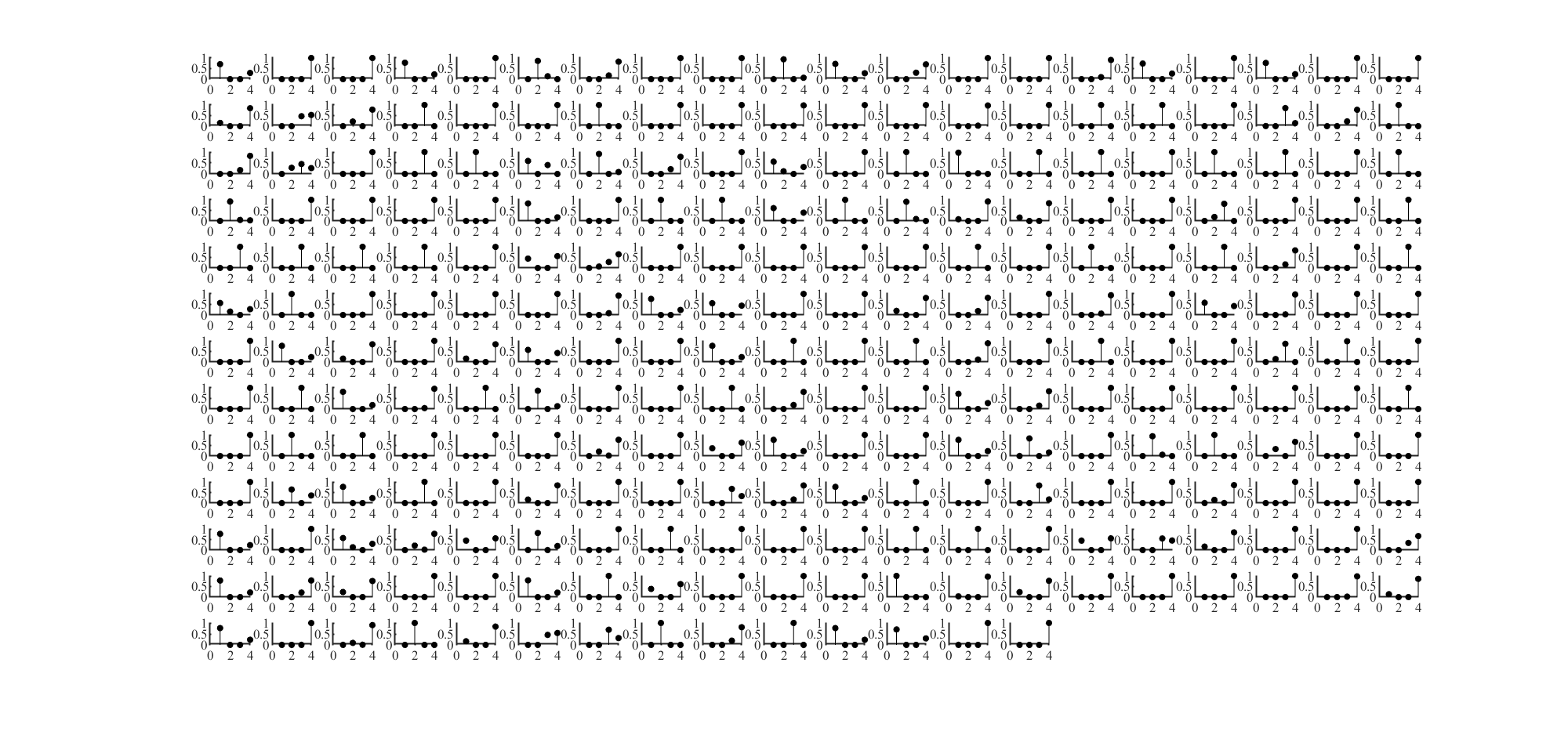}}
}
\resizebox{\columnwidth}{!}{
\subfigure[$\hat{\Pi}_{c}$ for column nodes side]{\includegraphics[width=1\textwidth]{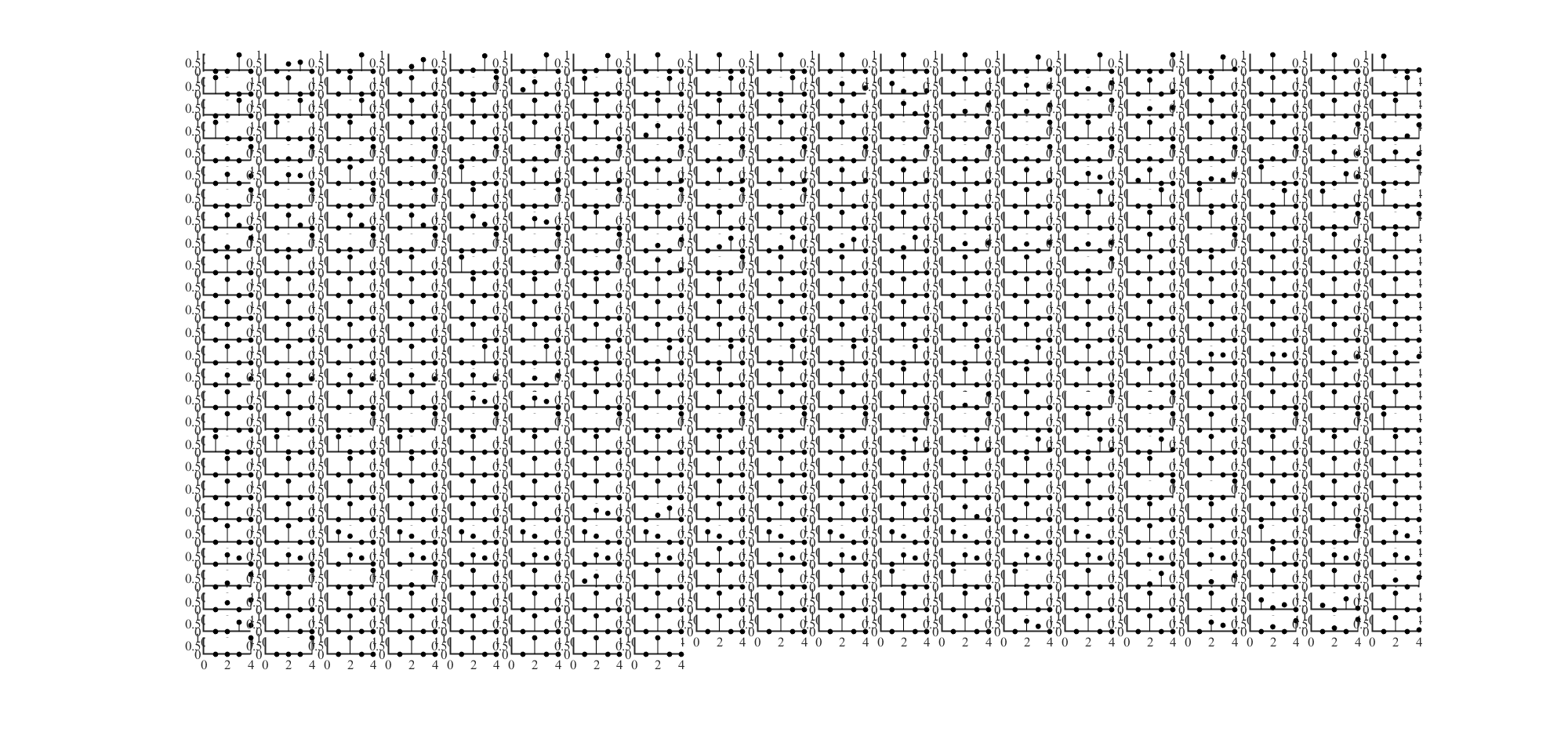}}
}
\caption{The estimated mixed membership matrices $\hat{\Pi}_{r}$ and $\hat{\Pi}_{c}$ detected by DiSP for Unicode languages network. For panel (a), each sub-panel denotes a membership vector of a country; for panel (b), each sub-panel denotes a membership vector of a language; we order the communities 1 to 4 on the X axis, and the node's membership belonging to each cluster is on the Y axis.}
\label{unicodePirc} 
\end{figure}

\begin{figure}
\centering
\resizebox{\columnwidth}{!}{
\subfigure[Crisis in a Cloister]{\includegraphics[width=0.25\textwidth]{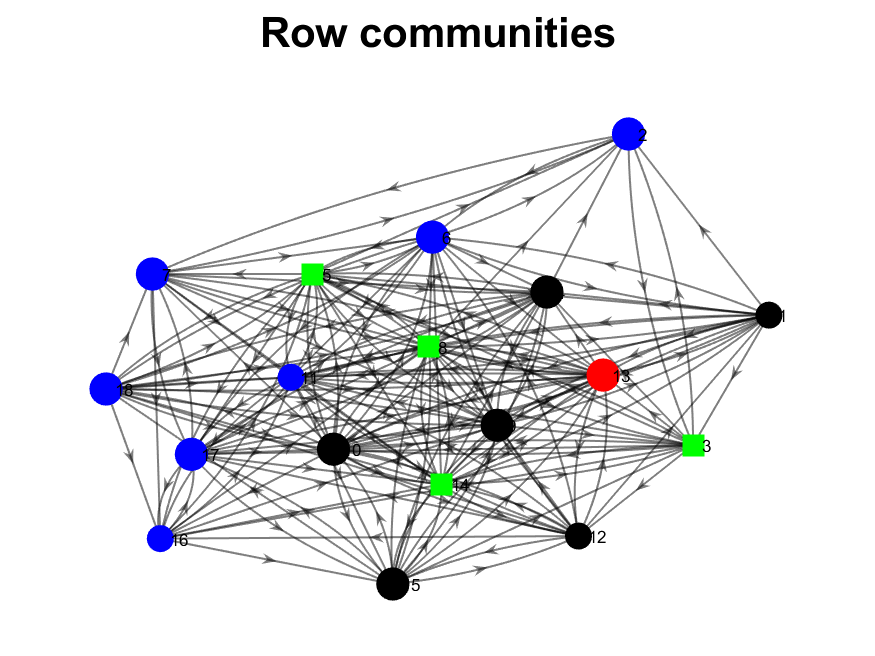}}
\subfigure[Crisis in a Cloister]{\includegraphics[width=0.25\textwidth]{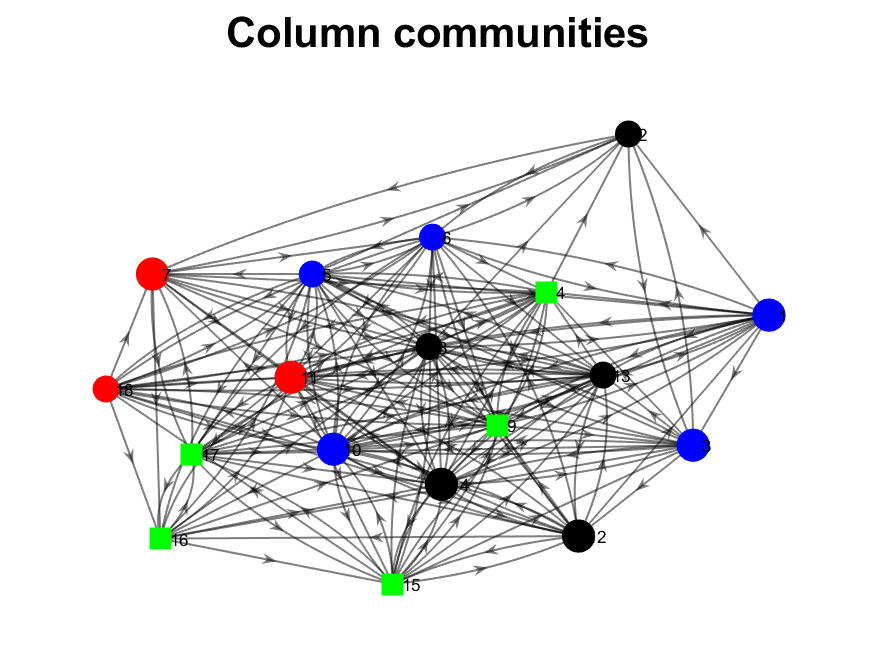}}
\subfigure[Highschool]{\includegraphics[width=0.25\textwidth]{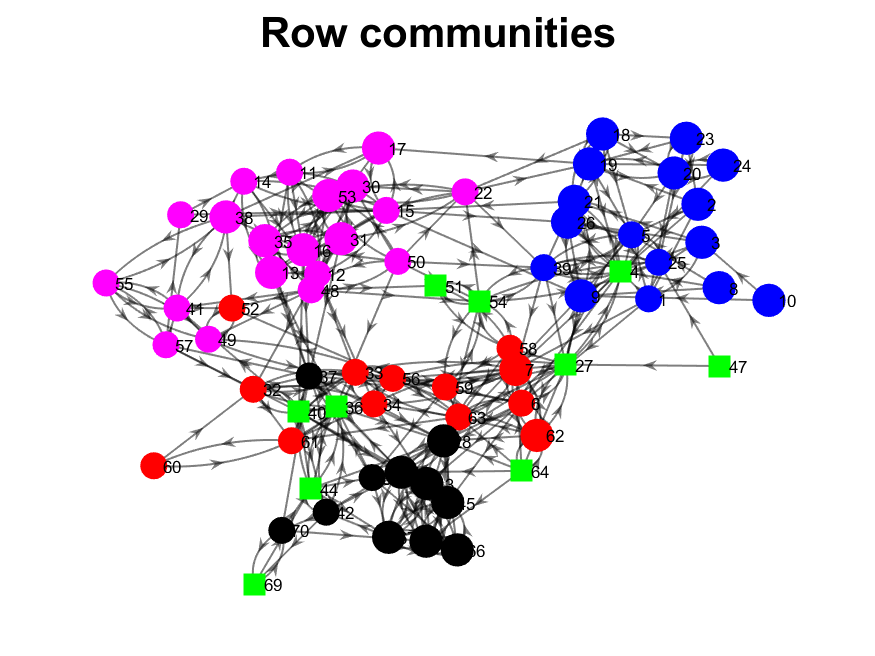}}
\subfigure[Highschool]{\includegraphics[width=0.25\textwidth]{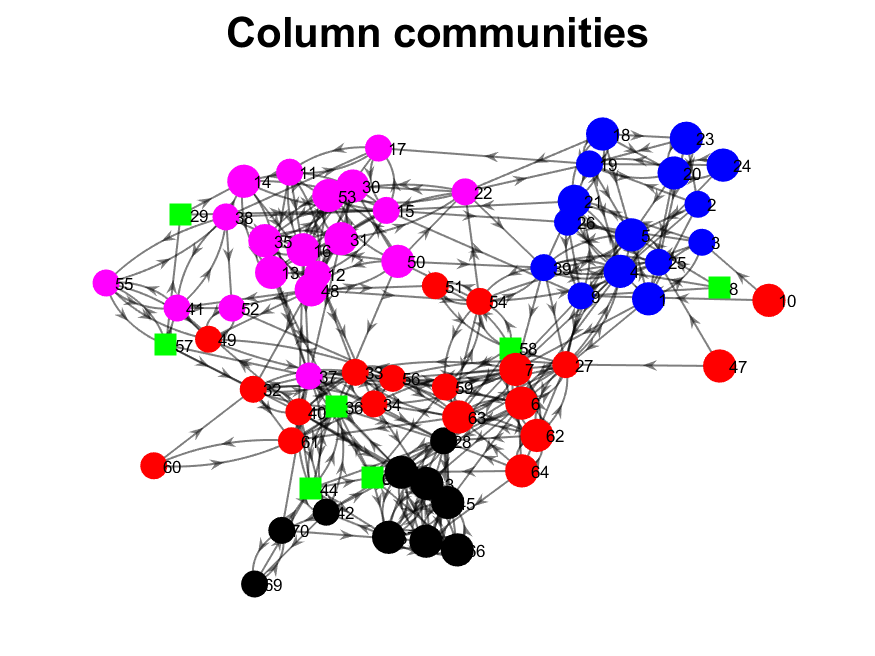}}
}
\caption{Row and column communities are detected by DiSP for Crisis in a Cloister and Highschool. Colors indicate communities, green square indicate highly mixed nodes and highly pure nodes are enlarged, where the row and column communities are obtained by $\hat{\mathcal{C}}_{r}$ and $\hat{\mathcal{C}}_{c}$, respectively. For visualization, we do not show edge weights here.}
\label{NetDiReal} 
\end{figure}

For the visualization of community membership of each node, we show the estimated membership matrices $\hat{\Pi}_{r}$ and $\hat{\Pi}_{c}$ detected by DiSP for three small scale networks, Crisis in a Cloister, Highschool, and Unicode languages in Figures \ref{crisisPirc}, \ref{highschoolPirc}, and \ref{unicodePirc}, respectively. For the visualization of the community structure of both row nodes side and column nodes side, Figure \ref{NetDiReal} depicts row and column communities identified by DiSP on Crisis in a Cloister and Highschool.
\section{Conclusion and future work}\label{sec7}
In this paper, we investigate the problem of estimating community membership in overlapping bipartite weighted networks. A novel model, named the Bipartite Mixed Membership Distribution-Free (BiMMDF) model, is proposed. An efficient spectral algorithm with a theoretical guarantee of estimation consistency is used to infer the community membership for networks generated from BiMMDF. The separation condition of BiMMDF for different distributions is analyzed in Examples \ref{Bernoulli}-\ref{Signed}. We also model large-scale bipartite weighted networks with many missing edges by combining BiMMDF with a model for bipartite un-weighted networks. Our theoretical results are verified by substantial computer-generated bipartite weighted networks. We also apply the algorithm to eight real-world networks with encouraging and interpretable results in understanding community structure of real-world bipartite weighted networks. Our BiMMDF is useful to generate overlapping bipartite weighted networks with true node memberships under different distributions. We expect that BiMMDF will have wide applications in studying the community structure of bipartite weighted networks, just as the mixed membership stochastic blockmodels has been widely studied in recent years.

For future research, first, rigorous methods should be developed to estimate $K$ for overlapping bipartite weighted networks generated from BiMMDF. Actually, more than our BiMMDF, estimating $K$ for all models in Table \ref{table-network-model} is a challenging, interesting, and prospective topic. Second, it is possible to design new algorithms based on the ideas of nonnegative matrix factorization or likelihood maximization, or tensor methods mentioned in \cite{mao2020estimating} to estimate node memberships for networks generated from BiMMDF. Third, like \cite{rohe2011spectral,joseph2016impact,DISIM}, it is possible to design spectral algorithms based on applications of modified Laplacian matrix or regularized Laplacian matrix to fit BiMMDF. Forth, DiSP can be accelerated by some random-projection techniques \citep{zhang2022randomized} to handle large-scale bipartite weighted networks. Fifth,  besides the community structure, identifying the influential nodes to spread information in networks is an appealing topic for researchers. However, BiMMDF can not generate influential nodes in bipartite weighted networks and DiSP only estimates community memberships for overlapping bipartite weighted networks. Thus DiSP can not identify the influential nodes. Though substantial algorithms have been developed to find influential nodes, see algorithms developed in \citep{zhang2019groups,molaei2020identifying,yang2021improved,shang2021identifying,boroujeni2022role,curado2023novel} and references therein, these algorithms only work for un-directed\&directed un-weighted\&weighted networks with positive edge weights. It is a critical issue to develop algorithms to identify the influential nodes in bipartite weighted networks with negative edge weights. We leave these open problems to future work.
\section*{CRediT authorship contribution statement}
\textbf{Huan Qing:} Conceptualization, Methodology, Software, Data
curation, Writing – original draft, Writing – review \& editing. \textbf{Jingli Wang:} Data curation, Writing-reviewing \& editing, Funding acquisition.
\section*{Declaration of competing interest}
The authors declare no competing interests.
\section*{Data availability}
Data and code will be made available on request.
\section*{Acknowledgements}
Qing's work was supported by the High level personal project of Jiangsu Province NO.JSSCBS20211218. Wang's work was supported by the Fundamental Research Funds for the Central Universities, Nankai Univerity, 63221044 and the National Natural Science Foundation of China (Grant 12001295).
\bibliographystyle{model5-names}\biboptions{authoryear}
\bibliography{refBiMMDF}
\appendix
\section{Proof of theoretical results for DiSP}
\subsection{Proof of Theorem \ref{Main}}
\begin{proof}
Let $H_{\hat{U}}=\hat{U}'U$, and $H_{\hat{U}}=U_{H_{\hat{U}}}\Sigma_{H_{\hat{U}}}V'_{H_{\hat{U}}}$ be the top-K SVD of $H_{\hat{U}}$. Define $\mathrm{sgn}(H_{\hat{U}})=U_{H_{\hat{U}}}V'_{H_{\hat{U}}}$.
Let $H_{\hat{V}}=\hat{V}'V$, and $H_{\hat{V}}=U_{H_{\hat{V}}}\Sigma_{H_{\hat{V}}}V'_{H_{\hat{V}}}$ be the top-K SVD of $H_{\hat{V}}$. Define $\mathrm{sgn}(H_{\hat{V}})=U_{H_{\hat{V}}}V'_{H_{\hat{V}}}$. Based on our model BiMMDF, Assumption \ref{a1}, and Condition \ref{c1}, we have below results:
\begin{itemize}
  \item $\mathbb{E}[A(i,j)-\Omega(i,j)]=0$ under BiMMDF.
  \item $\mathbb{E}[(A(i,j)-\Omega(i,j))^{2}]\leq \rho\gamma$.
  \item Let $\mu=\mathrm{max}(\frac{n_{r}\|U\|^{2}_{2\rightarrow\infty}}{K},\frac{n_{c}\|V\|^{2}_{2\rightarrow\infty}}{K})$ be the incoherence parameter defined in Definition 3.1. \cite{chen2021spectral}. By Lemma 8 \cite{qing2021DiMMSB} and Condition \ref{c1}, we have $\mu=O(1)$.
  \item Let $c_{b}=\frac{\tau}{\sqrt{\rho\gamma \mathrm{min}(n_{r},n_{c})/(\mu \mathrm{log}(n_{r}+n_{c}))}}$. By Condition \ref{c1}, we have $\mathrm{min}(n_{r},n_{c})=O(\mathrm{max}(n_{r},n_{c}))$. By the fact that $\mu=O(1)$ given in the last bullet and Assumption \ref{a1}, we have $c_{b}\leq O(1)$.
  \item By $\kappa(P)=O(1)$ in Condition \ref{c1} and Lemma 10 \cite{qing2021DiMMSB}, we have $\kappa(\Omega)=O(1)$.
\end{itemize}
The first four bullets suggest that the settings and assumptions of Theorem 4.4 \cite{chen2021spectral} are satisfied, so by Theorem 4.4. \cite{chen2021spectral}, with probability at least $1-O(\frac{1}{(n_{r}+n_{c})^{5}})$, we have
\begin{align*}
\mathrm{max}(\|\hat{U}\mathrm{sgn}(H_{\hat{U}})-U\|_{2\rightarrow\infty},\|\hat{V}\mathrm{sgn}(H_{\hat{V}})-V\|_{2\rightarrow\infty})&\leq C\frac{\sqrt{\rho\gamma K\mathrm{log}(n_{r}+n_{c})})}{\sigma_{K}(\Omega)},
\end{align*}
provided that $\sigma_{K}(\Omega)\gg \sqrt{\rho\gamma(n_{r}+n_{c})\mathrm{log}(n_{r}+n_{c})}$.

For convenience, let $\varpi=\mathrm{max}(\|\hat{U}\hat{U}'-UU'\|_{2\rightarrow\infty}, \|\hat{V}\hat{V}'-VV'\|_{2\rightarrow\infty})$ be the row-wise singular vector error. Since $\|\hat{U}\hat{U}'-UU'\|_{2\rightarrow\infty}\leq2\|U-\hat{U}\mathrm{sgn}(H_{\hat{U}})\|_{2\rightarrow\infty}$ and $\|\hat{V}\hat{V}'-VV'\|_{2\rightarrow\infty}\leq2\|V-\hat{V}\mathrm{sgn}(H_{\hat{V}})\|_{2\rightarrow\infty}$, we have
\begin{align*}	
\varpi\leq C\frac{\sqrt{\rho\gamma K\mathrm{log}(n_{r}+n_{c})})}{\sigma_{K}(\Omega)}.
\end{align*}
Lemma 10 \cite{qing2021DiMMSB} gives $\sigma_{K}(\Omega)\geq\rho\sigma_{K}(P)\sigma_{K}(\Pi_{r})\sigma_{K}(\Pi_{c})$, so we have
\begin{align*}	
\varpi\leq C\frac{\sqrt{\gamma K\mathrm{log}(n_{r}+n_{c})}}{\sigma_{K}(P)\sigma_{K}(\Pi_{r})\sigma_{K}(\Pi_{c})\sqrt{\rho}}.	
\end{align*}
By Assumption \ref{a1}, we have $\sigma_{K}(\Pi_{r})=O(\sqrt{\frac{n_{r}}{K}})$ and $\sigma_{K}(\Pi_{c})=O(\sqrt{\frac{n_{c}}{K}})$, which gives
\begin{align}\label{rowwise}	
\varpi\leq C\frac{K^{1.5}\sqrt{\gamma \mathrm{log}(n_{r}+n_{c})}}{\sigma_{K}(P)\sqrt{\rho n_{r}n_{c}}}.	
\end{align}
By Theorem 2 \cite{qing2021DiMMSB} where the proof is distribution-free, there exist two permutation matrices $\mathcal{P}_{r},\mathcal{P}_{c}\in\mathbb{R}^{K\times K}$ such that for $i\in[n_{r}],j\in[n_{c}]$,
\begin{align}\label{MainBoundActually}	&\|e'_{i}(\hat{\Pi}_{r}-\Pi_{r}\mathcal{P}_{r})\|_{1}=O(\varpi\kappa(\Pi'_{r}\Pi_{r})K\sqrt{\lambda_{1}(\Pi'_{r}\Pi_{r})}),\|e'_{j}(\hat{\Pi}_{c}-\Pi_{c}\mathcal{P}_{c})\|_{1}=O(\varpi\kappa(\Pi'_{c}\Pi_{c})K\sqrt{\lambda_{1}(\Pi'_{c}\Pi_{c})}).
\end{align}
By Assumption \ref{a1} and Equation (\ref{rowwise}), we have
\begin{align*}	&\|e'_{i}(\hat{\Pi}_{r}-\Pi_{r}\mathcal{P}_{r})\|_{1}=O(\frac{K^{2}\sqrt{\gamma \mathrm{log}(n_{r}+n_{c})}}{\sigma_{K}(P)\sqrt{\rho n_{c}}}),\|e'_{j}(\hat{\Pi}_{c}-\Pi_{c}\mathcal{P}_{c})\|_{1}=O(\frac{K^{2}\sqrt{\gamma \mathrm{log}(n_{r}+n_{c})}}{\sigma_{K}(P)\sqrt{\rho n_{r}}}).
\end{align*}
\begin{rem}
Without Assumption \ref{a1}, the error bounds of DiSP are always the same as that of Equation (\ref{MainBoundActually}). In this paper, we consider Assumption \ref{a1} mainly for theoretical convenience.
\end{rem}
\end{proof}
\subsection{Proof of Corollary \ref{PhaseTransistion}}
\begin{proof}
For $BiMMDF(n,2,\Pi_{r},\Pi_{c},\alpha_{\mathrm{in}},\alpha_{\mathrm{out}},\mathcal{F})$, since $K=2$, by basic algebra, we have $\sigma_{K}(\rho P)=\sigma_{2}(\rho P)=||p_{\mathrm{in}}|-|p_{\mathrm{out}}||=\rho\sigma_{2}(P)$. Recall that we let $\mathrm{max}_{k,l\in[K]}|P(k,l)|=1$ in Definition \ref{BiMMDF}, we have $\mathrm{max}_{k,l\in[K]}|\rho P(k,l)|=\rho=\mathrm{max}(|p_{\mathrm{in}}|,|p_{\mathrm{out}}|)$, which gives that $\frac{||p_{\mathrm{in}}|-|p_{\mathrm{out}}||}{\sqrt{\mathrm{max}(|p_{\mathrm{in}}|,|p_{\mathrm{out}}|)}}=\sqrt{\rho}\sigma_{K}(P)$ should shrink slower than $\sqrt{\frac{\gamma\mathrm{log}(n)}{n}}$ for small error rates with high probability by Theorem \ref{Main}. Since $\frac{||p_{\mathrm{in}}|-|p_{\mathrm{out}}||}{\sqrt{\mathrm{max}(|p_{\mathrm{in}}|,|p_{\mathrm{out}}|)}}=\frac{||\alpha_{\mathrm{in}}|-|\alpha_{\mathrm{out}}||}{\sqrt{\mathrm{max}(|\alpha_{\mathrm{in}}|,|\alpha_{\mathrm{out}}|)}}\sqrt{\frac{\mathrm{log}(n)}{n}}$, $\frac{||\alpha_{\mathrm{in}}|-|\alpha_{\mathrm{out}}||}{\sqrt{\mathrm{max}(|\alpha_{\mathrm{in}}|,|\alpha_{\mathrm{out}}|)}}$ should shrink slower than $\sqrt{\gamma}$ for small error rates with high probability, i.e.,
\begin{align}\label{Sharp0}
\frac{||\alpha_{\mathrm{in}}|-|\alpha_{\mathrm{out}}||}{\sqrt{\mathrm{max}(|\alpha_{\mathrm{in}}|,|\alpha_{\mathrm{out}}|)}}\gg\sqrt{\gamma}.
\end{align}

Meanwhile, since $\rho=\mathrm{max}(|p_{\mathrm{in}}|, |p_{\mathrm{out}}|)=\frac{\mathrm{log}(n)}{n}\mathrm{max}(|\alpha_{\mathrm{in}}|,
|\alpha_{\mathrm{out}}|)$ under $BiMMDF(n,2,\Pi_{r},\Pi_{c},\alpha_{\mathrm{in}},\alpha_{\mathrm{out}},\mathcal{F})$, we have $\frac{\mathrm{log}(n)}{n}\mathrm{max}(|\alpha_{\mathrm{in}}|,
|\alpha_{\mathrm{out}}|)\gamma n\geq \tau^{2}\mathrm{log}(2n)$ by Assumption \ref{a1}, which gives
\begin{align}\label{Sharp1}
 \gamma\mathrm{max}(|\alpha_{\mathrm{in}}|,
|\alpha_{\mathrm{out}}|)\geq \tau^{2}\frac{\mathrm{log}(2n)}{\mathrm{log}(n)}.
\end{align}
Now, when Equation (\ref{Sharp1}) holds, Equation (\ref{Sharp0}) can be released as
\begin{align}\label{Sharp2}
||\alpha_{\mathrm{in}}|-|\alpha_{\mathrm{out}}||\gg\tau\sqrt{\frac{\mathrm{log}(2n)}{\mathrm{log}(n)}}.
\end{align}
Since $\frac{\mathrm{log}(2n)}{\mathrm{log}(n)}=\frac{\mathrm{log}(2)}{\mathrm{log}(n)}+1=1+o(1)$ when $n$ is not too small, Equations (\ref{Sharp1}) and (\ref{Sharp2}) can be released as
\begin{align}\label{SharpGeneral}
\gamma\mathrm{max}(|\alpha_{\mathrm{in}}|,
|\alpha_{\mathrm{out}}|)\geq \tau^{2}+o(1)\mathrm{~and~}||\alpha_{\mathrm{in}}|-|\alpha_{\mathrm{out}}||\gg\tau.
\end{align}
Next, we show that the lower bound requirement of $\sigma_{K}(\Omega)$ in Theorem \ref{Main} and $\kappa(P)=O(1)$ in Condition \ref{c1} hold naturally as long as Equation (\ref{SharpGeneral}) holds. By Lemma 10 \cite{qing2021DiMMSB}, we know that $\sigma_{2}(\Omega)\geq\rho\sigma_{2}(P)\sigma_{2}(\Pi_{r})\sigma_{2}(\Pi_{c})$ holds under $BiMMDF(n,2,\Pi_{r},\Pi_{c},\alpha_{\mathrm{in}},\alpha_{\mathrm{out}},\mathcal{F})$ without other assumptions. Therefore, to guarantee that the condition $\sigma_{2}(\Omega)\gg \sqrt{\rho\gamma n\mathrm{log}(2n)}$ in Theorem \ref{Main} always holds when $n_{r}=n_{c}=n$ and $K=2$ under $BiMMDF(n,2,\Pi_{r},\Pi_{c},\alpha_{\mathrm{in}},\alpha_{\mathrm{out}},\mathcal{F})$,  we need $\rho\sigma_{2}(P)\sigma_{2}(\Pi_{r})\sigma_{2}(\Pi_{c})\gg \sqrt{\rho\gamma n\mathrm{log}(2n)}=\sqrt{\rho\gamma n(\mathrm{log}(2)+\mathrm{log}(n))}=O(\sqrt{\rho\gamma n\mathrm{log}(n)})$, i.e.,
\begin{align}\label{sigmaKPlowerbound}
\sigma_{2}(P)\gg \sqrt{\frac{\gamma n\mathrm{log}(n)}{\rho \lambda_{2}(\Pi'_{r}\Pi_{r})\lambda_{2}(\Pi'_{c}\Pi_{c})}}.
\end{align}
Since $\lambda_{2}(\Pi'_{r}\Pi_{r})=O(\frac{n}{2})=O(n),
\lambda_{2}(\Pi'_{c}\Pi_{c})=O(\frac{n}{2})=O(n)$ under $BiMMDF(n,2,\Pi_{r},\Pi_{c},\alpha_{\mathrm{in}},\alpha_{\mathrm{out}},\mathcal{F})$, Equation (\ref{sigmaKPlowerbound}) gives that $\sigma_{2}(P)$ should shrink slower than $\sqrt{\frac{\gamma\mathrm{log}(n)}{\rho n}}$, which matches with the consistency requirement on $\sigma_{2}(P)$ obtained from Theorem \ref{Main} under $BiMMDF(n,2,\Pi_{r},\Pi_{c},\alpha_{\mathrm{in}},\alpha_{\mathrm{out}},\mathcal{F})$. Therefore, under $BiMMDF(n,2,\Pi_{r},\Pi_{c},\alpha_{\mathrm{in}},\alpha_{\mathrm{out}},\mathcal{F})$, the lower bound requirement on $\sigma_{K}(\Omega)$ in Theorem \ref{Main} holds naturally as long as Equation (\ref{SharpGeneral}) holds. Finally, since $\kappa(P)=\kappa(\rho P)=\frac{||\alpha_{\mathrm{in}}|+|\alpha_{\mathrm{out}}||}{||\alpha_{\mathrm{in}}|-|\alpha_{\mathrm{out}}||}$, we see that $\kappa(P)=O(1)$ in Condition \ref{c1} holds immediately when Equation (\ref{SharpGeneral}) holds.
\end{proof}
\section{Extra simulation results}
\begin{figure}
\centering
\resizebox{\columnwidth}{!}{
\subfigure[]{\includegraphics[width=0.33\textwidth]{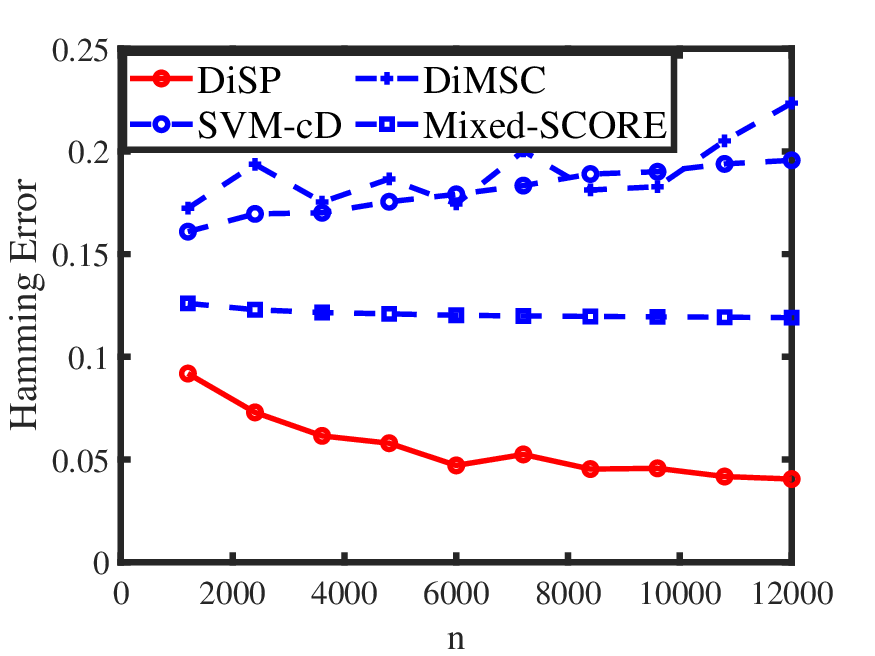}}
\subfigure[]{\includegraphics[width=0.33\textwidth]{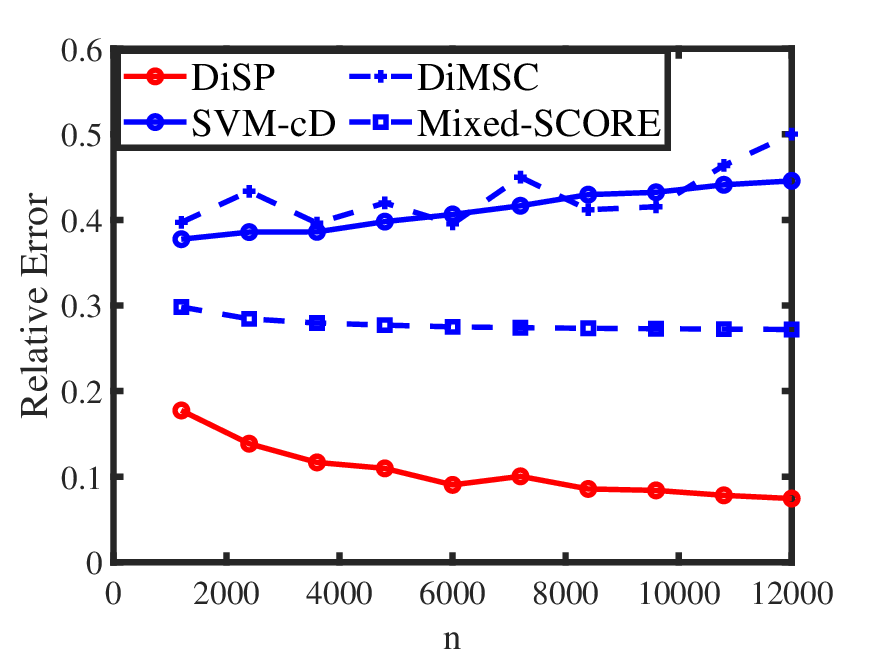}}
\subfigure[]{\includegraphics[width=0.33\textwidth]{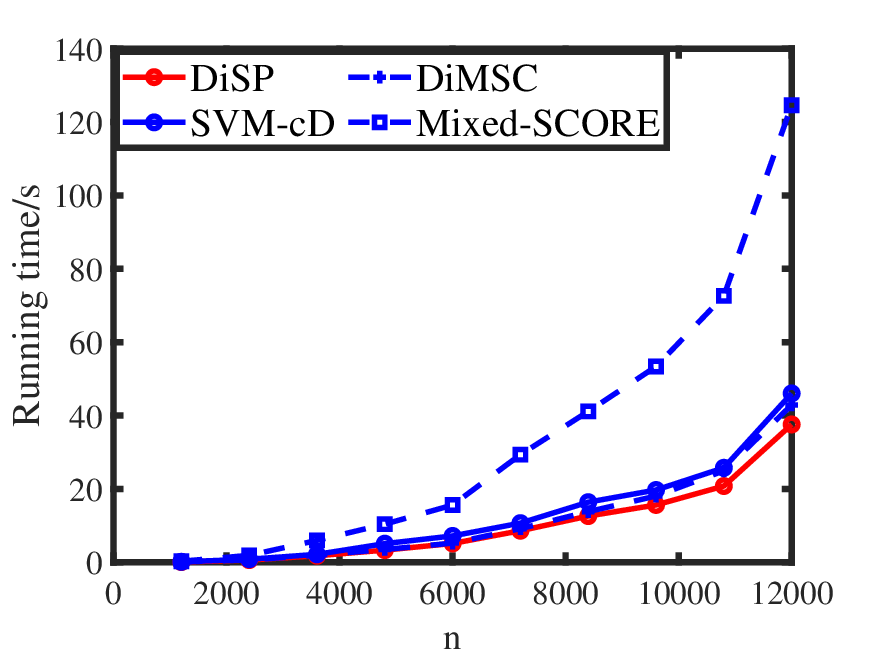}}
}
\caption{Panel (a): Hamming Error against increasing $n$. Panel (b): Relative Error against increasing $n$. Panel (a): Running time against increasing $n$.}
\label{ChangeN} 
\end{figure}

\textbf{Changing $n$}: When $n_{r}=n_{c}=n$, we record the running time for each approach when $n$ increases. For simplicity, we only consider Normal distribution here. Let $K=2, \rho=1, p=0.9, \sigma^{2}_{A}=1, n_{r,0}=n/4, n_{c,0}=n/3$, and all mixed nodes have mixed membership $(1/2,1/2)$. Set the connectivity matrix $P$ as $P_{2}$ in Section \ref{sec5Synthetic}. We vary $n$ in the range $\{1200,2400,\ldots,12000\}$. We report the averaged Hamming Error, the averaged Relative Error, and the averaged running time over 50 repetitions for each $n$ for each method. We see in Figure \ref{ChangeN} that DiSP outperforms its competitors both in estimation errors and running time. Meanwhile, for a bipartite weighted network with 12000 row nodes and 12000 column nodes, DiSP takes around 40 seconds to process a standard personal computer (Thinkpad X1 Carbon Gen 8) using MATLAB R2021b.
\end{document}